\documentclass[reprint,accepted=2023-02-16,a4paper]{quantumarticle}
\pdfoutput=1

\usepackage[authoryear, round]{natbib}
\renewcommand{\cite}{\citep}

\usepackage{qip}
\usepackage{graphicx}
\usepackage[colorlinks=true,citecolor=blue,urlcolor=blue,linkcolor=black]{hyperref}
\usepackage{amsfonts}
\usepackage{amsmath,amssymb}
\usepackage{dsfont}
\usepackage{euscript}
\usepackage{float}
\usepackage{amsthm}
\usepackage{bbm}
\usepackage{ragged2e}
\usepackage{color}
\usepackage{xcolor}
\usepackage{pgf}
\usepackage{color,soul}
\usepackage{tikz}
\usetikzlibrary{backgrounds}
\usepackage{float}
\usepackage{multirow}
\usepackage{array}
\usepackage{mathtools}
\usepackage{comment}
\usepackage[utf8]{inputenc}

\usepackage{array}
\usepackage{booktabs}
\usepackage{pifont}
\newcommand{\cmark}{\ding{51}}%
\newcommand{\xmark}{\ding{55}}%

\newcommand{\sA}{\mathsf{A}}
\newcommand{\sB}{\mathsf{B}}
\newcommand{\sX}{\mathsf{X}}
\newcommand{\sY}{\mathsf{Y}}
\newcommand{\sE}{\mathsf{E}}

\newcommand{\iwp}[1]{}
\newcommand{\gkt}[1]{}
\newcommand{\cl}[1]{}
\newcommand{\sg}[1]{}
\newcommand{\jk}[1]{}
\newcommand{\et}[1]{}
\newcommand{\revision}[1]{{\color{black} #1}}

\newcommand{\expectation}[1]{\left\langle #1 \right\rangle}
\newcommand{\ceil}[1]{\left\lceil #1 \right\rceil}

\newcommand{\abs}[1]{\left\lvert #1 \right\rvert}
\newcommand{\tr}{\mathrm{tr}}
\newcommand{\vb}[1]{\mathbf{#1}}
\newcommand{\transcript}{\vb{P}}
\newcommand{\norm}[1]{\left\lVert#1\right\rVert}

\begin{document}
\title{Security of device-independent quantum key distribution protocols: a review}

\author{Ignatius W. Primaatmaja}
\affiliation{Department of Electrical \& Computer Engineering, National University of Singapore, Singapore}
\affiliation{Centre for Quantum Technologies, National University of Singapore, Singapore}

\author{Koon Tong Goh}
\affiliation{Department of Electrical \& Computer Engineering, National University of Singapore, Singapore}

\author{Ernest Y.-Z. Tan}
\affiliation{Institute for Quantum Computing and Department of Physics and Astronomy, University of Waterloo, Canada}

\author{John T.-F. Khoo}
\affiliation{Department of Electrical \& Computer Engineering, National University of Singapore, Singapore}
\affiliation{Department of Computer Science, National University of Singapore, Singapore}

\author{Shouvik Ghorai}
\affiliation{Department of Electrical \& Computer Engineering, National University of Singapore, Singapore}

\author{Charles Lim}
\affiliation{Department of Electrical \& Computer Engineering, National University of Singapore, Singapore}
\affiliation{Centre for Quantum Technologies, National University of Singapore, Singapore}
\affiliation{Global Technology Applied Research, JPMorgan Chase \& Co, Singapore}

\begin{abstract}
Device-independent quantum key distribution (DI-QKD) is often seen as the ultimate key exchange protocol in terms of security, as it can be performed securely with uncharacterised black-box devices. The advent of DI-QKD closes several loopholes and side-channels that plague current QKD systems. While implementing DI-QKD protocols is technically challenging, there have been recent proof-of-principle demonstrations, resulting from the progress made in both theory and experiments. In this review, we will provide an introduction to DI-QKD, an overview of the related experiments performed, and the theory and techniques required to analyse its security. We conclude with an outlook on future DI-QKD research.
\end{abstract}
\maketitle

\tableofcontents

\section{Introduction}

Cryptosystems today are designed with the requirement that they should be secure even if everything about the system is publicly known except for the input key. This idea is known as \emph{Kerckhoffs’s principle}~\cite{Petitcolas2011}\footnote{In Shannon's seminal work on secrecy~\cite{Shannon1949}, the same design principle is expressed in the assumption ``the enemy knows the system being used". }. However, a sophisticated enemy today can do more than just \emph{knowing the system} --- the enemy could also exploit his system knowledge to create side-channels through active engineering efforts. Indeed, these vulnerabilities can be introduced through a wide variety of avenues, from implementation flaws to supply chain attacks. The possibility of performing cryptography with untrustworthy devices is a tantalising solution to these challenges.

At first glance, this task seems impossible, since the security of cryptography is typically analysed in very concrete terms, assuming secure and characterised devices. Remarkably, it turns out the answer can be found in the foundations of quantum theory. This approach, called \emph{device-independent quantum cryptography}, uses \emph{nonlocality}~\cite{bell1964} to certify the security of cryptosystems. 
The basic idea is to randomly subject the quantum devices to a \emph{Bell test}, where a Bell inequality is evaluated using the input-output measurement data, and the degree of nonlocality is quantified by the observed Bell inequality violation.  
The higher the violation, the lower the possible degree of correlation with any other system due to the monogamy of nonlocality~\cite{Toner2009}. Importantly, this trade-off can be formalised, and one can rigorously bound the adversary's information about the devices using the observed Bell violation.

The beauty of quantifying security using nonlocality is that the method is agnostic to the physics of the devices. In particular, the only information needed to evaluate the Bell inequality is the joint input-output probability distribution of the devices. Therefore, the physics behind the distribution is not relevant and the devices can be seen as black boxes that return some output when given an input.

Device-independent quantum cryptography also has a nice physical interpretation in the context of \emph{self-testing}~\cite{mayers1998quantum,mayers2004self}\footnote{For a recent review of the subject, we refer the interested reader to~\cite{supic2020selftesting}.}.
This is an elegant phenomenon where, from the input-output distribution of a system alone, it can be deduced that the underlying quantum system of an uncharacterised device is, up to some local isometries, close to some ideal system. Indeed, when seen in this light, it is not surprising that one can also perform cryptography with uncharacterised devices --- their input-output statistics can indicate that the actual system's physics is very close to the ideal secure system's physics.\footnote{The connection between self-testing and quantum cryptography was first proposed by Mayers \& Yao in \cite{mayers1998quantum}, where they proposed a self-testing approach for quantum key distribution. Interestingly, the work of Mayers and Yao did not draw any connection to nonlocality despite their similarities.}

In this review article, the cryptographic task under consideration is the key exchange problem, in which two honest users establish a shared secret key over an insecure channel. The use of quantum cryptography to perform key exchange is referred to as \emph{quantum key distribution} (QKD). \textit{Device-independent} QKD (DI-QKD) specifically uses the methods of device-independent quantum cryptography to prove security, in contrast to conventional \textit{device-dependent} QKD (DD-QKD), which requires the faithfulness of its implementation to some ideal specifications in order to be secure.
Consequently, the possibility of miscalibrations within a DD-QKD system, which would cause its devices to deviate from their specifications, implies a larger attack surface that can be exploited by the adversary. The device-independent approach to QKD eliminates such concerns by accounting for the devices' faithfulness organically within its framework. This review paper aims to summarise recent theoretical developments in DI-QKD, both in protocol design and in techniques for analysing the security of QKD in the device-independent framework.

In the next two subsections, we shall give a brief introduction to QKD and the device-independent framework. For a more in-depth overview of the security of QKD, we refer the reader to~\cite{scarani2009security, xu2020secure}. Additionally, as the device-independent framework relies crucially on nonlocality, the reader can refer to the detailed review of Bell nonlocality in~\cite{brunner2014bell}. 
A more pedagogical presentation of background material can also be found in~\cite{tan_thesis, valerio_book}.

\subsection{Basics of quantum key distribution}
Quantum key distribution~\cite{scarani2009security, xu2020secure} is a key exchange protocol between two remote users via an \emph{insecure} quantum channel, where the adversary can perform arbitrary quantum operations on transmitted quantum systems, and an \emph{authenticated} classical channel, where messages can be read by the adversary but not modified. Under a well-defined set of assumptions (Section~\ref{sec: assumptions}), the secret keys exchanged using QKD can be proven to be \textit{information-theoretically secure}, which implies in particular that the optimal strategy for an adversary with unbounded computational resources to recover the exchanged key is a uniformly random guess. This means that the security of QKD cannot be threatened by any technological advancement or new algorithm, unlike classical cryptography, which relies on the computational difficulty of solving certain problems. The analogous task in classical cryptography, which is to establish an information-theoretically secure key using an insecure classical channel and an authenticated classical channel, is not possible. Such a key can be pre-agreed upon, for example by the parties physically meeting up to exchange this key, but this pre-agreement itself requires a secure classical channel, such as that of physical contact, which makes its practical deployment rather cumbersome. Quantum cryptography is strictly necessary to exchange an information-theoretically secure key over an insecure channel, and offers a more practical means of achieving this ultimate level of security.

In this review, we shall focus on QKD in the sense of key exchange between two users only. The extension of QKD to three or more users is known as quantum \textit{conference key agreement} (CKA). For a review on this topic, we refer the reader to~\cite{murta2020qcka}. Analogously, when the devices are treated as black boxes, the task is known as device-independent CKA (DI-CKA)~\cite{ribeiro2018fully,grasselli2021entropy}. While this task would not be considered in this review, many notions and techniques presented here are also relevant in DI-CKA.

\subsubsection{Generic setting}
\begin{figure}[ht]
\centering
\includegraphics[width=0.48\textwidth]{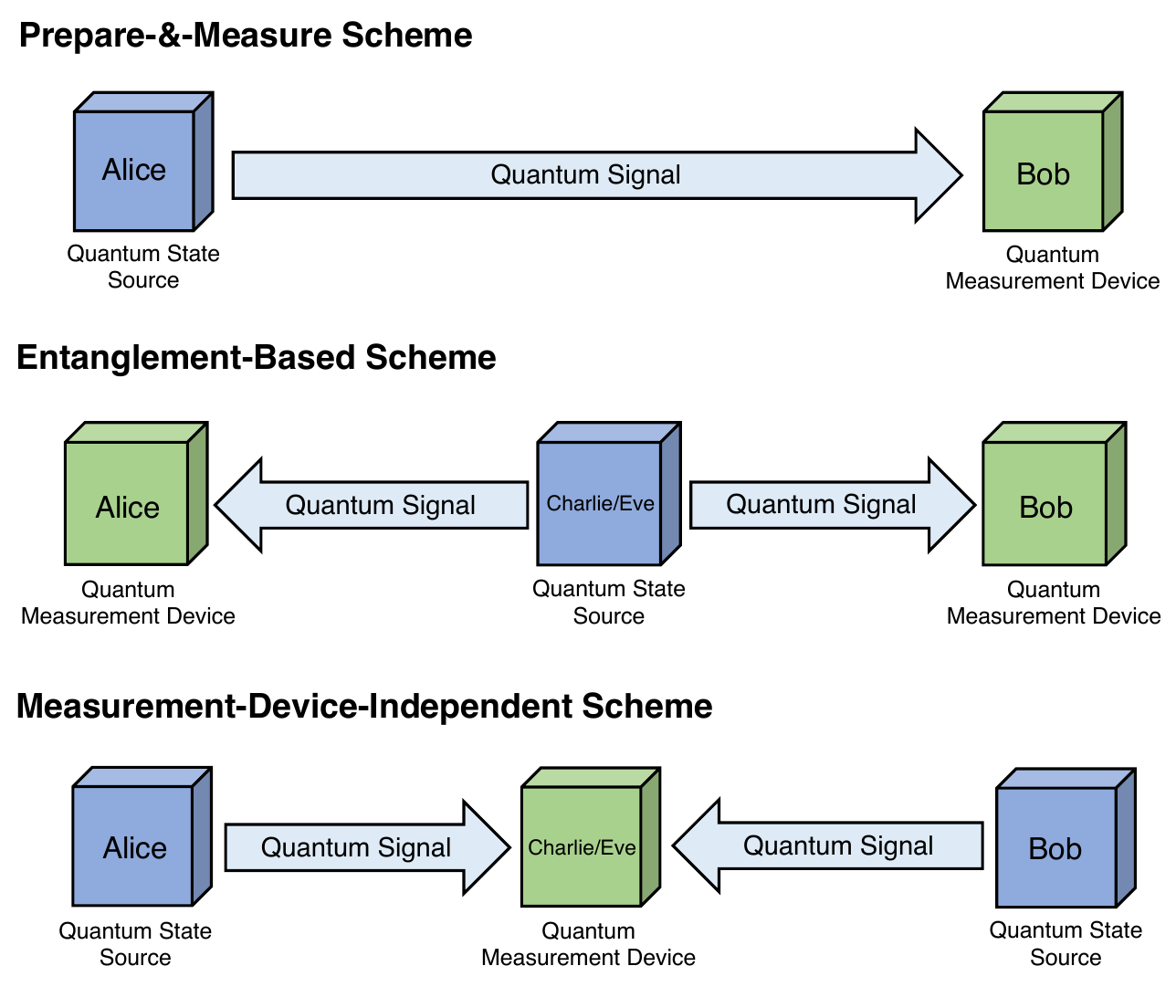}
\caption{Various QKD schemes: Prepare-\&-measure (P\&M), entanglement-based (EB) and measurement-device-independent (MDI) protocols.}
\end{figure}

The first QKD protocols were in the form of the prepare-\&-measure (P\&M) scheme \cite{bennett1984quantum} where one party (Alice) prepares and sends a quantum state to the other party (Bob), who will measure the quantum state received. The entanglement-based (EB) QKD scheme~\cite{ekert1991quantum,bennett1992quantum} was introduced shortly after, where Alice and Bob each measure one part of an entangled system. Subsequently, the measurement-device-independent (MDI) QKD scheme~\cite{lo2012measurement, braunstein2012side} was developed, where Alice and Bob each send a quantum state to an untrusted, and presumably malicious, measurement device. While QKD protocols can take these various forms, the security proofs of P\&M- and MDI-QKD protocols are often performed by reducing them to an equivalent EB-QKD protocol~\cite{bennett1992quantum}. Hence, one can usually speak of the security definition of any QKD protocol in the context of the EB-QKD scheme without loss of generality.
\begin{figure*}[t]
    \centering
    \includegraphics[width = \linewidth]{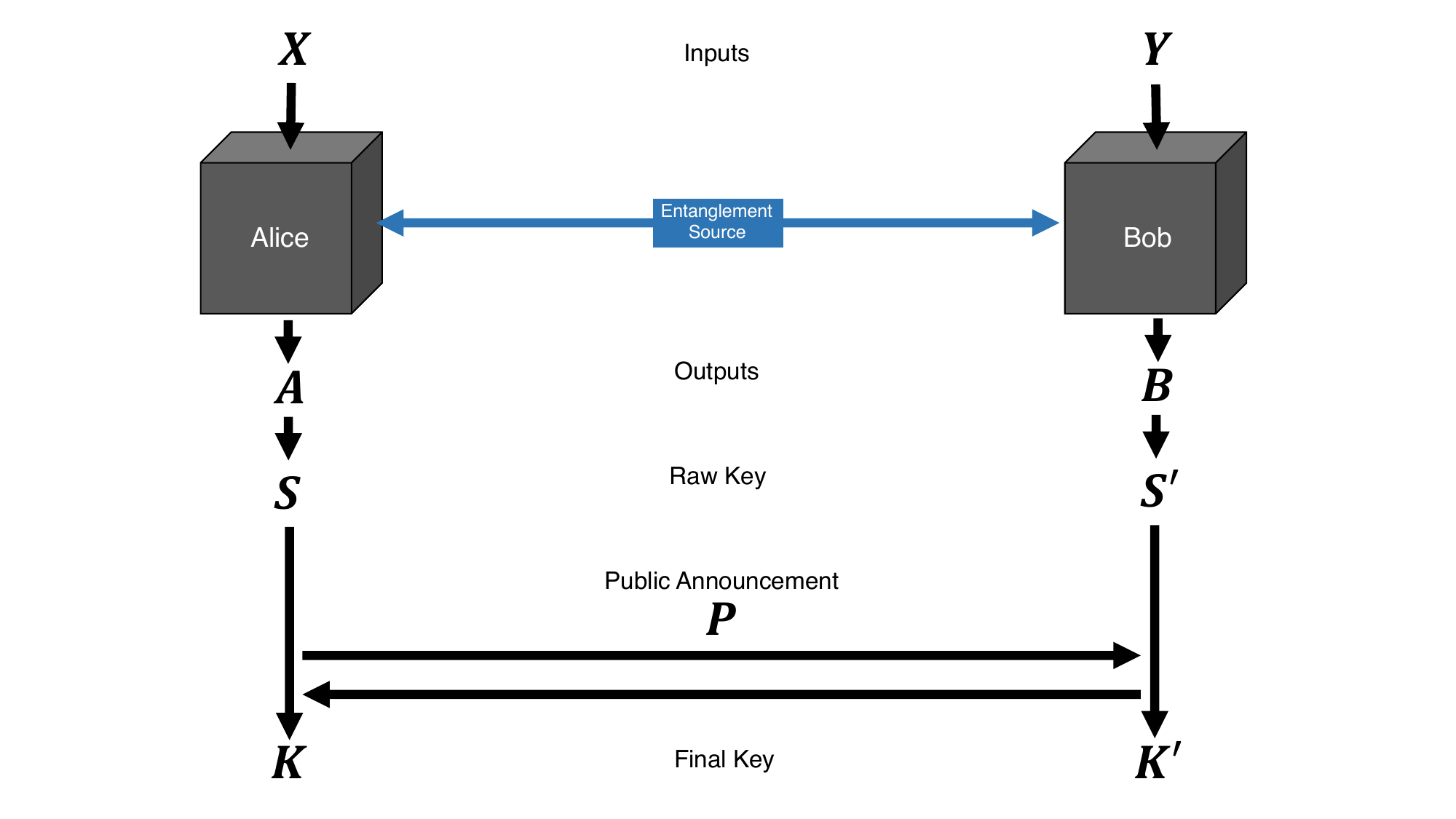}
    \caption{The typical setup for entanglement-based QKD protocols. Alice and Bob receive the inputs, $\vb{X}$ and $\vb{Y}$, to their devices and obtain the outputs $\vb{A}$ and $\vb{Y}$. From these data, they would then create their respective raw key $\vb{S}$ and $\vb{S}'$ (possibly, by performing some classical processing on their raw data). Thereafter, they perform some classical post-processing over an authenticated classical channel on their respective raw key to obtain the final secret key $\vb{K}$ and $\vb{K}'$. In this step, the transcript of the public communication is denoted by $\vb{P}$. Depending on each specific protocol, the classical post-processing can be based on one-way or two-way classical communication.}
    \label{fig: notation}
\end{figure*}

Regardless of whether the protocol employs the P\&M scheme, the EB scheme, or the MDI scheme, a typical QKD protocol can be divided into two layers --- the quantum communication layer (in which quantum information is distributed via the insecure quantum channels to form a pair of raw keys) and the classical post-processing layer (in which the raw keys are converted into secure keys). For the quantum communication layer, we consider a generic EB-QKD protocol \revision{(illustrated in Figure~\ref{fig: notation}, which also summarises our notation)} where Alice's (resp. Bob's) QKD device takes the inputs $\vb{X}$ (resp. $\vb{Y}$) and gives the outputs $\vb{A}$ (resp. $\vb{B}$) by making appropriate measurements (depending on their inputs) on the quantum states 
distributed to the devices. 
In the most conservative scenario, the adversary Eve holds a purification $E$ of the distributed quantum state.
We let the number of rounds (and hence the length for the inputs and outputs of each party) be $n$. 
From their local data and by communicating their inputs for each round to perform sifting, Alice and Bob can then create their raw keys, denoted by $\vb{S}$ and $\vb{S}'$, respectively. 

These raw keys can be thought of as two strings of random numbers that are weakly correlated to each other, and to Eve's quantum side information, $E$. Alice and Bob then use an authenticated (but public) classical channel to announce a random subset of their data for the purpose of estimating the correlation\footnote{We do not specify the measure of correlation here. Many QKD protocols use the quantum bit-error rate as a measure of correlation but other measures of correlation are available. In the context of device-independent QKD, a Bell violation is typically used to measure correlations.} between their data. This step is typically termed \emph{parameter estimation}. Typically, the rounds in which the data are announced for parameter estimation are called ``test rounds'', while the rounds in which the data are kept private (for producing the secret key later) are called the ``generation rounds''. Broadly speaking, there are two ways to choose whether a given round is a test round or a generation round. The first method is to fix the total number of test rounds, and to have Alice and Bob randomly select that number of rounds from their recorded data to serve as the test rounds. The second method is to assign a fixed probability for each round in the protocol to be a test round, and then to announce the data for the rounds which are tagged as test rounds in the parameter estimation step.
In either case, if the exchanged information indicates that there is sufficient correlation between their data, they continue the protocol; otherwise, they abort it.

If they decide to continue the protocol, they use the authenticated classical channel again to perform \emph{information reconciliation}, which is often convenient to analyse in two parts: \emph{error correction} (in which the parties try to make their strings identical) and \emph{error verification} (in which they check whether they succeeded). At the end of this step, they would share a pair of identical strings, but these may still be correlated to Eve's side-information (which now also includes any public announcements that Alice and Bob have made; we denote the transcript of the public classical communication by $\transcript$). To finally obtain a pair of identical strings that are secret (i.e.~uniformly random from Eve's perspective), Alice and Bob perform \textit{privacy amplification}. This can be achieved by randomly choosing a hash function from a suitable family that maps their strings to shorter ones, which we shall denote as $\vb{K}$ and $\vb{K}'$ for Alice and Bob respectively, each with length $\ell$. The secret key rate $r$ is then defined as  $r \coloneqq \ell / n$.

\subsubsection{Security definition}
The qualitative goal of QKD is to distribute a pair of keys securely between two remote users. Given this, it is important to first formalise what ``security'' means in the context of this task. 
To that end, let us first consider what we might reasonably require from an ``ideal'' key distribution protocol. Note that, for any physical QKD implementation, it is always possible for Eve to perform a \textit{denial-of-service} attack --- an attack that prevents the protocol from successfully producing the shared secret keys (e.g.~by always distributing completely insecure states, or by simply blocking the communication channel). In light of this, when considering the ideal key distribution protocol, we should allow some possibility of it aborting. That aside, however, we can reasonably impose that whenever the ideal protocol does not abort, its output to the honest parties should be a pair of identical strings of a fixed length $\ell$, and Eve gets no usable information about these strings.

A full formalisation of this notion in terms of a ``real-versus-ideal-world'' paradigm can be found in e.g.~\cite{portmann14,portmann21}, which also discusses its relationship to \textit{composable security}: the requirement that a protocol remains secure when composed with other protocols. Here, we shall simply state the security condition arising from that formalism, which can be seen to qualitatively correspond to the requirements sketched above. To begin with, let us consider the output of a real QKD protocol carried out with arbitrary states, supplied by Eve. Accounting for the fact that it may abort with some probability (and using the convention that Alice and Bob both set their keys to an abort symbol $\perp$ in that case), it would be in the form
\begin{equation}
    \rho_{\vb{K}\vb{K}'\transcript E} = \ketbra{\perp, \perp}{\perp, \perp}_{\vb{K}\vb{K}'} \otimes \tilde{\rho}_{\transcript E}^\text{abort} + \tilde{\tau}_{\vb{K}\vb{K'}\transcript E},
\end{equation}
where $\tilde{\rho}^\text{abort}$, $\tilde{\tau}$ are sub-normalised states corresponding to the protocol aborting and accepting respectively. The normalisation of $\rho_{\vb{K}\vb{K}'\transcript E}$ gives
\begin{equation}
\begin{split}
    \tr[\tilde{\rho}_{\transcript E}^\text{abort}] &= \Pr[\text{abort}],\\
    \tr[\tilde{\tau}_{\vb{K}\vb{K'}\transcript E}] &= 
    \Pr[\text{accept}] = 
    1 - \Pr[\text{abort}],
\end{split}
\end{equation}
and furthermore the latter state has the form
\begin{equation}
    \tilde{\tau}_{\vb{K}\vb{K'}\transcript E} = \sum_{k,k' \in \{0,1\}^\ell} 
    \ketbra{k,k'}{k,k'}_{\vb{K} \vb{K}'} \otimes \tilde{\tau}^{k,k'}_{\transcript E},
\end{equation}
where each $\tilde{\tau}^{k,k'}$ is the sub-normalised state conditioned on Alice and Bob having key values $k$ and $k'$ respectively.

Then, for a fixed security parameter $\varepsilon \in (0,1)$, a QKD protocol is $\varepsilon$-secure\footnote{This is sometimes instead referred to as $\varepsilon$-sound in later works, e.g.~\cite{portmann14,portmann21}.} if~\cite{renner_thesis}
\begin{equation} \label{eq: eps secure}
    \frac{1}{2} \norm{\rho_{\vb{K}\vb{K}'\transcript E} - \rho^{\text{ideal}}_{\vb{K}\vb{K}'\transcript E}}_1 \leq \varepsilon,
\end{equation}
where $\rho^{\text{ideal}}_{\vb{K}\vb{K}'\transcript E}$ is an ``ideal state''\footnote{Note that $\rho^{\text{ideal}}_{\vb{K}\vb{K}'\transcript E}$ is constructed in terms of the real output state $\rho_{\vb{K}\vb{K}'\transcript E}$; we are not defining a specific $\rho^{\text{ideal}}_{\vb{K}\vb{K}'\transcript E}$ that $\rho_{\vb{K}\vb{K}'\transcript E}$ must be close to. Further explanation of this definition choice can be found in~\cite{portmann14,portmann21}.}
\begin{multline}
    \rho^{\text{ideal}}_{\vb{K}\vb{K}'\transcript E} \coloneqq  \ketbra{\perp, \perp}{\perp, \perp}_{\vb{K}\vb{K}'} \otimes \tilde{\rho}_{\transcript E}^\text{abort} \\
    + \chi^{\ell}_{\vb{K}\vb{K'}} \otimes \tilde{\tau}_{\transcript E},
\end{multline}
with
\begin{equation}
    \chi^{\ell}_{\vb{K}\vb{K'}} \coloneqq \sum_{k \in \{0,1\}^\ell} 2^{-\ell} \ketbra{k,k}{k,k}_{\vb{K}\vb{K'}}
\end{equation}
being a uniformly distributed and perfectly correlated bit-string of length $\ell$. In other words, the output of the real QKD protocol is $\varepsilon$-close in trace distance to some ``ideal state'' that is (apart from the abort component) perfectly correlated between Alice and Bob, uniformly random, and independent from Eve's side-information. 

Noting that the abort components in $\rho_{\vb{K}\vb{K}'\transcript E}$ and $\rho^{\text{ideal}}_{\vb{K}\vb{K}'\transcript E}$ are exactly the same, the condition~\eqref{eq: eps secure} can also be written in the following equivalent form:
\begin{equation} \label{eq: eps secure accept}
    \frac{1}{2} \norm{\tilde{\tau}_{\vb{K}\vb{K'}\transcript E} - \chi^{\ell}_{\vb{K}\vb{K'}} \otimes \tilde{\tau}_{\transcript E}}_1 \leq \varepsilon,
\end{equation}
keeping in mind that $\tilde{\tau}$ is the \emph{sub}-normalised state conditioned on the protocol accepting.\footnote{It is impossible to ensure that~\eqref{eq: eps secure accept} holds with the normalised conditional states instead --- this is because Eve can always trivially implement a ``classical strategy'' that gives her perfect knowledge of the outputs, in which case the accept probability is typically exponentially small, but the normalised state conditioned on accepting would still be very far from the normalised ``ideal term'' $\chi^{\ell}_{\vb{K} \vb{K}'} \otimes \Pr[\text{accept}]^{-1} \tilde{\tau}_{\transcript E}$.} We stress that this security definition is for protocols where the key length $\ell$ is a fixed parameter. For protocols with adaptive key length as a function of the observed statistics, 
the security condition should be modified to~\cite{benor05}
\begin{equation}
    \frac{1}{2} \norm{\tilde{\tau}_{\vb{K}\vb{K'}\transcript E} - \sum_{\ell} \chi^{\ell}_{\vb{K}\vb{K}'} \otimes \tilde{\tau}^{\ell}_{\transcript E}}_1 \leq \varepsilon,
\end{equation}
where $\tilde{\tau}^{\ell}$ is the sub-normalised state conditioned on the protocol producing a key of length $\ell$. However, we will not discuss protocols with adaptive key length in detail.

It is often convenient in security analyses to break down the security condition~\eqref{eq: eps secure accept} into two slightly simpler criteria:  \textit{correctness} and \textit{secrecy}.
We shall discuss the correctness criterion first. A QKD protocol is said to be $\varepsilon_\text{cor}$-correct if 
\begin{equation} \label{eq:correct}
    \Pr[\vb{K} \neq \vb{K}' \text{ and protocol accepts}] \leq \varepsilon_\text{cor}.
\end{equation}
Under the convention of setting $\vb{K} = \vb{K}' = \perp$ whenever the protocol aborts, this can be rewritten equivalently in the more succinct form $\Pr[\vb{K} \neq \vb{K}'] \leq \varepsilon_\text{cor}$.
On the other hand, for the secrecy criterion, a QKD protocol is $\varepsilon_\text{sec}$-secret (with respect to Alice's key) if
\begin{equation} \label{eq:secret}
    \frac{1}{2} \norm{\tilde{\tau}_{\vb{K}\transcript E} - \chi^{\ell}_{\vb{K}} \otimes \tilde{\tau}_{\transcript E}} \leq \varepsilon_\text{sec},
\end{equation}
where $\chi^{\ell}_{\vb{K}} \coloneqq \tr_{\vb{K}'}[\chi^{\ell}_{\vb{K} \vb{K}'}]$ is a uniformly distributed bit-string of length $\ell$.

A protocol that is $\varepsilon_\text{cor}$-correct and $\varepsilon_\text{sec}$-secret in the sense described above (i.e.~with the secrecy condition only involving Alice's key) is also $(\varepsilon_\text{cor} + \varepsilon_\text{sec})$-secure, as shown in~\cite{portmann14,portmann21}. \revision{To briefly outline the proof (see~\cite{portmann14,portmann21} for details), the idea is to construct a specific ``intermediate'' state that satisfies the correctness criterion perfectly, in such a way that its distance to the actual state $\rho_{\vb{K}\vb{K}'\transcript E}$ is bounded by $\varepsilon_{\text{cor}}$. Importantly, for this intermediate state, it is irrelevant whether the secrecy criterion is analysed based on Alice's or Bob's key, since for this state those keys are identical whenever the protocol accepts. The claim that the protocol is $(\varepsilon_\text{cor} + \varepsilon_\text{sec})$-secure can then be obtained by applying the triangle inequality.}
With this in mind, we can analyse the security of a QKD protocol by proving its correctness and secrecy separately. In particular, it is not necessary to explicitly analyse the secrecy of Bob's key since it will be taken care of by the above property.

We shall discuss in more detail how the correctness and secrecy conditions can be proven in Section~\ref{sec: finite key}. To give a short overview here, the correctness condition can usually be enforced straightforwardly by choosing an error verification procedure based on two-universal hashing (although some early works used other approaches). As for the secrecy condition, it can be ensured by using an appropriate privacy amplification scheme in the protocol. For example, one could use a family of two-universal hash functions in the privacy amplification step. In this case, one could invoke the leftover hash lemma against quantum side-information~\cite{tomamichel2011leftover} to prove that the secrecy condition is met as long as the output length $\ell$ of the protocol is chosen to be slightly less than the \textit{conditional smooth min-entropy} 
of the string on which Alice performs privacy amplification\footnote{\revision{This claim holds regardless of whether one-way or two-way classical post-processing is used --- the type of classical post-processing simply affects how the input to the privacy-amplification step is related to Alice and Bob's raw data and the classical register $\transcript$. That being said, in the case of two-way classical post-processing, the input to the privacy-amplification step may depend on the original input/output data in a complicated manner (for example, it may depend on multiple rounds of the raw data in a highly correlated way), and hence its smooth min-entropy is typically not straightforward to evaluate in such protocols (for the DI case at least), unless further simplifying assumptions are made.}} (we can ignore Bob's side since the secrecy condition only involves Alice's key), which is defined as follows: for a given smoothing parameter $\epsilon_s$ and a classical-quantum state $\rho_{\sA E}$, its conditional smooth min-entropy is
\begin{equation}
    H_\text{min}^{\epsilon_s}(\sA|E)_{\rho_{\sA E}} = \max_{\sigma_{\sA E} \in \cB_{\epsilon_s}(\rho_{\sA E})} H_\text{min}(\sA|E)_{\sigma_{\sA E}}.
\end{equation}
Here, the maximisation is taken over the set
\begin{equation*}
    \cB_{\epsilon_s}(\rho_{\sA E}) = \left\{ \sigma_{\sA E} \in \mathrm{D}_{\leq}(\cH_{\sA E}): \Delta_p(\sigma_{\sA E}, \rho_{\sA E}) \leq \epsilon_s \right\},
\end{equation*}
where $\mathrm{D}_{\leq}(\cH_{\sA E})$ denotes the set of sub-normalised states in the Hilbert space $\cH_{\sA E}$. The definition of $\cB_{\epsilon_s}$ is based on the purified distance
\begin{equation}
    \Delta_p(\rho,\sigma) = \sqrt{1 - F^2(\rho, \sigma)},
\end{equation}
where $F(\rho, \sigma) = \tr[|\sqrt{\rho} \sqrt{\sigma}|] + \sqrt{(1- \tr[\sigma])(1-\tr[\rho])}$ is the generalised fidelity. Consequently, proving the secrecy of a QKD protocol is often reduced to finding a lower bound on the smooth min-entropy of the raw key conditioned on all the information gathered by Eve over the course of the protocol. We defer further discussion of how such bounds are derived to Section~\ref{sec: finite key}.

However, the security criterion~\eqref{eq: eps secure accept} is not the only requirement for a QKD protocol~\cite{portmann14,portmann21}.
Recall that~\eqref{eq: eps secure accept} is defined using the sub-normalised states conditioned on accepting. 
As one consequence, a protocol that always aborts would satisfy this criterion trivially.
Since such a protocol is undesirable, we also impose the requirement that the protocol should succeed in producing a pair of secret keys with high probability, in the presence of a realistic amount of noise. This is formalised as a requirement of \emph{completeness}~\cite{portmann14,portmann21}: A protocol is said to be $\varepsilon_\text{com}$-complete if 
the honest implementation (which might be noisy) satisfies
\begin{equation}
    \Pr[\text{abort}]_\text{honest} \leq \varepsilon_\text{com}.
\end{equation}
Here, the probability of aborting under the honest behaviour, which we denote by $\Pr[\text{abort}]_\text{honest}$, is calculated for a certain modelled behaviour of the device, since it is just to ensure that the protocol accepts with high probability when everything behaves as expected. This is unlike $\Pr[\text{abort}]$ encountered previously, which depends on the real behaviour of the implemented protocol that is subjected to Eve's attacks (this abort probability can, and \emph{should}, be high whenever Eve performs an attack that gives her enough information to compromise the key).

We briefly remark that the above security definitions were originally developed for the context of device-dependent QKD, where they also have significant operational implications~\cite{portmann14,portmann21}. In the context of DI-QKD, 
it is still possible to treat these as 
purely mathematical conditions and prove that they are indeed fulfilled (as we shall discuss in Section~\ref{sec: finite key}); however, their operational implications become more subtle due to issues regarding device reuse~\cite{barrett2013memory,portmann21}.
Still, those issues are restricted entirely to the operational interpretations (and there is ongoing work on resolving this point), so these security definitions are still well-posed, and in this work we shall apply them directly to DI-QKD.

\subsubsection{Side-channels and quantum hacking}
Although QKD provides information-theoretic security on paper, exploitable side-channels could exist in its implementations. The security of practical QKD systems depends not only on what Eve can do in the quantum channel but also the \emph{side-channels} in their implementations: channels that are not modelled in the security proof, but through which information can still be leaked, compromising security. Hacking attacks can be performed in the quantum communication layer, the classical post-processing of the protocol or even after the keys have been produced. Side-channels from classical information processing systems are also an issue in classical cryptography. However, the fact that QKD is susceptible to hacking attacks in the quantum communication layer (which we shall refer to as ``quantum hacking'') might be surprising to some, especially in the earlier days of QKD where the claim of ``unconditional security'' was commonly touted. This has to be understood in the context of how security is proven in QKD.

Typically, to prove the security of a QKD protocol, it is often necessary to specifically model the devices (e.g., the measurement operators of Alice and Bob in a EB-QKD protocol, the quantum states emitted by Alice's source in a P\&M-QKD protocol, etc.) that are used in the protocol. Such a security proof is \textit{device-dependent} (DD); it depends on the assumption that the devices behave according to the model. Quantum hacking consists of attacks that cause the behaviour of the devices to deviate from the model that is used in the security proof.

For example, to prove the security of many P\&M QKD protocols, it is often necessary to assume the quantum states being emitted by the source are well-characterised. In the so-called Trojan horse attack~\cite{vakhitov2001large,gisin2006trojan}, Eve makes use of the reflectivity of practical sources to inject a bright light into the source used in the protocol, effectively modifying the emitted signals, and extracts additional information about the modulation from the reflected light. Another example of quantum hacking would be the blinding attack~\cite{makarov2009controlling,lydersen2010hacking} which invalidates the assumption that whether or not a measurement device registers a click is independent of the basis choice. In this attack, Eve controls the detector by sending bright light into the detector such that it would only click if the receiver chooses the same basis that is chosen by Eve.

While these attacks can be mitigated using some \textit{ad hoc} counter-measures, it is hard to assess the effectiveness of these counter-measures against more sophisticated attacks. Furthermore, there are side-channels that are opened without any active attacks by Eve. For example, it is known that in high-speed QKD systems, correlation in the modulation might arise between successive rounds~\cite{nagamatsu2016security, yoshino2018quantum, pereira2020quantum}. In such cases, the security proof, which typically assumes that the devices behave identically\footnote{This should not be confused with assuming that Eve attacks identically (and independently) in each round (i.e., she performs a collective attack). When proving security of DD-QKD against general attacks, Eve can perform any attack that she wants, but the devices of the legitimate parties --- which she does not control --- are assumed to behave identically in each round according to the specified model.} for each round may not hold.

Giving an exhaustive list of possible quantum hacking attacks or side-channels is not the goal of this review. For interested readers, a list of known attacks and side-channels is given in a recent review paper~\cite{xu2020secure}. As we shall explain in the next section, the goal of \textit{device-independent} QKD is to eliminate all side-channels in the quantum communication layer in a conclusive (i.e., robust against future discoveries of more sophisticated attacks) and elegant way. As explained in this section, quantum hacking and side-channels are consequences of our modelling of the devices when proving the security of a QKD protocol. By devising a security proof that is agnostic to the device modelling, device-independent QKD removes this vulnerability.

\subsection{Device-independent security}
\subsubsection{Motivation}
Given the possibility of compromising the security of QKD when the devices that implement the protocol do not behave as advertised (due to an oversight by the manufacturer, degradation of the devices, or quantum hacking by an adversary), we may want to rule out these scenarios by devising a security proof that is valid under minimal assumptions (which we list in detail in Section~\ref{sec: assumptions}). In particular, the device-independent framework aims to prove the security of the protocol without specifying the states and measurements that are used in the protocol (hence the term ``device-independent''), 
and QKD protocols that can be proven secure in this framework are referred to as ``device-independent QKD'' protocols.
With device-independent security, all side-channels 
that can be formalised as the devices performing ``incorrect'' measurements 
are eliminated\footnote{This does not mean that the protocol is immune against all hacking, as we still need to make assumptions about the classical post-processing step as well as the key management system. For more information, see Section~\ref{sec: assumptions}.}. To achieve it, we rely on Bell nonlocality to certify that the uncharacterised devices are producing outputs that are genuinely ``random'' to an adversary. In light of this, Bell nonlocality is a necessary condition\footnote{Any local correlations, by definition, can be explained by a local-hidden-variable model. An adversary is allowed to possess a copy of these hidden variables, in which case she could then predict the outputs of the QKD devices perfectly.} for DI-QKD's security, though recent work suggests it may not be a sufficient condition \cite{farkas2021bell, christandl2021upper}.

\subsubsection{Towards the notion of device-independence}
The idea of using Bell nonlocality \cite{bell1964,brunner2014bell} to certify the security of a key distribution protocol first appeared in Ekert's re-invention of QKD~\cite{ekert1991quantum}, although the notion of device-independence was not emphasised there. It would later re-appear in the seminal work of Mayers and Yao \cite{mayers1998quantum, mayers2004self} on \emph{self-testing}, where they found that, when a specific nonlocal behaviour is observed, it is possible to certify that, up to some local isometries, a quantum device consists of the state $\ket{\Phi^+}$ and the Pauli measurements, $\sigma_Z$ and $\sigma_X$ (as well as a third measurement $(\sigma_Z + \sigma_X)/\sqrt{2}$), which can then be used to implement the BBM92 protocol~\cite{bennett1992quantum}. However, the work of Mayers and Yao only discussed the situation where the ideal statistics are observed, which would never be the case in practice. The first step towards a security proof was taken when Barrett \textit{et al}. proposed a protocol and proved its security based only on the no-signalling assumption~\cite{barrett2005nosignalling}.

The term ``device-independence'' was finally coined in the work of Ac\'{i}n \textit{et al}. \cite{acin2007device} with the emphasis that the security claim is decoupled from the quantum states and measurements with which the protocol is implemented. Security against collective attacks\footnote{In device-dependent QKD, this means that Eve distributes states of the form $\rho_{AB} = (\rho_{A_i B_i})^{\otimes n}$, where $\rho_{A_i B_i}$ is the state in any single round. In the context of device-independent QKD, we further require the uncharacterised devices to behave independently and identically in each round.} in the asymptotic limit was then proven~\cite{acin2007device,pironio2009device}. However, the collective attack assumption is against the spirit of device-independence since it assumes that the devices are working in an independent-and-identically-distributed (i.i.d.) manner. Complete security proofs without the collective attack assumption were given later \cite{vazirani2014fully,miller2016robust,jain2020parallel, vidick2017parallel}, though the asymptotic bounds were significantly less robust against noise than the one obtained under the collective attack assumption. An improved bound \cite{arnon2018practical,arnon2019simple} was then obtained via the entropy accumulation theorem \cite{dupuis2020entropy, dupuis2019entropy}, which achieves the same asymptotic key rate as the collective-attacks scenario.

\subsubsection{Beyond fully device-independent security}
While DI-QKD offers information-theoretic security despite the protocol being implemented using an uncharacterised source of quantum states and uncharacterised measurement devices, its implementation is still extremely challenging. Inspired by DI-QKD, several QKD protocols with different levels of device characterisation have been proposed. Such protocols are commonly referred to as ``\textit{semi-device-independent}'' protocols, as their security does not require full characterisation of the device, but they still rely on some assumptions about the physical systems. For example, there are semi-device-independent frameworks based on assumptions about the dimension of the Hilbert space~\cite{woodhead2015secrecy, woodhead2016semi, pawlowski2011sdi, goh2016measurement}, energy of the source~\cite{vanHimbeeck2017semidevice}, etc.

There have also been proposals for QKD protocols where the device (source or measurement device) of one party is completely characterised while the other party's is not~\cite{tomamichel2012tight,branciard2012onesided, walk2016experimental,ioannou2021receiver, ioannou2021receiver_protocols}. This framework is referred to as the ``\emph{one-sided-device-independent}'' scenario. MDI-QKD \cite{lo2012measurement} is a related class of protocols where Alice and Bob both hold characterised sources (or equivalently, a measurement device and a source of entangled states~\cite{braunstein2012side}) and send their quantum states to a third party --- which can be assumed to be the eavesdropper. Note that in MDI-QKD, all the honest parties hold fully characterised devices, although there have been recent efforts to relax the characterisation requirements of the devices~\cite{navarrete2021practical, zhang2021securing}.

Finally, there have also been proposals to achieve device-independence from computational assumptions~\cite{metger2021device}. In this scenario, in place of the no-signalling scenario of the standard device-independent framework (see Section~\ref{sec: assumptions}), Alice and Bob's quantum channel is part of their devices (that are prepared by Eve) and not directly accessible by Eve. Further, the devices are assumed to be computationally bounded, in the sense that they are not able to break post-quantum cryptographic protocols (more precisely, the learning-with-errors problem) during the execution of the protocol. Under these assumptions, Eve is allowed to be entangled with Alice and Bob but she is not allowed to help the devices to violate the computational assumption. If these assumptions are satisfied, the generated key is information-theoretically secure even against a computationally unbounded adversary.

In the spirit of device-independence, these protocols aim to minimise side-channels introduced by our modelling of the devices that implement them, while being more practically achievable than fully device-independent QKD. These frameworks remain an active area of research. However, these protocols are not the focus of this review and we shall devote the remainder of this review paper to fully device-independent QKD protocols. 
\section{Assumptions} \label{sec: assumptions}
While the security of DI-QKD does not rely on the characterisation of the quantum state and the measurement devices, some assumptions are still needed to prove its security. In this section, we shall list these assumptions and discuss their implications.

The assumptions are as follows:

\begin{enumerate}
    \item Quantum theory is correct.
    \item The honest parties operate within secured locations using only trusted devices and adhering strictly to the protocol. The devices may be uncharacterised but cannot maliciously broadcast their inputs and outputs.\footnote{When considering DI-QKD implementation with multiple pairs of devices, it was shown in~\cite{curty2019foiling, zapatero2021secure} that DI-QKD can still be secure with the aid of secret sharing if some \textit{but not all} of the devices are malicious.}
    \item The honest parties have access to an authenticated classical channel.
    \item The honest parties each have a trusted and private random number generator to choose the inputs of their devices.
    \item The honest parties can perform any Bell test  required by the protocol in a manner that is free of various relevant \textit{loopholes} \cite{brunner2014bell,larsson2014loopholes}.
\end{enumerate}

The first assumption is usually taken for granted as quantum theory is the most accurate known scientific model for physical phenomena at the subatomic scale to date. That being said, DI-QKD may also be feasible even if quantum theory were to be superseded by another physical theory that respects the no-signalling principle \cite{barrett2005nosignalling}. As quantum theory's validity remains unchallenged today, it shall be assumed to be so for the rest of the discussion in this paper.

Like any other cryptographic protocol, DI-QKD is no longer secure once the private key is conceded to the adversary. Hence, the second assumption is necessary to prevent private information pertaining to the secret key from leaking to the adversary. While this private information must be stored securely, it is also crucial to exclude any malicious elements within the working spaces of the honest parties. This is in contrast to the initial belief that DI-QKD could employ devices that ``are entirely untrusted and provided by the eavesdropper'' \cite{acin2007device,pironio2009device}. When used in DI-QKD, such untrusted devices could broadcast obfuscated private key information to the adversary through various avenues: side-channels, back-doors \cite{scarani2009security} or even through the honest parties themselves \cite{barrett2013memory}. Thus, the integrity of the DI-QKD protocol remains intact when using uncharacterised devices but not with untrusted devices\footnote{Untrusted measurement devices can be employed securely in the case of MDI-QKD because the security analysis allows the eavesdropper to hold any information processed and produced by the measurement device. Any QKD protocol proven secure under that condition is secure with the use of untrusted measurement devices. (To achieve security in such a scenario, MDI-QKD requires that the honest parties can instead perform trusted state \emph{preparation}, rather than measurement.)}. As for the issue of ensuring the devices do not broadcast the inputs, we briefly defer this until the loophole discussion below.

The third assumption is crucial in preventing a possible man-in-the-middle attack, where the adversary simply impersonates an honest party to retrieve the secret key by following the QKD protocol. 
Fortunately, it is not too difficult in principle to establish information-theoretically secure authentication for a classical channel, by expending a small amount of pre-shared key~\cite{carter1979universal,wegman1981new}. (From this perspective, a QKD protocol using a channel authenticated this way would technically be a protocol for key \textit{expansion}, rather than a protocol for key generation without pre-shared resources.)

The fourth assumption is necessary because trusted randomness is required in DI-QKD protocols. Most DI-QKD protocols are executed in rounds, and in each round, Alice and Bob must decide if it is a generation or a test round. In the test rounds (and also in the generation rounds, for some protocols), Alice and Bob need to choose inputs to their devices. For the purposes of the security proof, these choices need to be independent of Eve, and (typically) also of the outputs in previous rounds. To ensure this, we impose the condition that these choices are made using trusted and private random number generators. This point is also closely related to the issue of closing various loopholes, as we shall now describe.

Finally, the fifth assumption concerns the fact that classical (local) resources can be used to ``fake'' a Bell violation if several \textit{loopholes} are not addressed. Since DI-QKD protocols rely on Bell violations to certify quantum behaviour, if these loopholes are not closed, these violations could be faked and the protocols would no longer be secure. We briefly highlight, however, that these loopholes manifest slightly differently in the context of Bell tests as compared to DI security proofs --- the former is concerned with what can be achieved by purely \textit{classical} resources in the presence of loopholes, while the latter is concerned with what Eve can achieve using side-information on \textit{quantum} resources in the presence of loopholes. Still, without further discussing this distinction, we shall broadly outline the most relevant loopholes, namely the 
\textit{detection loophole}, the \textit{measurement dependence loophole}, and the \textit{locality loophole} \cite{pearle1970hidden,bell2004speakable,brunner2014bell,larsson2014loopholes}. 

\begin{figure*}[t]
    \centering
    \includegraphics[width = \textwidth]{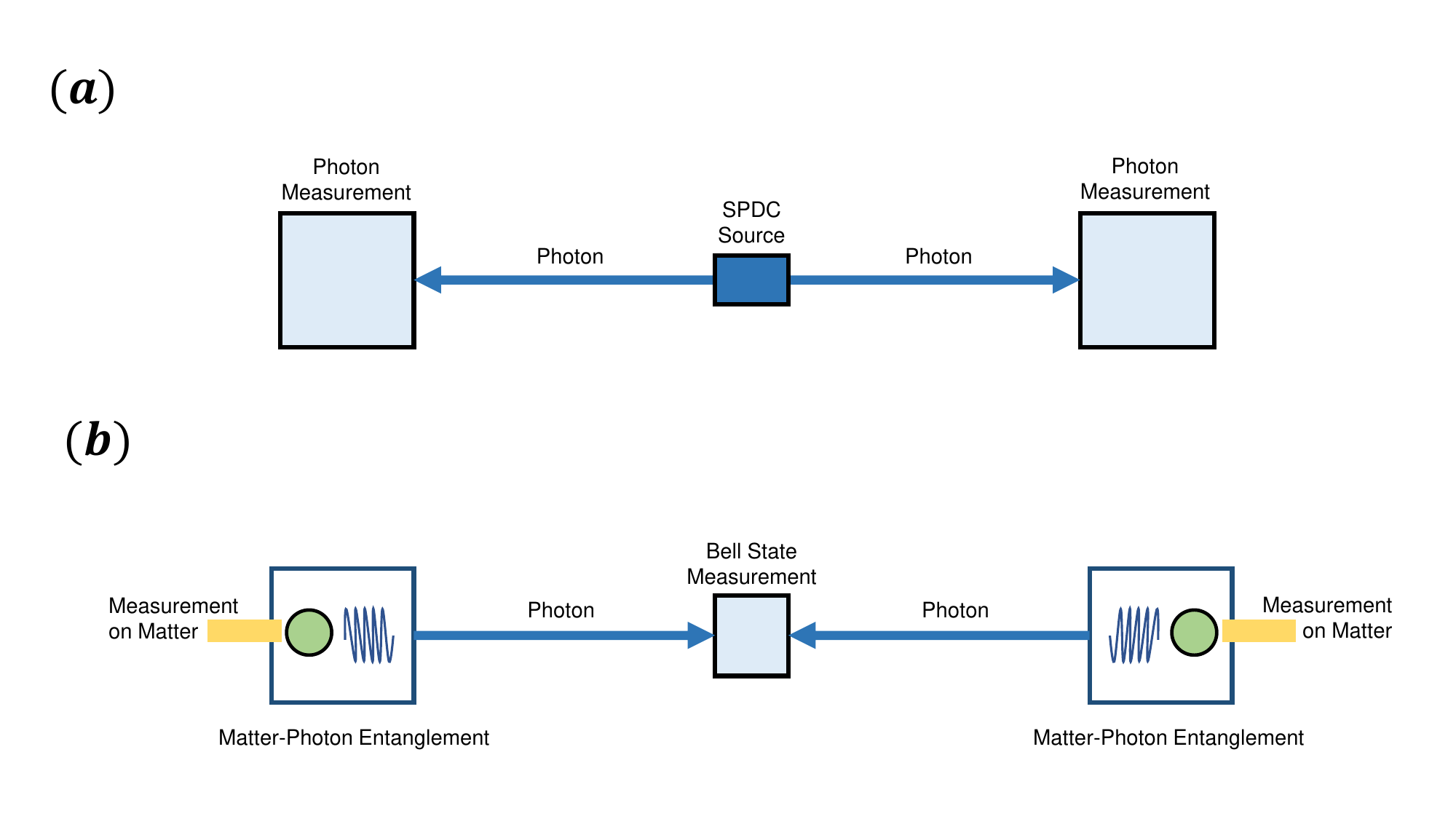}
    \caption{Schematics for two DI-QKD implementations. (a) Fully-photonic systems typically involve direct measurements of entangled photon-pairs coming from an SPDC source. In long distance implementations, qubit amplifier schemes can be integrated to the photon measurements to herald the arrival of the photons. (b) In heralded entanglement systems, each party typically generates entanglement locally between a matter (which serves as a long-lived quantum memory) and a photon. Bell state measurement on the emitted photons can be used to herald matter-matter entanglement via entanglement-swapping. Afterwards, each party can perform their measurements on the long-lived quantum memories located in their respective secure locations.}
    \label{fig: experiments}
\end{figure*}

The detection loophole refers to the fact that classical devices can fake a Bell violation if, in some of the rounds, the measurement devices do not give a conclusive outcome (the typical example of this would be a no-detection event in a photonic setup) and one chooses to discard such rounds in the security analysis. Fortunately, there are fairly straightforward methods to handle this loophole --- for instance, one can simply assign all inconclusive outcomes to a fixed value (as determined by the protocol; it could be one of the ``standard'' outcome values or a separate ``null'' symbol to be accounted for in the statistical analysis). 

As for the measurement dependence loophole and locality loophole, these are somewhat related issues (though it is convenient to formalise them separately in some proofs). The former concerns the question of whether the underlying resource (shared randomness, entangled states, etc) could be correlated to the inputs. The latter is the question of whether Alice's input might be communicated to Bob's device before it has to produce an output (or vice versa). If either of these loopholes is not closed, then the devices could again fake a Bell violation without any entanglement. 
We hence see the relation to the fourth assumption, in that one aspect of addressing these loopholes is to require that the inputs are chosen using a sufficiently trusted randomness source.
Furthermore, in the context of Bell tests, the issue of inputs leaking between the devices is usually addressed by ensuring spacelike separation between the generation of one party's input and the generation of the other party's output. In the context of DI protocols, however, it is worth considering whether this approach is strictly necessary. Given that we have already required (in the second assumption) that measures have been taken to ensure the \emph{outputs} remain private, it may be reasonable to suppose that these measures could also ensure the \emph{inputs} do not leak. This would come down to a question of whether the level of trust/characterisation of the devices in a particular setup is sufficient to justify such an assumption.

\section{Experiments}
Compared to its device-dependent counterpart, implementing DI-QKD is more technically demanding: it involves performing an adequately loophole-free Bell test over a meaningfully large distance, while achieving a significant Bell inequality violation and sufficiently low \emph{quantum bit-error rate}\footnote{Roughly speaking, 
this is the probability that any given bit in the honest parties' raw keys differs between them.} (QBER). Moreover, the devices have to be operating at a decent clockrate to suppress the finite-size effect to an acceptable level. While practical DI-QKD implementation is still a work-in-progress, there have been numerous recent significant developments in experimental DI-QKD \cite{zhang2021experimental,liu2021high,nadlinger2021device}, placing said goal in our sights. \revision{An overview of the result of these experiments can be found in Table~\ref{tab: proof of principle experiments}.}
\begin{table*}
    \centering
    \caption{The list of proof-of-principle experiments of DI-QKD.}
    \begin{tabular}{*7c}
    \toprule
        & & maximal & & & & \\
         & experimental & transmission & key rate\footnotemark[1] & event & non-i.i.d. & finite-key \\
        Reference & platform & distance &  (bits/round) & rate & security & analysis\\
        \midrule
        \hline
        \cite{liu2021high} & fully-photonic & 220~km & $2.33 \times 10^{-4}$ & 2~MHz & \xmark\footnotemark[2] & \xmark\\
        \hline
        \cite{nadlinger2021device} & trapped ions & 2~km & $0.0639$ &  119~Hz\footnotemark[3] & \cmark & \cmark\\
        \hline
        \cite{zhang2021experimental} & trapped atoms & 400~km & $0.07$ & 0.0122~Hz & \cmark & \xmark\\
        \bottomrule
    \end{tabular}
    
    \footnotetext[1]{Among these experiments, only \cite{nadlinger2021device} performed a full QKD experiment which include the classical post-processing. In \cite{liu2021high} and \cite{zhang2021experimental}, the authors only characterise the statistics generated in the experiment and estimate the achievable \textit{asymptotic} key rate if they perform the classical post-processing. We also note that the key rate reported in \cite{liu2021high} is with respect to i.i.d.~attacks.} 
    \footnotetext[2]{The security of the protocol being performed in this experiment has only been analysed under the assumption of i.i.d.~attacks. It is unclear if the security analysis can be extended against general attacks.}
    \footnotetext[3]{The experiment generated $1.5 \times 10^6$ rounds of data over $7.9$ hours (with a pause of 4.4 hours due to laser failure). We estimate the repetition rate of the system as $(1.5 \times 10^6 / 3.5~\mathrm{h}) \approx 119$~Hz.}
    \label{tab: proof of principle experiments}
\end{table*}

As DI-QKD experiments were built upon loophole-free Bell tests experiments, they can be divided into two broad groups (Figure \ref{fig: experiments}): the fully-photonic systems \cite{giustina2015significant,liu2021high,liu2021device, liu2018device, shalm2015strong,zhang2020experimental, li2018test, li2021experimental, bierhorst2018experimentally, shalm2021device, shen2018randomness} and the heralded entanglement systems \cite{hensen2015loophole,rosenfeld2017event, nadlinger2021device, zhang2021experimental}. Each choice of system carries a unique set of advantages and bottlenecks towards DI-QKD implementation. 

To date, all existing loophole-free Bell experiments (including all DI-QKD proof-of-principle demonstrations) are based on the CHSH (Clauser-Horne-Shimony-Holt) inequality~\cite{clauser1969proposed}, where the \textit{CHSH value} $S$ is measured:
\begin{equation} \label{eq: CHSH}
    S \coloneqq \expectation{\sA_0 \sB_0} + \expectation{\sA_0 \sB_1} + \expectation{\sA_1 \sB_0} - \expectation{\sA_1 \sB_1}.
\end{equation}
Here, $\sA_x$ and $\sB_y$ are the measurement outcomes that Alice and Bob obtain when Alice chooses the input $x \in \{0,1\}$ and Bob chooses the input $y \in \{0,1\}$. We denote the corresponding self-adjoint operators associated to these measurements by $A_x$ and $B_y$.
As shown in~\cite{clauser1969proposed}, all CHSH values achievable by local-hidden-variable models must obey the inequality $S\leq2$, but this inequality can be violated by quantum devices.
Hence, in this section, we shall focus on protocols that use the CHSH value in their statistical tests. However, it is worth noting that in certain cases, estimating other Bell inequalities in the parameter estimation routine can lead to a better performance of the protocol~\cite{sekatski2021device, woodhead2021device} (see also Section~\ref{sec: protocols}).

If the secret key rate is considered as the figure-of-merit for QKD implementations, any discussion on the performance of existing DI-QKD experiments can be summarised in its QBER $Q$, CHSH value $S$ and clockrate. However, as the existing DI-QKD experiments have been performed with different distances between the honest parties, the distance is another figure-of-merit that can be taken into account when comparing the experiments.

\subsection{Fully-photonic experiments}
\subsubsection{General features}
Fully-photonic loophole-free Bell experiments involve the preparation of photon-pair with entangled degrees-of-freedom (usually polarisation via spontaneous parametric down-conversion (SPDC) by pumping non-linear crystals) and subsequently, distant parties Alice and Bob will each measure a part of the entangled photons using single-photon detectors with high efficiencies (usually superconducting nanowire single-photon detectors (SNSPDs)). 

Adopting the fully-photonic system allows the user to enjoy high clockrate and low QBER at the expense of a low CHSH value. For example, in a recent proof-of-principle DI-QKD experiment~\cite{liu2021high}, the experimental setup could win the CHSH game with a probability of 0.7559 (or equivalently, a CHSH value of $S \approx 2.0472$) across a transmission distance of 20~m was reported. Indeed, any fully-photonic Bell experiment that is implemented by measuring a pair of two-mode squeezed vacuum states with single-photon detectors has its CHSH value limited by $S\lesssim 2.31$ \cite{vivoli2015challenging,vivoli2015comparing,tsujimoto2018optimal} even when ideal single-photon detectors (i.e., perfect detection efficiency and zero dark-count rate) are used. Clearly, the contributions from channel noise and loss, coupling loss at each interface, and imperfect detector efficiencies would account for such low CHSH values in past experiments. 

\subsubsection{Losses in fully-photonic experiments}
The main bottleneck of fully-photonic implementations of DI-QKD is the channel loss (which translates to weaker nonlocal correlations). When simulating a fully-photonic DI-QKD experiment, channel losses are typically modelled as a beam-splitter with transmittivity $\eta$, corresponding to the overall efficiency of the quantum channel.

Generally, channel losses can be classified into two categories: local losses and transmission losses. With this classification, the overall transmittivity of the channel is given by $\eta = \eta_l \eta_t$ where $\eta_l$ is the local efficiency and $\eta_t$ is the transmission efficiency. Local losses refer to those losses that are attributed to the local surroundings of the legitimate parties. These include losses due to detector inefficiencies and coupling losses. On the other hand, transmission losses refer to losses that occur in the optical channel during the transmission between the source and the receivers' lab. When an optical fibre is used as a transmission medium, the loss in the channel scales exponentially with distance. Hence, the transmission efficiency $\eta_t$ is given by
\begin{equation}
    \eta_t = 10^{- \xi L/10},
\end{equation}
where $L$ is the transmission length and $\xi$ is a coefficient that quantifies the attenuation of signal power in the fibre. The attenuation also depends on the wavelength of the signal. Standard optical fibres typically have $\xi \approx 0.2$~dB/km at 1550~nm, but there are ultra-low-loss fibres with lower $\xi$.

\subsubsection{Qubit amplifiers} \label{sec: qubit amplifiers}
Due to the detection loophole, the no-click events that are typically discarded in DD-QKD cannot be safely discarded in the case of DI-QKD. At the heart of the detection loophole is violation of the fair-sampling assumption; in the presence of a malicious attack, whether the detector would click or not would depend on the random inputs given by the trusted parties. This is in contrast to the honest implementation of the devices, where the main reason the detectors do not click is due to the photons being lost in the channel. Such a process is clearly consistent with the fair-sampling assumption. The purpose of \textit{qubit amplifiers}~\cite{gisin2010proposal, pitkanen2011efficient, curty2011heralded,meyer2013entanglement, seshadreesan2016progress, zapatero2019long, kolodynski2020device} is to herald the arrival of a photon without disturbing its qubit state\footnote{As such, an ideal qubit amplifier simulates a quantum non-demolition measurement. However, practical qubit amplifiers would introduce noise in the amplification process when there are imperfections (in the optical circuits or in the detectors) or when the input state contains multi-photon components.}, and hence allow post-selection to be done securely.

To illustrate the idea, let us consider a normalised pure quantum state $\ket{\psi}$ of the following form
\begin{equation}
    \ket{\psi} = \alpha \ket{v} + \beta \ket{\varphi},
\end{equation}
where $\ket{v}$ denotes the vacuum state and $\ket{\varphi}$ is a single-photon qubit state
\begin{equation}
     \ket{\varphi} = \left(\beta_H a_H^\dagger + \beta_V a_V^\dagger \right) \ket{v}.
\end{equation}
Here, $a_H$ and $a_V$ are the annihilation operators for the horizontally and vertically polarised-mode, respectively. The state $\ket{\psi}$ is essentially a coherent superposition between the vacuum state and a single-photon state defined across two orthogonal modes --- which defines a qubit. A qubit amplifier is an optical circuit that, conditioned on the heralding signal (which is typically based on a Bell state measurement), transforms the state $\ket{\psi}$ into $\ket{\psi'} = \alpha' \ket{v} + \beta' \ket{\varphi}$ such that the relative weight for the qubit state is higher, i.e., $ \abs{\beta'}^2 > \abs{\beta}^2$. Importantly for DI-QKD applications, qubit amplifiers can also be used for mixed quantum states.

The first qubit amplifier scheme based on teleportation with two single-photon sources is presented by Gisin \textit{et al.}~\cite{gisin2010proposal} which builds on the proposal of photon amplifiers~\cite{ralph2009nondeterministic}. It was noted that on-demand single-photon sources would give better performance but heralded single-photon sources are sufficient to implement the scheme. The original scheme can be improved by adding two 50/50 beam-splitters to the optical circuit~\cite{pitkanen2011efficient}. In particular, if only the vacuum and the single-photon component is present in the input state, the modified amplifier can perform perfect heralded projection to the single-photon space (conditioned on the heralding signal, the amplitude of the vacuum component is zero). Secondly, even in the presence of multi-photon components, the modified amplifier can amplify the single-photon component at the expense of lower success probability. Subsequently, schemes with entanglement-swapping relay were proposed \cite{curty2011heralded, meyer2013entanglement}. Importantly, these entanglement-swapping based qubit amplifiers can still be useful even when implemented with SPDC sources instead of an ideal source of entangled photon pairs. The analysis of the practical performance DI-QKD with qubit amplifiers in the asymptotic limit was then presented for the two single-photons architecture~\cite{kolodynski2020device} and for the entanglement-swapping relay architecture~\cite{seshadreesan2016progress}. The finite-key performance of the protocol using both architectures was also analysed~\cite{zapatero2019long}.

As a qubit amplifier allows us to simulate non-demolition projection onto the single-photon subspace, the qubit amplifier allows us to safely discard rounds in which the amplifier does not herald the arrival of the photon. Therefore, qubit amplifiers minimise the effect of transmission losses, though the effect of local losses is still present. Unfortunately, at the moment of writing, the existing local losses are still prohibitively large for most DI-QKD protocols\footnote{With the random post-selection protocol~\cite{xu2021device} and the noisy pre-processing protocol~\cite{ho2020noisy,woodhead2021device} being notable exceptions. See Section~\ref{sec: protocols} for more details.} to be implemented even at a short distance. Consequently, the use of qubit amplifiers is not yet relevant for DI-QKD with the current level of detector efficiencies.

\subsubsection{Relevant experiments}
Unfortunately, the most well-studied DI-QKD protocol  
has extremely demanding requirements on the overall losses ($\eta \gtrsim 80\%$~\cite{brown2021device,masini2021simple}) 
to obtain a positive secret key rate. Due to this, despite the fact that there are a number of fully-photonic loophole-free Bell experiments~\cite{giustina2015significant,liu2021high,liu2021device, liu2018device, liu2018high, shalm2015strong,zhang2020experimental, li2018test, li2021experimental, bierhorst2018experimentally, shalm2021device, shen2018randomness} with some of them even achieving sufficiently high CHSH values for device-independent randomness generation~\cite{bierhorst2018experimentally, li2021experimental, shen2018randomness, shalm2021device, liu2018device, liu2021device, zhang2020experimental}, a conclusive demonstration of fully-photonic DI-QKD is still missing.

However, there is a recent proof-of-principle fully-photonic DI-QKD experiment~\cite{liu2021high} which is asymptotically secure when Eve is restricted to i.i.d.~attacks. In this experiment, single-photon detectors with an efficiency of $\gtrsim 87\%$ were used (higher than the detector efficiency in any of previous fully-photonic loophole-free Bell experiments). Note that the experiment was enabled by a specific DI-QKD protocol that has strong robustness against channel losses~\cite{xu2021device} (see Section~\ref{sec: protocols}).

\subsection{Heralded entanglement experiments}

\subsubsection{General features}
The heralded entanglement loophole-free Bell experiments involve each party preparing an entangled state between a long-lived quantum system (e.g., trapped ions, atoms, NV-centre, etc) and a photon (typically in its polarisation degree-of-freedom). The long-lived systems are stored in each party's laboratories while the photonic systems are then sent for a Bell-state measurement. A successful Bell-state measurement then heralds a successful entanglement swapping, after which the long-lived systems can then be measured. The target state is typically the maximally-entangled qubit state and the noise is well-approximated by the depolarising noise model.

As the detection efficiencies for these long-lived systems are typically very high as compared to typical photon detectors, the fraction of no-detection outcomes (which then need to be assigned to deterministic outcomes) is smaller in the case of heralded entanglement systems (after post-selection on the heralded events). Consequently, contrary to their fully-photonic counterparts, heralded entanglement systems provide users with high CHSH value and low QBER, though at the expense of low clockrate. For example, a recent DI-QKD experiment using a trapped-ions-based heralded entanglement system~\cite{nadlinger2021device} reported a CHSH value of $S \approx 2.64$ and QBER of $Q \approx 1.8\%$ across a 2~m transmission distance. Similar to the case of fully-photonic systems with qubit amplifiers, the effect of transmission loss can be minimised in heralded entanglement systems. In heralded entanglement systems, transmission loss would lower the probability of successful Bell state measurement, but the Bell violation can still be high as long as the dark-count rates of the detectors are low.

However, the main bottleneck of heralded entanglement systems is their clockrates. For example, while the recent demonstration of fully-photonic DI-QKD~\cite{liu2021high} operated at a 2~MHz repetition rate, the recent trapped-ions-based heralded entanglement system~\cite{nadlinger2021device} performed $1.5 \times 10^6$ rounds of the DI-QKD protocol over $7.9$ hours (with a pause of 4.4 hours due to laser failure). This corresponds to a clockrate of about 119~Hz even when the pause is neglected. Indeed, previous efforts to increase the entangling rate in trapped-ions system~\cite{hucul2015modular}, and NV-centre systems~\cite{humphreys2018deterministic, kalb2017entanglement} result in lower fidelities to the point that no Bell violation was reported.

\subsubsection{Relevant experiments}
Due to its ability to exhibit high CHSH value, a heralded entanglement system (based on NV-centre) was used for the first loophole-free Bell experiment~\cite{hensen2015loophole}. Following that experiment, another loophole-free Bell experiment using heralded entangled atoms~\cite{rosenfeld2017event} was done, exhibiting a CHSH value of $S \approx 2.221$. Despite these earlier loophole-free Bell tests exhibiting significant Bell violation, the QBER is not sufficiently low to demonstrate DI-QKD.

The above-mentioned experiment using heralded entanglement of trapped ions~\cite{nadlinger2021device} was able to generate secret keys secure against general attacks with a security parameter of $\varepsilon = 10^{-10}$. The resulting key rate was estimated to be about 0.0639 bit per entanglement generation event. Another proof-of-principle demonstration using heralded entangled atoms was able to generate \textit{asymptotic} secret key rate of about 0.07 bit per entanglement generation event over 400~m~\cite{zhang2021experimental}. However, the block size was too small to generate a secret key when finite-size effects are considered.
\section{Protocols}
\label{sec: protocols}
\begin{table*}[t]
    \centering
    \caption{The list of variants of the standard DI-QKD protocol which is based on the CHSH inequality~\cite{acin2007device,pironio2009device} and their experimental requirements.}
    \begin{tabular}{ p{0.3\linewidth} >{\centering \arraybackslash}p{0.3\linewidth}  >{\centering \arraybackslash}p{0.3\linewidth} }
    \toprule
        Protocol  & Critical QBER & Critical Detection Efficiency\\
        \midrule
        \hline
        The standard protocol & 7.1\%~\cite{pironio2009device} & 86.5\%~\cite{woodhead2021device,sekatski2021device}\\
        \hline
        Generalised CHSH (gen. CHSH) & 7.4\%~\cite{woodhead2021device,sekatski2021device}& 86.5\%~\cite{woodhead2021device,sekatski2021device}\\
        \hline
        Noisy pre-processing (NPP) & 8.1\%~\cite{woodhead2021device,sekatski2021device} & 82.6\%~\cite{woodhead2021device,sekatski2021device}\footnotemark[1] \\
        \hline
        Random key-basis (RKB) & 8.4\%~\cite{masini2021simple} & 92.5\%~\cite{schwonnek2021device}\footnotemark[2]\\
        \hline
        Random post-selection\footnotemark[3] & (unknown) & 68.5\%~\cite{xu2021device}\\
        \hline
        Advantage distillation\footnotemark[3] & 9.1\%~\cite{tan2020advantage} & 89.1\%\footnotemark[4]~\cite{tan2020advantage}\\
        \hline
        NPP + gen. CHSH (+ bias) & 8.3\%~\cite{woodhead2021device,sekatski2021device} & 80.3\%~\cite{masini2021simple}\footnotemark[5]\\
        \hline
        RKB + NPP & 9.3\%~\cite{tan2020improved} & (unknown)\\
        \bottomrule
    \end{tabular}
    \footnotetext[1]{\cite{ho2020noisy} discovered the critical detection efficiency of 83.2\% when an SPDC source is used instead of partially entangled qubits.}
    \footnotetext[2]{In the Supplementary Note 2 of~\cite{schwonnek2021device}, an SPDC source was assumed instead of partially entangled qubit pairs. Furthermore, efficient error correction~\cite{ma2012efficient} was not used.}
    \footnotetext[3]{A full security proof of these protocols has not yet been provided.}
    \footnotetext[4]{To obtain this figure, the CHSH value was optimised instead of the key rate.}
    \footnotetext[5]{\cite{brown2021device} discovered the critical detection efficiency can be reduced to 80.0\% if the full input/output behaviour is used to test for nonlocality.}
    \label{tab: protocols}
\end{table*}
In this review article, we shall focus on the DI-QKD protocol proposed by Ac\'in \textit{et al.}~\cite{acin2007device,pironio2009device} as it is the most well-studied DI-QKD protocol in the literature. We will also discuss some of its variants which are designed to improve the robustness of the protocol against noisy/lossy experimental implementations with minimal changes to the quantum devices. Some of these modifications can also be combined and applied to other protocols. We note that there are other DI-QKD protocols in the literature which we do not discuss in detail here; to name a few: DI-QKD with local Bell tests~\cite{lim2013device}, DI-QKD based on measurement inputs~\cite{DIQKDmeasureinput}, and DI-QKD based on high-dimensional entangled states~\cite{brown2021computing,gonzales2021device}.

\revision{When comparing the robustness of different protocols, many works typically use the asymptotic key rate against i.i.d.~attacks (see Eq.~\eqref{eq: devetak winter} for the expression for the key rate and the discussion in Section~\ref{sec: techniques}) to benchmark their performance. This simplifies the analysis as compared to performing a full security analysis of the protocol. For a wide range of protocols (which include all the protocols in this section except for Subsections~\ref{sub: rand postselection protocol} and \ref{sub: advantage distillation}), the i.i.d.~attack assumption can be relaxed using the \emph{entropy accumulation theorem} (see Section~\ref{sec: finite key} for further discussion). The protocols that we present in this section were designed to maximise this asymptotic key rate and/or to minimise the experimental requirements (quantified by a suitably chosen figure-of-merit). A summary of the experimental requirements of each protocol can be found in Table~\ref{tab: protocols}

The two common figures-of-merit are the \emph{critical detection efficiency} and the \emph{critical QBER} under the depolarising noise model. The former refers to the minimum overall detection efficiency $\eta$ (which accounts for the total loss in the quantum channel with the entanglement source is placed in the middle) for the key rate to be positive, assuming the other sources of imperfections (such as dark counts, channel noise, etc) are eliminated. Furthermore, many works assume that the source can prepare any two-qubit state that will be optimised to minimise the detection efficiency\footnote{\revision{In practice, most photonic implementations use SPDC sources instead. Thus, the critical detection efficiency would be higher in practice.}}. This figure-of-merit is suitable for fully-photonic implementations. The latter refers to the maximum tolerated error rate $Q$ assuming that in each round, the state being measured is given by
\begin{equation}
    \rho_{AB} = (1-2Q) \ketbra{\Phi^+}{\Phi^+}_{AB} + 2Q \frac{\1_{AB}}{4},
\end{equation}
and the key generating measurement is taken to be $A_0 = B_2 = \sigma_Z$ (and $A_1 = B_3 = \sigma_X$, for the protocol with random key basis). The depolarising noise model is more suitable for the heralded entanglement implementation. We remind the reader that these models are simply used to benchmark the experimental requirements and are not assumed in the security analysis.}

\subsection{The standard DI-QKD protocol} \label{sub: standard protocol}
The first DI-QKD protocol with a security proof specialised for quantum (rather than no-signalling) adversaries, under a collective-attacks assumption, was proposed in the seminal work of Ac\'{i}n~\textit{et~al}.~\cite{acin2007device, pironio2009device}, and was inspired by Ekert's entanglement-based protocol~\cite{ekert1991quantum}. In this review paper, we shall refer to it as the ``standard'' DI-QKD protocol. In each round of the standard protocol, the two honest parties Alice and Bob receive the two parts of an entangled state in each round. Alice then randomly chooses an input $x \in \{0, 1\}$ while Bob randomly chooses an input $y \in \{0,1,2\}$, corresponding to different measurement choices. All measurements have binary outcomes labelled by $a, b \in \{-1, +1\}$. For photonic implementations of the protocols where there might be inconclusive outcomes due to no-detection or double-detection events, it is customary to assign such inconclusive outcomes to a deterministic outcome (e.g.~$+1$) to ensure that the measurements are binary. Here, we suppose that the protocol uses direct reconciliation, where Bob tries to guess Alice's key (based on her syndrome) in the error correction step.

In the generation rounds (i.e.~those used to produce raw data for the secret key), Alice chooses the setting $x = 0$ and Bob chooses the setting $y = 2$. The error-correction step in the protocol is based on the value of the QBER $Q$\footnote{In lossy channels, the fine-grained information about which bits were inconclusive can be used to improve the efficiency of the error correction step.}
\begin{equation}
    Q = \Pr[\sA_0 \neq \sB_2],
\end{equation}
where $\sA_x$ and $\sB_y$ denote the random variables corresponding to Alice's outcome with setting $x$ and Bob's outcome with setting $y$, respectively.\footnote{As we shall discuss in Section~\ref{sec: finite key}, an appropriately designed error-correction procedure can be based on the QBER of the \emph{honest implementation} of the devices, though earlier versions required Alice and Bob to estimate the QBER during the protocol itself.} On the other hand, in the test rounds, 
Alice and Bob use the inputs $x \in \{0, 1\}$ and $y \in \{0,1\}$ to estimate the CHSH value~\cite{clauser1969proposed}, i.e.~the quantity $S$ presented in Eq.~\eqref{eq: CHSH}.

An equivalent statistical test can be formulated in terms of the probability $\omega$ of winning the \textit{CHSH game}. In this game, in each test round, Alice (resp. Bob) chooses between the input values $x = 0$ and $x = 1$ (resp. $y = 0$ and $y = 1$) with uniform probability, and we re-label the outcomes $a, b$ from $\{-1, +1\}$ to $\{0, 1\}$. Then, Alice and Bob win the CHSH game if
\begin{equation}
    a \oplus b = x \cdot y
\end{equation}
where $\oplus$ denotes summation modulo 2. With this, the winning probability is given by
\begin{multline} \label{eq: winning probability}
    \omega = \frac{1}{4} \Big( \Pr[a = b|x=0,y=0] \\
    + \Pr[a = b | x = 0, y = 1] \\
    + \Pr[a = b | x = 1, y = 0] \\
    + \Pr[a \neq b|x = 1, y = 1] \Big).
\end{multline}
The latter formulation is more suitable for security analysis via the entropy accumulation theorem.

The original work~\cite{acin2007device, pironio2009device} analysed the asymptotic security of the protocol under the assumption of collective attacks. The authors managed to discover an explicit attack that saturates their bound, which implies that the bound derived for this protocol in these works is tight. Security against general attacks in the finite-key regime was then proven later \cite{vazirani2014fully,miller2016robust}. However, in the asymptotic limit, those results have less noise tolerance than the one derived using the collective attack assumption. A tighter bound that asymptotically recovers the collective attack result was subsequently obtained via the entropy accumulation theorem~\cite{arnon2018practical}.

\subsection{DI-QKD based on other constraints}
Interestingly, the state used in the optimal attack found in~\cite{pironio2009device} for the standard DI-QKD protocol exhibits an asymmetry with regards to the correlators. For a given CHSH value $S$, the optimal attack yields the following correlators
\begin{equation}
    \begin{split}
    \expectation{\sA_0 \sB_0} &= 2/S,\\
    \expectation{\sA_0 \sB_1} &= 2/S, \\
    \expectation{\sA_1 \sB_0} &= \frac{S^2/2-2}{S},\\
    \expectation{\sA_1 \sB_1} &= -\frac{S^2/2-2}{S},
    \end{split}
\end{equation}
which reflects the asymmetry in the standard protocol: the measurement $A_0$ is used for both key generation and testing, while the measurement $A_1$ is only used for testing.

Based on this observation, one can modify the protocol to use a generalised CHSH inequality that reflects this asymmetry as well~\cite{woodhead2021device, sekatski2021device}:
\begin{equation} \label{eq: asymmetric CHSH}
    S_\alpha = \alpha \expectation{\sA_0 \sB_0} + \alpha \expectation{\sA_0 \sB_1} + \expectation{\sA_1 \sB_0} - \expectation{\sA_1 \sB_1},
\end{equation}
where $\alpha \in \mathbb{R}$ is a free parameter that can be optimised according to the expected behaviour of the devices. The quantum communication layer of the protocol is identical to that of the standard DI-QKD protocol. The difference lies in the parameter estimation step, where the asymmetric CHSH value $S_\alpha$ is estimated in place of the usual CHSH value $S$.

The asymptotic security of the protocol against collective attacks has been proven analytically~\cite{woodhead2021device} and numerically~\cite{sekatski2021device}. In both works, it was discovered that the improvement that one can obtain from using this asymmetric CHSH inequality is more significant in implementations where the noise can be modelled by depolarising channels. However, for fully-photonic implementations, one can only get a marginal improvement in the critical detection efficiency\footnote{This refers to the minimal efficiency $\eta$ (assuming it is identical for all measurements) for the protocol to produce non-zero key rate.}. \revision{Greater improvement can be obtained by considering the bias of the key, quantified by $\expectation{\sA_0}$~\cite{masini2021simple}}.

In the same spirit, one could also modify the protocol such that, for parameter estimation, Alice and Bob estimate the full ``behaviour'', i.e.~all the conditional probabilities
\begin{equation}
    P(a,b|x,y) := \Pr[\sA = a, \sB=b|X=x, Y=y]
\end{equation}
for all $a \in \cA, b\in \cB, x \in \cX, y \in \cY$. Such constraints contain more fine-grained information compared to a single Bell inequality, and hence might give a higher asymptotic secret key rate and noise tolerance as compared to a DI-QKD protocol based on one Bell inequality. It was shown that by using the full behaviour as constraints, the critical detection efficiency for the fully-photonic implementations can be lower as compared to the standard DI-QKD protocol which uses the CHSH value as a constraint~\cite{tan2021computing,brown2021computing,brown2021device}. A related approach is to use a linear programming technique to choose a Bell inequality that is maximally violated by the expected behaviour~\cite{datta2021device}.

\subsection{DI-QKD with noisy pre-processing}
\label{sub: noisy preprocessing}
In the context of DD-QKD, it is known that the robustness of some protocols to experimental imperfections (such as channel loss or noise) can be improved by randomly flipping some of the bits in the raw key before performing error correction and privacy amplification --- a step that is known as \textit{noisy pre-processing}~\cite{kraus2005lower,renner2005information}. The reason for this is that although such random flips reduce the correlation between Alice and Bob (hence increasing the error correction cost), they would also increase Eve's uncertainty about the raw key. Importantly, in some parameter regimes, Eve is ``penalised'' more than Bob, resulting in an overall increase in the key rate.

It was shown in~\cite{ho2020noisy,woodhead2021device} that the same effect is present in DI-QKD, and hence this method can be used to improve its key rates.
More specifically, consider a protocol with the same quantum layer as the standard DI-QKD protocol. However, after estimating the CHSH value $S$ in the parameter estimation step, Alice and Bob perform a noisy pre-processing step (if the protocol did not abort). In this step, for each bit in her raw key, Alice will randomly and independently flip the bit with probability $p$. After the noisy pre-processing step, Alice and Bob will continue the protocol with information reconciliation and privacy amplification. It was found in~\cite{ho2020noisy,woodhead2021device} that this noisy pre-processing yields significant improvements for the photonic implementation of the protocol; for instance, in the case where there is no additional noise in the channel, the critical detection efficiency in an SPDC model is reduced from 90.9\% (in the standard protocol) to 83.2\% (in the protocol with noisy pre-processing). Noisy pre-processing was also studied for DI-QKD protocols employing the asymmetric CHSH inequality~\cite{woodhead2021device, sekatski2021device}. \revision{It was shown that it is possible to reduce the critical detection efficiency of the noisy pre-processing protocol to roughly $80\%$ by accounting for the bias $\expectation{\sA_0}$~\cite{masini2021simple} or using the full behaviour~\cite{brown2021device}.}

\subsection{DI-QKD with random key basis}
In the original proposal of the BB84 protocol, Alice and Bob use both $X$ and $Z$-basis measurements to generate their raw keys~\cite{bennett1984quantum}. There, each basis is chosen with probability $1/2$, and hence the probability of both parties choosing the same basis is also $1/2$. Since there is no correlation when Alice and Bob measure different bases, these rounds are discarded, and hence half of the rounds in the original BB84 protocol are discarded.

To improve the key rate, in most protocols, Alice and Bob single out one of their measurements as a key generating measurement while the other measurements are only used to test the channels~\cite{lo2005efficient}. In these protocols, the key generating measurements are chosen with high probability whereas the test measurements are chosen with low probability. In this way, the proportion of rounds used for key generation is maximised.

However, for the standard DI-QKD protocol, the optimal attack discovered in~\cite{pironio2009device} has the interesting feature that (taking Alice's key-generating measurement to be $A_0$) we have
\begin{equation}
    H(\sA_0|E) \leq H(\sA_1|E),
\end{equation}
\revision{where $H(\sA_x|E)$ is the conditional von Neumann entropy of the outcome of the measurement $A_x$, given Eve's (single-round) quantum side-information $E$.}

In other words, because the key generating measurement was fixed to be $A_0$, Eve could focus on minimising her uncertainty about that measurement at the expense of having higher uncertainty about the other measurement $A_1$. Based on this observation, Schwonnek \textit{et al}. modified the standard DI-QKD protocol to one that uses both of Alice's measurements to generate the raw key~\cite{schwonnek2021device}, to exploit Eve's uncertainty about $\sA_1$. With this proposal, Bob needs to perform an additional measurement to obtain outcomes that are better correlated to Alice's second measurement.

In this protocol, Alice chooses a measurement setting $x \in \{0,1\}$ while Bob chooses a measurement setting $y \in \{0,1,2,3\}$. To generate the raw key, in each key generation round, Alice chooses $x=0$ with probability $p$ and $x=1$ with probability $1-p$. Similarly, Bob chooses $y = 2$ with probability $p$ and $y = 3$ with probability $1-p$. 
They will subsequently apply a sifting step in which they only keep the rounds in which $x=0$ and $y=2$, or $x=1$ and $y=3$. Since there are essentially two pairs of generation inputs in this case, there are hence two QBER values (which can potentially be different) relevant for error correction:
\begin{equation}
\begin{split}
    Q_0 &= \Pr[\sA_0 \neq \sB_2],\\
    Q_1 &= \Pr[\sA_1 \neq \sB_3].
\end{split}
\end{equation}
As for parameter estimation in the test rounds, Alice and Bob estimate the CHSH value $S$, the same as in Eq.~\eqref{eq: CHSH}. 

Conditioned on the round being a key generation round, the overall key rate suffers a factor of $p_s$ penalty due to sifting, where
\begin{equation}
    p_s = p^2 + (1-p)^2.
\end{equation}
We note that when $p = 1$, the protocol is reduced to the standard DI-QKD protocol~\cite{acin2007device, pironio2009device} and we have $p_s = 1$. One then expects that for sufficiently high CHSH value $S$, the penalty due to sifting outweighs the benefit of increasing Eve's uncertainty about the raw key, and hence it would be optimal to choose $p \rightarrow 1$. (Alternatively, some techniques based on pre-shared keys have been proposed to bypass the sifting factor; see~\cite{tan2020improved,bhavsar21} for details.)

The use of the random key basis protocol is most suitable when the channel noise is depolarising, and hence $Q_0 = Q_1$. In light of this, it is better suited for the heralded entanglement implementation of DI-QKD. For fully-photonic implementations, non-maximally-entangled states are normally used and these states have strong correlations in one measurement basis and weaker correlations in other bases. Accordingly, we have $Q_0 > Q_1$ (or vice versa) and the error correction cost for one of the measurement basis is higher than the other, which seems to limit the improvement that one can obtain by random key basis protocols in fully-photonic implementations.

The security of the protocol was first analysed numerically~\cite{schwonnek2021device} in the asymptotic limit, assuming collective attacks. An analytical security bound for random key basis protocols under the same assumptions with $p = 1/2$ was then derived~\cite{masini2021simple}\revision{, where it was noted that the CHSH inequality is an optimal measure of nonlocality for $p = 1/2$}. A detailed finite-key security proof for the random key-basis protocol which also incorporates noisy pre-processing was given in~\cite{tan2020improved}. \revision{By combining the random key-basis and noisy pre-processing, the critical QBER can be increased to 9.33\%~\cite{tan2020improved}.} An implementation of the protocol was recently demonstrated~\cite{zhang2021experimental}.

\subsection{DI-QKD with random post-selection}\label{sub: rand postselection protocol}
In DD-QKD, post-selection is a common practice as photons are occasionally lost in the quantum channel, and hence in some rounds, the receiver's detectors would not click. These rounds are naturally discarded in these QKD protocols as no secret correlation can be derived from these rounds. Discarding these ``no-click'' rounds increases the correlation between Alice and Bob, and hence decreases the cost of error correction.

However, in DI-QKD, some care is needed when post-selection is employed, as discarding some events might open up the detection loophole. For example, when one naively discards the ``no-click'' events, it is possible to violate the CHSH inequality using a classical strategy. In light of this, most DI-QKD protocols would simply assign a deterministic outcome for rounds in which the detectors do not click (instead of discarding these events). This would, in turn, decrease the amount of the certified randomness produced by the measurement and consequently poses a challenge to the implementation of photonic DI-QKD between remote users without qubit amplifiers.

With this caveat in mind, a DI-QKD protocol with a random post-selection step was proposed~\cite{xu2021device, liu2021high}. Similar to the standard protocol, Alice chooses between two binary-output measurements $A_0$ and $A_1$ while Bob chooses between three binary-output measurements $B_0$, $B_1$ and $B_2$. As usual, for both parties, should an inconclusive outcome be obtained, they would map it to ``$1$''. Moreover, as usual, $A_0$ and $B_2$ are used both to generate the key and to test for nonlocality while the other measurements are only used for testing, from which the full behaviour $\{P(a,b|x,y)\}_{a,b,x,y}$ is estimated. Random post-selection consists of the following step: for key generation rounds (rounds in which $x = 0$ and $y = 2$ are chosen), if Alice obtains outcome ``$1$'', she will choose to discard the round with probability $1-p$, while she will choose to keep all the rounds in which she obtains outcome ``$0$''. Similarly, Bob will independently apply the same post-selection strategy. A given round is kept only if both parties agree to keep that round. Using this strategy, a round is kept with probability
\begin{equation}
    P_\cV = \sum_{a,b \in \{0,1\}} P(a,b|x=0, y=2) p^{a+b}.
\end{equation}
Note that to avoid complications from the detection loophole, none of the data from the test rounds are discarded. 

The security of the protocol against collective attacks is proven in the asymptotic limit \cite{xu2021device, liu2021high} using numerical techniques based on the Navascu{\'e}s-Pironio-Ac{\'{i}}n (NPA) hierarchy~\cite{navascues2007bounding, navascues2008convergent, pironio2010convergent} via conditional min-entropy~\cite{xu2021device} and the quasi-relative entropies method~\cite{liu2021high}. Interestingly, it was shown (with the collective attack caveat) that with an ideal source of two-qubit entangled states and pure-loss channel, the critical detection efficiency is as low as 68.5\%~\cite{xu2021device}. An experimental demonstration of the protocol (combined with noisy pre-processing) using spontaneous parametric down-conversion (SPDC) and superconducting single-photon detector of efficiency 87.49\% was performed~\cite{liu2021high}. \revision{The authors showed that in principle, such a setup} could achieve an asymptotic key rate (assuming i.i.d.~attacks) of 446 bit/s over 20 m of standard fibre and an asymptotic key rate of 2.6 bit/s over 220 m, though the classical post-processing of the protocol was not actually implemented in the experiment. \revision{However, at the moment of writing, a full security analysis of the protocol (in particular, one that accounts for non-i.i.d.~attacks) is not yet available.}

\subsection{DI-QKD with advantage distillation} \label{sub: advantage distillation}
All of the protocols that we have listed so far use one-way classical post-processing to convert their raw keys into a pair of secret key. However, in some DD-QKD protocols, it is known that noise tolerance can be significantly improved by adopting two-way classical communication~\cite{chau2002practical,gottesman2003proof,ma2006decoy,bae2007key,watanabe2007key,khatri2017numerical}. Based on this,~\cite{tan2020advantage} studied the so-called \textit{repetition code protocol} under the assumption that Eve performs collective attacks. 

Consider a protocol where in each round, Alice randomly chooses a measurement $A_0, A_1, ..., A_{\abs{\cX}-1}$ while Bob randomly chooses a measurement $B_0, B_1, ..., B_{\abs{\cY}-1}$ with $A_0$ and $B_0$ being the binary-output key-generating measurements. Now, after Alice and Bob have gathered all their raw data from their devices, they divide the outcomes from the key-generating rounds into $m$ blocks, each of size $k$. Focusing on one block, denote the raw output strings in that block as $\vb{A}_0$ and $\vb{B}_0$ respectively. Alice would randomly generate a bit $C$ and send the message $\vb{M} = \vb{A}_0 \oplus (C, C, ..., C)$. Here, $\oplus$ denotes bit-wise summation modulo 2. Bob would reply with a bit $D$, where $D=1$ (indicating the block is ``accepted'') if $\vb{B}_0 \oplus \vb{M} = (C', C', ..., C')$ for some bit $C'$, and otherwise $D=0$ (indicating the block is ``rejected'') in which case they overwrite the values of $C,C'$ with some erasure symbol. Alice and Bob repeat this procedure for every block, thus obtaining some strings $\vb{C} = (C_1, C_2, ..., C_m)$ and $\vb{C}' = (C'_1, C'_2, ..., C'_m)$ respectively. These strings are then used to produce their final key, by applying one-way error correction followed by privacy amplification.

When applied to the standard CHSH-based DI-QKD protocol under the depolarising noise model, advantage distillation using the repetition code protocol increases the optimal critical QBER to $Q_{\text{crit}}^\text{AD} \approx 9.1\%$ from $Q_{\text{crit}}^\text{std} \approx 7.1\%$ in the standard case. 
As for the lossy photonic channel model with no additional noise,  
if the states and measurements are chosen to maximise the CHSH value, the critical detection efficiency for this advantage distillation protocol is  $\eta_{\text{crit}}^\text{AD} \approx 89.1\%$~\cite{tan2020advantage}, which is better than the value of $90.7\%$ obtained by the standard protocol for those parameters. However, it was later observed that if the states and measurements are instead chosen to maximise the key rate directly, the value for the standard protocol can in fact be improved to $88.4\%$~\cite{brown2021computing}. The critical value for advantage distillation in this case has not been computed yet. See~\cite{tan_thesis} for a listing of some of the known detection-efficiency thresholds for the standard protocol, under different choices for the states and measurements.
\section{Security analyses}
\label{sec: techniques}
\newcommand{\rawbit}{\mathsf{S}}
\newcommand{\rDD}{r_{\mathrm{DD}}}
\newcommand{\conv}[1]{\mathop{\mathrm{conv}}\!\left\{#1\right\}}

In this section, we review the techniques to find lower bounds on the asymptotic key rates. 
As a starting point, we focus on the scenario of collective attacks~\cite{devetak2005distillation, tomamichel2009fully} (we shall discuss the situation for general attacks afterward). For this scenario, in every round there is a well-defined Alice-Bob quantum state $\rho_{AB}$ and set of possible measurements that could be performed on it. 
Eve also has an extension $E$ of the state $\rho_{AB}$ (conservatively, we allow this to be a purification), which we refer to as her quantum side-information.\footnote{In the collective-attacks scenario, it does not really matter whether we assume the i.i.d.~structure on only the Alice-Bob states or require it for Eve's purification as well --- this is because all purifications are isometrically equivalent, hence given i.i.d.~Alice-Bob states, we can focus only on an i.i.d.~purification without loss of generality.} Note that throughout this section, since we are focused on single-round analyses, we shall simplify notation by omitting the indices that indicate the round associated to each register (e.g.~we write $A, B, E$ instead of $A_i, B_i, E_i$ for the $i$-th round registers; this is in contrast with other sections where e.g.~$E$ was used to denote the side-information across all rounds). These indices are implicit for all quantum and classical registers throughout this section but they would be made explicit in the rest of the review article.

We shall simplify the overview further by focusing only on protocols where each individual round has the following structure: some public announcements $T$ are made over the classical channel (e.g.~the basis choice of each party, whether a given round is discarded or kept, etc.), and then Alice and Bob generate raw key bits denoted by $\rawbit, \rawbit'$ (this should be understood to refer to the values after any relevant pre-processing).
After all these single-round procedures are performed, Alice sends Bob a single string $\transcript_{\text{EC}}$ for error correction and verification (note that this string typically cannot be included in the single-round announcements $T$, because it may depend on Alice's entire raw key string), which he uses to produce a guess for Alice's raw bits; finally, they apply privacy amplification to produce the final key. This covers most of the protocols we have described above, with the exception of the advantage-distillation protocol (which processes the data in blocks of multiple rounds).

For such protocols, the asymptotic key rate against collective attacks is given by the Devetak-Winter bound~\cite{devetak2005distillation,tomamichel2009fully}
\begin{equation} \label{eq: devetak winter}
    r \propto H(\rawbit|T, E) - H(\rawbit|\rawbit', T),
\end{equation}
where the constant of proportionality depends on the probability that a given round is kept, e.g.~due to sifting or post-selection (alternatively, another way to formalise such processes is to set $\rawbit,\rawbit'$ to a deterministic value for all ``discarded'' rounds, in which case the sifting factor is automatically included in the values of the entropies and the proportionality factor can be set to 1). 
Strictly speaking, the original Devetak-Winter formula was derived for a slightly different context where the states are fully characterised. However, as we shall discuss later in Section~\ref{sec: finite key}, essentially the same formula works for DI-QKD, albeit with slight differences in the interpretations of some terms.

Typically, the $H(\rawbit|\rawbit', T)$ term in the formula is fairly easy to analyse, because it only depends on Alice and Bob's data (and for appropriately designed protocols, can be based on the value in the honest case only; see Section~\ref{sec: finite key}). Hence, the main quantity of interest is the term $H(\rawbit|T,E)$\footnote{In most protocols, the announced classical information is the choice of bases, $X$ and $Y$. Then, the sifting process typically accepts only the rounds in which the bases are ``matched'' (i.e., the measurement outcomes are strongly correlated) --- this allows Eve to deduce $Y$ from $X$ (for the sifted rounds). For protocols where the raw bit is constructed based on Alice's measurement outcome $\sA$ without additional processing, we then have $H(\rawbit|T,E)_\rho = p_s H(\sA|X,E)_{\rho_s}$ where $p_s$ is the sifting rate and $H(\sA|X,E)$ is evaluated on the sifted state.}.
The security analysis basically comes down to solving the following optimisation problem:
\begin{equation}\label{eq:vNopt}
\begin{split}
\inf \quad &H(\rawbit|T,E) \\
\text{s.t.}\quad &\tr[\rho_{AB} \Gamma_j] = \gamma_j, \quad \forall j.
\end{split}
\end{equation}
Here, each $\Gamma_j$ is a linear operator whose expectation value is estimated in the protocol (e.g.~in the standard protocol, there is a single $\Gamma_j$, which is the CHSH operator), and each value $\gamma_j$ can be informally interpreted as the corresponding estimated value (putting aside all finite-statistics considerations). Typically, $\Gamma_j$ is a non-commutative polynomial in the measurement operators. Recall that in the device-independent setting, the quantum state and the measurements (and the Hilbert spaces in which they are living) are all unknown, and hence the infimum must be taken over all these. (Though as a small simplification, for the purposes of this optimisation it usually suffices to restrict to projective measurements only, by constructing a suitable \textit{joint} Naimark dilation --- see e.g.~\cite{harris16,tan_thesis}.)

Remarkably, even though we have initially presented the optimisation~\eqref{eq:vNopt} in the context of collective attacks,
an optimisation of basically the same form is also the quantity of interest
when proving security against general attacks using the entropy accumulation theorem~\cite{dupuis2019entropy, dupuis2020entropy, arnon2018practical, arnon2019simple}. Hence, solving this optimisation also essentially suffices to handle general attacks as well --- we shall discuss further details in Section~\ref{sec: finite key}. That being said, there are also proof frameworks that are not directly based on bounding the single-round conditional von Neumann entropy (e.g.~the quantum probability estimation framework~\cite{knill2018quantum, zhang2020efficient, zhang2020experimental}). For the remainder of this section, though, our focus will be on bounding the optimisation~\eqref{eq:vNopt}.

\subsection{Classes of approaches}
\subsubsection{Jordan's lemma}
The main challenge of analysing the security of DI-QKD is the fact that the measurement devices, including their respective Hilbert space dimensions, are uncharacterised. As such, we cannot exclude \emph{a priori} the possibility that it would take an unbounded number of parameters to parameterise the measurements. Interestingly, for DI-QKD protocols whose security relies only on two binary-outcome measurements, one can reduce the calculations from an unknown Hilbert space to the qubit setting. More precisely, let $A_0$ and $A_1$ be Hermitian operators on an arbitrary Hilbert space $\cH$ with eigenvalues $\pm 1$. There exists a basis in which both $A_0$ and $A_1$ can be written as
\begin{equation}
\begin{split}
    A_0 = \bigoplus_\alpha \vec{a}_{0|\alpha} \cdot \vec{\sigma}\\
    A_1 = \bigoplus_\alpha \vec{a}_{1|\alpha} \cdot \vec{\sigma}\\
\end{split}    
\end{equation}
where $\vec{\sigma} = (\1, \sigma_X, \sigma_Y, \sigma_Z)$ is a vector of Pauli matrices. Note that the argument can be applied separately to each party.

In other words, they are block-diagonal, in blocks of dimension $2 \times 2$. This result is commonly referred to as \textit{Jordan's lemma}. A simple proof of the lemma is given in the work of Pironio \textit{et al}.~\cite{pironio2009device}. It is worth noting that even when one party uses more than two binary-outcome measurements (e.g., in the standard DI-QKD protocol, Bob has one measurement to generate the key, and two measurements to estimate the CHSH value), the lemma can still be useful if the security of the protocol (that is, the bound on Eve's uncertainty about $\rawbit$) only relies on two binary-outcome measurements for each party.

The block-diagonal structure of the measurement allows simplification of the calculations to qubits, because one can interpret each measurement as consisting of a projective measurement to determine which block is being measured, followed by a qubit measurement in the appropriate block. We could then conservatively assume that Eve is the one performing the initial projective measurement herself. Since such a measurement removes the coherence between different blocks, we can conclude that it is sufficient to assume that Eve distributes states which are block-diagonal with blocks of dimension $4 \times 4$\footnote{Applying Jordan's lemma on each party's measurements would yield blocks of dimension $2 \times 2$ for each party. Then, the tensor product $A_x \otimes B_y$ would be block diagonal with each block having dimension of $4 \times 4$ as claimed.}. Hence, one could imagine Eve sharing some common random variable $\Lambda$ with Alice's and Bob's measurement devices, which takes the value $\lambda$ with probability $p_\lambda$\footnote{The parameter $\lambda$ can be thought of as the pair $(\alpha, \beta)$ which specifies the $2 \times 2$ block associated with Alice's and Bob's local measurements}. Depending on the value of $\lambda$, Eve distributes the two-qubit state $\rho^\lambda$ to Alice and Bob, who perform the qubit measurements $A_x^\lambda$ and $B_y^\lambda$, respectively.

Therefore, to prove the security of DI-QKD protocols in which Jordan's lemma applies, it is sufficient to analyse the case in which Alice and Bob are measuring two-qubit states, and to find a \textit{convex} lower bound on the single round conditional von Neumann entropy $H(\rawbit|T,E)$. The convexity requirement arises from the fact that Jordan's lemma allows Eve to adopt a convex combination of qubit strategies instead of picking a single qubit strategy. Thus, to complete the security proof via Jordan's lemma, one has to perform a convex analysis of the bound and/or convexify the bound by taking a \emph{convex hull} (that is, finding the greatest convex function upper-bounded by this bound) of the lower bound derived under the assumption that Alice and Bob receive two qubit states.

While Jordan's lemma can greatly simplify the security analyses of DI-QKD protocols, its applicability is severely limited to protocols which only use two binary-outcome measurements on each party for test rounds. At the time of writing, there has been no extension of the lemma to cover the cases in which more measurements are being performed or when the measurements produce more than two outcomes. In the literature, Jordan's lemma has been applied to analyse the security of DI-QKD protocols both analytically~\cite{acin2007device,pironio2009device,ho2020noisy,woodhead2021device,masini2021simple} and numerically~\cite{schwonnek2021device,sekatski2021device,tan2020improved}. A version of this qubit reduction suitable for multipartite DI protocols was also developed in~\cite{grasselli2021entropy}.

\subsubsection{Hierarchy of semi-definite programs}

Performing an optimisation over quantum states and operators in the device-independent scenario is a challenging task as its feasible set does not admit a simple characterisation and the Hilbert space involved has potentially unbounded dimension. In the context of DI-QKD, the purification of the quantum state measured by the honest parties $\rho_{AB}$ is given to Eve and the joint quantum state can be written as $\ket{\psi}_{ABE}$ that resides in a Hilbert space $\mathcal{H}$ of arbitrary dimension. Hence, we can define a set of operators $\mathcal{O}\coloneqq\{O_k\}_k$ such that $O_k$ resides in the same Hilbert space $\mathcal{H}$. 

We consider optimisations that minimise (or maximise) a linear combination of some \emph{moments} $\expectation{O_i^{\dagger}O_j}\coloneqq\bra{\psi}O_i^{\dagger}O_j\ket{\psi}$ of these operators over all possible quantum state and operators. These optimisations admit a relaxation to a semi-definite program (SDP), which can then be solved using well-known solvers e.g.~SeDuMi \cite{sturm1999using}, SDPT3 \cite{toh1999sdpt3}, MOSEK \cite{mosek2015mosek} etc. This formulation can be achieved by defining a Gram matrix $\Gamma$ where $\Gamma_{ij} \coloneqq \expectation{O_i^{\dagger}O_j},\, O_k \in \mathcal{O}$, which implies that $\Gamma$ must be positive semi-definite, i.e.~$\Gamma \succeq 0$. Hence, the positive semi-definiteness of a matrix, whose elements represent various moments of operators, imposes a necessary condition in the optimisation of interest, thereby resulting in a relaxed optimisation over quantum states and operators. 

Exploiting the properties of this Gram matrix, the Navascu\'{e}s, Pironio and Ac\'{i}n (NPA) hierarchy of SDPs provides a systematic framework to formulate relaxations on optimisation over the set of quantum correlations~\cite{navascues2007bounding, navascues2008convergent}. The $n$-th level of the NPA hierarchy is defined by the positive semi-definiteness of the Gram matrix\footnote{In the case of the NPA hierarchy, the matrix $\Gamma^{(n)}$ contains the moments (up to the $2n$-th moment) of the operators in $\cM$. As such, $\Gamma^{(n)}$ is often referred to as the \textit{moment matrix}.} $\Gamma^{(n)}$ such that $\Gamma^{(n)}_{ij} \coloneqq \expectation{O_i^{\dagger}O_j},\, O_k \in \mathcal{O}_n$ where the set $\mathcal{O}_n$ contains all products of $l$ measurement operators drawn with replacement from the set $\mathcal{M}\coloneqq\{A_i,B_j,\dots\}_{i,j,\dots}$ and the identity operator $\1$, with $1\leq l\leq n$. The measurement operators of different parties are then constrained to be commuting operators as a consequence of the tensor product structure. If a behaviour $P(a,b,\dots|x,y,\dots)$ is compatible with some positive semi-definite $\Gamma^{(n)}$, we say that $P(a,b,\dots|x,y,\dots) \in \mathcal{Q}_n$. Since $\Gamma^{(n)}$ is a sub-matrix of $\Gamma^{(n+1)}$, the positive semi-definiteness of the latter imposes at least as many constraints as the former, which implies that $\mathcal{Q}_{n+1} \subseteq \mathcal{Q}_{n}\, \forall n$. Moreover, it has been established that $\lim_{n\rightarrow \infty}\mathcal{Q}_{n} = \mathcal{Q}'$, the set of behaviours compatible with quantum theory (the \emph{quantum set}), under the assumption that the measurement operators belonging to different parties are commuting.


In the DI-QKD setting this review focuses on, we are working with the set of quantum correlations for measurement operators obeying a tensor product structure, which we shall denote as $\mathcal{Q}$ (in contrast to the aforementioned set $\mathcal{Q}'$ for commuting-operator correlations). 
%
\revision{These two sets were initially claimed to be equal in~\cite{tsirelson1993some}, although it was later realised that this equality was not proven, and the question of whether $\mathcal{Q} = \mathcal{Q}'$ (or as a variant, the question of whether the {closures} of the two sets are equal) came to be known as \emph{Tsirelson's problem}.

While the sets coincide for finite-dimensional systems, Tsirelson's problem in the sense of whether $\mathcal{Q} = \mathcal{Q}'$ was resolved in the negative by~\cite{slofstra2020tsirelson}, who showed that infinite-dimensional systems can give rise to correlations in $\mathcal{Q}'$ but not $\mathcal{Q}$. In fact, it was then shown in~\cite{mip_re} that even $\bar{\mathcal{Q}} \subsetneq \mathcal{Q}'$, where $\bar{\mathcal{Q}}$ is the closure of $\mathcal{Q}$.} 
This means that, for a given problem, the NPA hierarchy might possibly converge to a value that is separated by a constant amount from the value of interest in our framework. Still, this only implies that the resulting bound might not be tight --- by construction, this approach will never give an ``insecure'' (i.e.~over-estimated) value for the entropy. Interested readers may find further detail in~\cite{tsirelson_physical, fritz2012tsirelson, q_corr_sets}, among other references.

More generally, a similar hierarchy of SDP relaxations can be defined for an arbitrary set of operators (not necessarily measurement operators). Importantly, the SDP hierarchy also admits operator inequality constraints by defining \textit{localising matrices}~\cite{pironio2010convergent}. In fact, the SDP hierarchy can be used to lower bound optimisation problems of the following form~\cite[Eq. 39]{pironio2010convergent}
\begin{equation} \label{eq: non-commutative polynomial optimisation}
\begin{split}
\min_{(\cH, \ket{\psi}, \vec{O})} \quad & \bra{\psi} p(\vec{O}) \ket{\psi}\\
\text{s.t.} \qquad & q_i(\vec{O}) \succeq 0, \quad \forall i \in \{1, ..., m\} \\
& r_i(\vec{O}) \ket{\psi} = 0, \quad \forall i \in \{1,\ldots,m'\}, \\
& \bra{\psi} s_i(\vec{O}) \ket{\psi} \geq 0, \quad \forall i \in \{1,\ldots,m''\},
\end{split}
\end{equation}
where the optimisation is taken over all Hilbert spaces $\cH$, all bounded operators $\vec{O} = (O_1, ..., O_k)$ in $\cH$ and all normalised states $\ket{\psi}$ living in $\cH$. \revision{The functions $p(\vec{O})$, $q_i(\vec{O})$, $r_i(\vec{O})$, and $s_i(\vec{O})$ are polynomials in the operators $\vec{O}$ and their adjoints}. Similarly, operators that are related by tensor products can be relaxed into commuting operators. Finally, the SDP hierarchy is shown to converge to the solution of \eqref{eq: non-commutative polynomial optimisation} if the polynomials $\{q_i\}_i$ that define the operator inequality constraints satisfy the so-called \textit{Archimedean} property~\cite{pironio2010convergent}.

\subsection{Techniques based on Jordan's lemma}
As mentioned previously, Jordan's lemma allows us to analyse DI-QKD protocols by deriving lower bounds on the conditional von Neumann entropy $H(\rawbit|T,E)$ assuming that Alice and Bob receive two-qubit states. One then has to show that the resulting bound is convex or, if the derived bound is found to be non-convex, take a convex hull to obtain a convex bound. Once a convex lower bound on the conditional von Neumann entropy for qubit attacks is obtained, by Jordan's lemma, the bound is valid for any attacks involving quantum systems of arbitrary dimension.

To obtain a lower bound on the conditional von Neumann entropy based on the reduction to two-qubit systems, three methods have been presented in the literature.

\subsubsection{Reduction to Bell-diagonal states}
This method is applicable for protocols whose security parameter is independent of the marginal correlators (e.g.~protocols which rely on the CHSH~\eqref{eq: CHSH} or the asymmetric CHSH value~\eqref{eq: asymmetric CHSH} to evaluate security). For these protocols, one could consider that, without loss of generality, the marginal correlators are unbiased, i.e., $\expectation{\sA_x} = \expectation{\sB_y} = 0$. This can be enforced by performing a symmetrisation step, where Alice and Bob randomly perform coordinated bit-flips. However, the security of the protocol is unchanged even if the symmetrisation step is omitted~\cite{scarani2008,pironio2009device}. 

\revision{To understand why the symmetrisation can be omitted in practice, let $\mathsf{R}$ be a uniform random variable that indicates whether or not the bit-flip is performed and let $\tilde{\rawbit}$ denote Alice's bit after the symmetrisation. We can assume that $\mathsf{R}$ is generated independently from a trusted random number generator, and hence is not controlled by Eve. To perform the coordinated bit-flip, we can imagine Alice announcing $\mathsf{R}$ to Bob via a public channel, and hence the conditional entropy that we need to bound is $H(\tilde{\rawbit}|E,\mathsf{R})$. But since the bit $\tilde{\rawbit}$ is a deterministic function of $\rawbit$ and $\mathsf{R}$, we also have $H(\tilde{\rawbit}|E,\mathsf{R}) = H(\rawbit|E)$, which is the original conditional von Neumann entropy that we need to bound if the symmetrisation was not performed. Therefore, the bound that is derived by incorporating the symmetrisation would remain valid for the case when the symmetrisation is not performed. Thus, the symmetrisation was only used to simplify the proof but does not need to be implemented.}

For a given block (specified by $\Lambda = \lambda$), we are free to label the axes of the Bloch sphere such that Alice's and Bob's measurement in that block lie on the $(X,Z)$-plane of the Bloch sphere~\cite{pironio2009device}. With this choice, the bit-flips in the symmetrisation step can be seen as a coordinated Pauli-$Y$ such that we can assume that Eve sends two-qubit states of the form
\begin{equation} \label{eq: Bell block diagonal}
    \bar{\rho}^\lambda = \frac{1}{2} \left[\rho^\lambda + (\sigma_Y \otimes \sigma_Y) \rho^\lambda (\sigma_Y \otimes \sigma_Y)  \right].
\end{equation}

States of this form are block-diagonal in the Bell basis with only two independent non-zero off-diagonal terms. In particular, there is enough freedom to choose the measurements such that 1) they lie in the $(X,Z)$-plane; 2) the off-diagonal terms in \eqref{eq: Bell block diagonal} are purely imaginary; and 3) we have
\begin{equation} \label{eq: Bell diagonal constraints}
    \begin{split}
        p_{\Phi^+} &\geq p_{\Psi^-},\\
        p_{\Phi^-} &\geq p_{\Psi^+},\\
    \end{split}
\end{equation}
where $\{\Phi^+, \Phi^-, \Psi^+, \Psi^- \}$ are the usual Bell states and we denote $p_{\Phi^\pm} = \bra{\Phi^\pm}\bar{\rho}^\lambda \ket{\Phi^\pm}$ and $p_{\Psi^\pm} = \bra{\Psi^\pm}\bar{\rho}^\lambda \ket{\Psi^\pm}$. Finally, since the state $\bar{\rho}^\lambda$ and its complex conjugate $(\bar{\rho}^\lambda)^*$ produce the same statistics for a fixed set of measurements and give Eve the same information, we can assume that Eve distributes states of the form
\begin{equation}
    \rho = \frac{1}{2} \left[\bar{\rho}^\lambda + (\bar{\rho}^\lambda)^*\right],
\end{equation}
which is diagonal in the Bell basis with eigenvalues satisfying \eqref{eq: Bell diagonal constraints}.

By showing that Bell-diagonal states suffice for Eve's optimal attack and focusing on them, we minimise the number of parameters, so that the optimisation of Eve's attack can be performed explicitly. This was done for protocols based on the CHSH value~\cite{pironio2009device, ho2020noisy} and for protocols that use a class of asymmetric CHSH inequalities~\cite{sekatski2021device}. However, the optimisation of Eve's attack is still a non-trivial task that may be difficult to solve analytically, as can be seen from the somewhat involved analysis in~\cite{sekatski2021device}.

\subsubsection{Entropic uncertainty relations}
It is well known that to observe Bell nonlocality, it is necessary for some of the underlying measurements to be incompatible. We can leverage this fact to relate the commutativity of the measurements of the key-generating party to the observed value of Bell violation. The pioneering work in this direction was done by Seevinck and Uffink~\cite{seevinck2007local}, with the assumption that the underlying quantum systems are qubits. A device-independent relation between the local overlap of the measurements and the achievable CHSH value was later derived by Lim \textit{et al.}~\cite{lim2013device} and also independently by Tomamichel and H\"anggi~\cite{tomamichel2013link}. Once the overlap of the measurements is bounded in terms of the CHSH value, one can then apply entropic uncertainty relations~\cite{coles2017entropic} to bound Eve's uncertainty about the outcome of one of the measurements~\cite{lim2013device}.

Extending this idea, given a measurement that generates the key, one can also consider a \textit{virtual} complementary measurement and bound the correlation between the virtual measurement and the other party based on the observed Bell violation~\cite{woodhead2021device, masini2021simple}. \revision{Eve's optimal attack was also studied by \cite{woodhead2021device,masini2021simple} using this approach for the protocol using the generalised CHSH inequality for evaluating security, and noisy pre-processing.} A similar idea was also presented in the work of Zhang \textit{et al}. with focus on complementarity and its extension to finite-key analysis~\cite{zhang2021quantum}. In contrast to the earlier methods where the uncertainty relations were applied on the actual measurements, here, one does not need to bound the overlap of the measurements, but rather bounds the hypothetical correlations that may arise from a virtual complementary measurement.

For the simplest example, suppose that Alice generates the raw key using her $A_0$ measurement, and no noisy pre-processing is applied. We have the following uncertainty relation~\cite{masini2021simple}
\begin{equation}\label{eq: bb84-type bound}
    H(\sA| X = 0, E) \geq 1 - \phi\left(\abs{\expectation{\bar{A}_0 \otimes B}}\right),
\end{equation}
where
\begin{equation}
    \phi(x) \coloneqq h_2\left(\frac{1+x}{2}\right)
\end{equation}
with the binary entropy function defined
\begin{equation}
    h_2(x) \coloneqq -x \log_2 x - (1-x) \log_2(1-x).
\end{equation}
Here, $\bar{A}_0$ is a (virtual, i.e.\ not actually measured) Pauli operator that is orthogonal to $A_0$, and $B$ is any $\pm 1$-observable on Bob's system, which can be optimised to maximise the key rate. Then, to obtain a lower bound on the conditional von Neumann entropy, one has to lower bound $\abs{\expectation{\bar{A}_0 \otimes B}}$ subject to the observed experimental statistics. As an illustration, when the CHSH value is measured in the protocol, we have~\cite{masini2021simple}
\begin{equation} \label{eq: correlation bound}
    \abs{\expectation{\bar{A}_0 \otimes B}} \geq \sqrt{\frac{S^2}{4}-1}.
\end{equation}
Lower bounds of the correlation between Alice's complementary measurement and Bob's virtual measurement for other Bell inequalities can also be derived~\cite{masini2021simple}.

Generally, security analysis via this method is modular, consisting of:
\begin{enumerate}
    \item Derivation of a lower bound on the appropriate measure of uncertainty in terms of the correlation between Alice's complementary measurement and Bob's virtual measurement. This bound depends only on how the raw key is generated (e.g., whether two bases are used to generate the key, whether noisy pre-processing is applied, etc). \revision{In the previous example, this lower bound is given by Eq.~\eqref{eq: bb84-type bound}.}
    
    \item Derivation of a lower bound on the correlation between Alice's complementary measurement and Bob's virtual measurement. This bound depends only on the Bell inequality that is used in the protocol. \revision{In the previous example, this lower bound is given by Eq.~\eqref{eq: correlation bound}.}
    
    \item Convexity analysis to take into account convex combination of qubit attacks. If necessary, the convex hull of the qubit lower bounds computed at discrete points is taken.
\end{enumerate}

This technique is versatile due to its modular design, as it can be easily adapted to different protocols that use different Bell inequalities or protocols with modified raw key generation processes (e.g., with noisy pre-processing or random key basis)~\cite{masini2021simple}. The authors also derived the correlation bounds in terms of both standard and asymmetric CHSH values. A bound that incorporates the marginal correlator related to the key generating measurement was also presented. Furthermore, the second step of the procedure can be done numerically when analytical bounds on the correlation are hard to obtain.

\subsubsection{Numerical analyses via Jordan's lemma}
The two methods mentioned earlier aim to derive analytical solutions to the optimisation of the conditional von Neumann entropy. However, such analytical solutions can be hard to obtain, and hence the security analyses can be rather involved. For DD-QKD, numerical approaches~\cite{winick2018reliable, coles2016numerical} are known to simplify the security analyses of DD-QKD and could (in some cases) provide tighter bounds than the one provided by analytical techniques. However, there are some roadblocks that have to be addressed before we can apply these numerical methods to DI-QKD.

While Jordan's lemma allows us to reduce the analysis to two-qubit states and qubit measurements, the qubit measurements performed by Alice and Bob are still unknown, and hence the standard numerical techniques for DD-QKD~\cite{coles2016numerical,winick2018reliable} cannot be directly applied. This is because these techniques are catered to solve an optimisation of the form
\begin{equation}
    \begin{split}
        \inf_\rho \quad  &H(\rawbit|T,E)\\
        \text{s.t.} \quad & \tr[\Gamma_j(M_{a|x}, M_{b|y})\rho] = \gamma_j
    \end{split}
\end{equation}
for fixed sets of measurements $\{M_{a|x}\}_{a,x}$ and $\{M_{b|y}\}_{b,y}$. Here, $\{\Gamma_j\}_j$ are functions of the measurement operators of Alice and Bob (for example, the Bell operator) with $\gamma_j$ being its expected value. On the other hand, the problem we are trying to solve has the following form
\begin{equation}
    \begin{split}
        \inf_{\rho,\{M_{a|x}\}_{a,x},\{M_{b|y}\}_{b,y}} \quad  &H(\rawbit|T,E)\\
        \text{s.t.} \quad & \tr[\Gamma_j(M_{a|x}, M_{b|y})\rho] = \gamma_j
    \end{split}
\end{equation}
which is more challenging as the measurements $\{M_{a|x}\}_{a,x}$ and $\{M_{b|y}\}_{b,y}$ are now optimisation variables as well. In particular, the $\tr[\Gamma_j(M_{a|x}, M_{b|y})\rho]$ terms are now nonlinear in the optimisation variables, making it harder to (for instance) express the problem as a convex optimisation.

To approach this, one can begin by noting that a constrained optimisation is always lower-bounded by its Lagrange dual problem, which in this case takes the form
\begin{multline} \label{eq: lagrange dual qubit}
   \sup_{\vec{\lambda}} \left(\inf_{\rho, \{M_{a|x}\}, \{M_{b|y}\}} H(\rawbit|T,E)\right) \\
    - \sum_j \lambda_j \left(\tr[\Gamma_j(M_{a|x}, M_{b|y})\rho] - \gamma_j\right).
\end{multline}

Furthermore, in the context of DI-QKD, this lower bound is in fact typically tight in a certain sense (see e.g.~\cite{tan2021computing} for further discussion). 
In principle, to get the tightest possible bound, one would have to solve the optimisation over the Lagrange multipliers $\vec{\lambda}$ in~\eqref{eq: lagrange dual qubit}; however, this may be challenging in practice. Instead, we can simply observe that since the optimisation over $\vec{\lambda}$ is a supremum, any specific choice of $\vec{\lambda}$ still yields a secure lower bound, and hence it is fine to simply use heuristic methods to find a good choice of $\vec{\lambda}$. While this might not yield a perfectly tight bound, it does have the convenient property that each choice of $\vec{\lambda}$
yields a lower bound on the conditional entropy $H(\rawbit|T,E)$ that is affine (and hence convex) in $\vec{\gamma}$. Therefore, it can be directly used with Jordan's lemma to prove the security against attacks using quantum systems in arbitrary Hilbert spaces.

To now apply the standard numerical techniques for DD-QKD to analyse the security of DI-QKD protocols, one can use the following procedure \cite{schwonnek2021device, tan2020improved}:
\begin{enumerate}
    \item Apply Jordan's lemma to reduce the calculations to two-qubit analysis.
    
    \item Parameterise the measurement of Alice and Bob. The following parameterisation can always be adopted due to the freedom to label the Bloch sphere axes
    \begin{equation}
        \begin{split}
            A_0 &= \sigma_Z, \quad A_1 = \cos(\alpha) \sigma_Z + \sin(\alpha) \sigma_X,\\
            B_0 &= \sigma_Z, \quad B_1 = \cos(\beta) \sigma_Z + \sin(\beta) \sigma_X,
        \end{split}
    \end{equation}
    for some $\alpha, \beta \in [0, \pi]$ with $\{M_{a|x}\}_{a,x}$ and $\{M_{b|y}\}_{b,y}$ being their respective projectors.
    
    \item For a fixed set of measurement angles $(\alpha, \beta)$, use the standard numerical techniques for DD-QKD \cite{coles2016numerical, winick2018reliable} to find a reliable lower bound on the conditional von Neumann entropy. The standard numerical techniques for DD-QKD \cite{coles2016numerical, winick2018reliable} formulate this problem as a non-linear convex optimisation. Alternatively, the problem can be re-cast as an SDP which could be more efficiently solved but at the cost of the tightness of the bound~\cite{schwonnek2021device}.
        
    \item For fixed $\alpha$, we optimise Bob's measurement. For convenience, we could write $b_Z \coloneqq \cos(\beta)$ and $b_X \coloneqq \sin(\beta)$. Then, for fixed value of $\alpha$, we write the Bell operator in terms of $\alpha, b_Z, b_X$. The feasible region is then characterised by the semi-circle $$\cS_\text{FR} = \{(b_Z, b_X)| b_Z^2 + b_X^2 = 1, b_X \geq 0\}. $$
    
    One could relax the above feasible region into a polytope that fully contains the semi-circle $\cS_\text{FR}$. Crucially, the objective function is affine with respect to $(b_Z, b_X)$, and hence the optimal solution of the relaxed problem is attained at an extremal point of the polytope. A simple algorithm is as follows \cite{schwonnek2021device}. We start with a polytope characterised by the following extremal points $\cV = \{(1,0), (1,1), (-1, 1), (-1,0) \}$ and evaluate the conditional entropy for each point using the method in Step 3 and then find the point $b_{\min}$ that minimises the conditional entropy. We cut the polytope by removing $b_{\min}$ and add a new edge which is tangential to the semi-circle $\cS_\text{FR}$. We iterate the process until the desired precision is achieved.
    
    \item Optimise Alice's measurement: we divide the interval $[0,\pi]$ into several (not necessarily uniform) intervals $\{[\alpha_j - \delta_j, \alpha_j + \delta_j] \}_j$ and then apply a continuity bound~\cite{schwonnek2021device, sekatski2021device, tan2020improved}. Specifically, for $\delta \in [0,\pi]$, we want to find a monotonically increasing function $\epsilon_\text{con}(\delta)$ that bounds the change in the objective function if we change $\alpha$ by at most $\delta$.
    
    At the end of this step, we want to obtain a lower bound of the form
    \begin{equation}
        \min_\alpha f(\alpha) \geq \min_j f(\alpha_j) - \epsilon_\text{con}(\delta_j)
    \end{equation}
    where $f(\cdot)$ is the result of the optimisation in Step 4.
\end{enumerate}

If Eq.~\eqref{eq: lagrange dual qubit} is solved directly for a fixed value of $\vec{\lambda}$, then the resulting bound is affine and we are done. However, if one instead had to re-formulate the problem, and has solved an auxiliary optimisation problem (e.g.~the approach in~\cite{schwonnek2021device} was formulated in terms of a trace norm instead of the conditional von Neumann entropy directly), additional steps to obtain a convex bound would be necessary.

\subsection{Techniques based on SDP hierarchies}

Approaches that leverage Jordan's lemma allow us to work with simple two-qubit systems, but their applicability is limited the two-input-two-output scenarios. Here, we present another approach for the security analysis of DI-QKD, via the SDP hierarchy that we have introduced earlier. These techniques are versatile, being applicable in any Bell scenario, but we have to formulate the objective functions as linear functions of elements of the moment matrix. Therefore, this first requires bounding the conditional von Neumann entropy $H(\rawbit|T,E)$ in terms of such functions.

\subsubsection{Conditional min-entropy}
We first consider protocols where the raw bit $\rawbit$ is obtained by simply taking Alice's measurement outcome $\sA$ (i.e., we do not consider noisy pre-processing or random post-selection). A rather convenient lower bound on the conditional von Neumann entropy (for a fixed state $\rho_{\sA X E}$ that depends on Eve's attack) is the conditional min-entropy\footnote{If $\sA$ is uniform and binary-valued, a tighter bound $H(\sA|X,E) \geq 2(1-P_g(\sA|X,E))$ holds~\cite{briet2009}.}
\begin{align}
    &H(\sA|X,E) \nonumber\\
    &\geq H_{\min}(\sA|X,E) \geq -\log_2 P_g(\sA|X,E),
\end{align}
where $P_g(\sA|X,E)$ is the (maximal) guessing probability, defined as
\begin{equation}
    P_g(\sA|X,E) = \max_{\{E_{a|x}\}_{a,x}} \sum_{a,x} \nu_x P(a|x) \tr \left[E_{a|x} \rho_E^{(a,x)} \right],
\end{equation}
where $\nu_x$ is the probability of Alice choosing measurement setting $X=x$ conditioned on successful sifting (i.e., on Bob also choosing the appropriate key-generating setting), $\rho_E^{(a,x)}$ is Eve's quantum side information conditioned on Alice choosing the setting $X=x$ and obtaining outcome $\sA = a$, and $\{E_{a|x}\}_{a,x}$ describes a projective measurement by Eve that can depend on $x$. Note that the last point accounts for the fact that Eve might adjust her measurement strategy depending on Alice's announcement. In particular, for protocols that generate keys from multiple bases, Eve could keep her ancillae until Alice announces her basis choice and adjust her measurement strategy accordingly. 

Equivalently, we can express the guessing probability in terms of the tripartite quantum state $\ket{\psi}$ shared between Alice, Bob, and Eve.
\begin{equation}
    P_g(\sA|X,E) = \max_{\{E_{a|x}\}_{a,x}} \sum_{a,x} \nu_x \bra{\psi} M_{a|x} \otimes \1 \otimes E_{a|x}  \ket{\psi}.
\end{equation}
The expression is clearly an expectation value of an operator polynomial, and hence can be computed using the NPA hierarchy\footnote{The SDP also simultaneously solves the optimisation over Eve's attack (the state being prepared as well as Alice's measurement).}.

For protocols that incorporate noisy pre-processing~\cite{ho2020noisy, tan2020improved} (say, the raw bit is obtained by flipping the measurement outcome with probability $p$), the guessing probability (assuming that $\sA \in \{0,1\}$) is given by
\begin{multline}
P_g(\rawbit|X,E) \\
= \max_{\{E_{a|x}\}_{a,x}} \bigg( (1-p) \sum_{a,x} \nu_x \bra{\psi} M_{a|x} \otimes \1 \otimes E_{a|x}  \ket{\psi} \\
+ p \sum_{a,x} \nu_x \bra{\psi} M_{a|x} \otimes \1 \otimes E_{a \oplus 1|x}  \ket{\psi} \bigg).
\end{multline}
Then, similarly, we can use the bound $H(\rawbit|X,E) \geq H_\text{min}(\rawbit|X,E) = -\log_2 P_g(\rawbit|X,E)$.

For the random post-selection protocol discussed in Subsection~\ref{sub: rand postselection protocol}, a simple modification to the definition of the guessing probability can also be made. Recall that in this protocol, Alice and Bob discard generation rounds with outcome `1' with probability $1-p$, while keeping all rounds with outcome `0'. Denoting the event in which both parties agree to keep a round by $\cV$ and letting the key generating settings for Alice (resp. Bob) be given by $x^*$ (resp. $y^*$), then the guessing probability is given by\revision{\footnote{This bound was first studied in~\cite{thinh2016randomness} where a deterministic post-selection scheme was used}}
\begin{multline}
    P_g(\sA|X = x^*, E, \cV) \\ = \max_{\{E_{a}\}_{a}} \frac{1}{P_\cV} 
    \sum_{a,b \in \{0,1\}} p^{a+b} \bra{\psi} M_{a|x^*} \otimes N_{b|y^*} \otimes E_{a} \ket{\psi},
\end{multline}
where 
\begin{equation}
    P_\cV = \sum_{a,b \in \{0,1\}} P(a,b|x^*, y^*) p^{a+b}
\end{equation}
is the probability of keeping a given round.

\subsubsection{Gibbs-Golden-Thompson method}
While the lower bound via conditional min-entropy is extremely versatile and can be used in any protocol in conjunction with the NPA hierarchy, the gap between conditional von Neumann entropy and the conditional min-entropy can be large in many situations. The work of~\cite{tan2021computing} aimed to provide another versatile bound that is compatible with the NPA hierarchy and also tighter than the conditional min-entropy.

Consider a protocol with a single key generating measurement (denoted by $x^*$ and $y^*$, for Alice and Bob, respectively). Suppose that there is no noisy pre-processing or postselection involved. Then, we are interested in finding a lower bound on $H(\sA| X = x^*, E)$. Suppose further that the quantities measured in the protocol can be represented as the expectation values of operator polynomials $\{\Gamma_j\}_j$, with:
\begin{equation}
    \Gamma_j = \sum_{a,b,x,y} c^{(j)}_{abxy} M_{a|x} \otimes N_{b|y},
\end{equation}
for some constants $c^{(j)}_{abxy}$. Let $\gamma_j$ be the expected value for the polynomial $\Gamma_j$ on the state $\rho_{AB}$.

Using Gibbs' variational principle and a generalisation of the Golden-Thompson inequality, it can be shown that the conditional von Neumann entropy $H(\sA|X = x^*, E)$ is lower-bounded by~\cite{tan2021computing}
\begin{equation} \label{eq: Gibbs-Golden-Thompson bound}
   H(\sA| X = x^*, E) \geq \expectation{\Gamma}_\rho - \ln \expectation{K}_\rho,
\end{equation}
with $\Gamma = \sum_j \lambda_j \Gamma_j$ for some $\{\lambda_j\}_j$ and
\begin{equation} \label{eq: K operator arbitrary}
    K = \cT^\dagger \cT \left[ \int_{\mathbb{R}} \mathrm{d}t \, \beta(t) \prod_{j} \abs{e^{\frac{1+it}{2} \lambda_j \Gamma_j} }^2 \right],
\end{equation}
where
\begin{align}
    &\cT[\rho_{AB}] \nonumber\\
    &= \sum_{a \in \cA} \left(\sqrt{M_{a|x^*}} \otimes \1_B \right) \rho_{AB} \left(\sqrt{M_{a|x^*}} \otimes \1_B \right)^\dagger
\end{align}
is the pinching channel associated to the key-generating measurement and \begin{equation}
    \beta(t) = \frac{\pi/2}{\cosh(\pi t) + 1}.
\end{equation}

As mentioned earlier, with a suitable Naimark dilation, the measurements can be assumed to be projective, and consequently the pinching channel $\cT$ is both self-adjoint and idempotent. Thus, $\cT^\dagger \cT[\rho] = \cT[\rho]$. Furthermore, if $\{X_{k|l}\}_{k,l}$ are projectors (not necessarily rank-1), we have
\begin{equation}
    \exp\left(\sum_{k} c_{k|l} X_{k|l}\right) = \sum_{k} e^{c_{k|l}} X_{k|l},
\end{equation}
and hence $K$ from Eq.~\eqref{eq: K operator arbitrary} can be simplified to
\begin{equation}
    K = \cT \left[\int_{\mathbb{R}} \mathrm{d}t \, \beta(t) \abs{\prod_{xy}  \sum_{ab} e^{\kappa_{abxy}} M_{a|x} \otimes N_{b|y}}^2 \right],
\end{equation}
with
\begin{equation}
    \kappa_{abxy} = \frac{1 + i t}{ 2}\sum_j \lambda_j c^{(j)}_{abxy}.
\end{equation}

For fixed $\lambda_j$ and $c^{(j)}_{abxy}$, the integral can be evaluated in closed form~\cite{tan2021computing}, and hence maximising $\expectation{K}_\rho$ is a non-commutative polynomial optimisation which can be evaluated using an SDP hierarchy. Once an upper bound on $\expectation{K}_\rho$ is obtained, we can plug it to Eq.~\eqref{eq: Gibbs-Golden-Thompson bound}. Note that we have a freedom to choose the operator polynomial $\Gamma$ (which would in turn, determine the operator polynomial $K$), and hence we could optimise our choice of $\Gamma$ to maximise $H(\sA|X = x^*, E)$.

One could also use the same bound for protocols that use multiple key-generating basis~\cite{schwonnek2021device}. The idea is that for any $\alpha_x > 0$, we have
\begin{equation} \label{eq:tangent}
    \ln \expectation{K_x} \leq \frac{\expectation{K_x}}{\alpha_x} + \ln \alpha_x - 1.
\end{equation}
The right hand side is simply a tangent line of $\ln \expectation{K_x}$ at point $\expectation{K_x} = \alpha_x$.

To obtain a lower bound on the conditional entropy
\begin{equation*}
    H(\sA|X,E) = \sum_{x} \nu_x H(\sA|X=x, E),
\end{equation*}
it is sufficient to evaluate the following instead
\begin{align}
    H(\sA|X,E) &\geq \sum_x \nu_x \left( \expectation{\Gamma}_\rho - \ln \expectation{K_x}_\rho \right) \nonumber\\
    &\geq \expectation{\Gamma}_\rho - \expectation{ \sum_x \frac{\nu_x}{\alpha_x} K_x}_\rho \nonumber \\
    & \hspace{3cm}-\sum_{x} \nu_x \ln \alpha_x + 1.
\end{align}
In the first line, we apply the Gibbs-Golden-Thompson method for each $x$, assuming that the same operator polynomial $\Gamma$ is used for all settings. Then, in the second line, we use the tangent bound~\eqref{eq:tangent}. One could then use the SDP hierarchy to maximise $\expectation{\sum_x (\nu_x/\alpha_x) K_x}_\rho$. Again, the choice of the operator polynomials $\Gamma$ and $K_x$ as well as the tangent points $\alpha_x$ can be optimised to maximise the key rate.

\subsubsection{Iterated mean divergences}
While the Gibbs-Golden-Thompson method~\cite{tan2021computing} provides a promising improvement over the conditional min-entropy method, it requires significantly more computational resources. For example, in the simplest scenario (i.e., two-input-two-output), the Gibbs-Golden-Thompson method requires optimisation of a sixth-degree polynomial whereas the conditional min-entropy method only requires optimisation of a second-degree polynomial.
Brown \textit{et al}.~\cite{brown2021computing} proposed another bound on the conditional von Neumman entropy that can be more efficiently computed than the one obtained via Gibbs-Golden-Thompson method.

To that end, they defined the so-called \textit{iterated mean divergences} $D_{(\alpha_k)}(\rho||\sigma)$ \cite{brown2021computing}, which are a family of R\'{e}nyi divergences. These divergences are characterised by the constant
\begin{equation}
    \alpha_k \coloneqq 1 + \frac{1}{2^k-1}, \quad (k \in \mathbb{N}).
\end{equation}

For a given state $\rho \in \mathrm{D}(\cH_{AB})$ (where $\mathrm{D}(\cH_{AB})$ is the set of normalised density matrices living in the Hilbert space $\cH_{AB}$), and a R\'{e}nyi divergence $\mathbb{D}$, we may define its associated conditional entropy
\begin{equation} \label{eq: iterated mean entropy}
    \mathbb{H}^\uparrow(A|B)_\rho = - \sup_{\sigma_B \in \mathrm{D}(\cH_B)} \mathbb{D}(\rho_{AB}||\1_A \otimes \sigma_B).
\end{equation}
For iterated mean divergences, the corresponding conditional entropies are~\cite{brown2021computing}
\begin{equation} \label{eq: BFF1 entropy}
    H_{(\alpha_k)}^\uparrow(A|B)_\rho = \frac{\alpha_k}{1 - \alpha_k} \log_2 Q_{(\alpha_k)}^\uparrow(\rho), 
\end{equation}
where $Q_{(\alpha_k)}^\uparrow(\rho)$ is defined as
\begin{equation} \label{eq: BFF1 Q}
    \begin{split}
    \max_{\{V_i\}_{i=1}^k} \quad & \tr\left[ \frac{(V_1 + V_1^\dagger)}{2} \rho \right] \\
    \text{s.t. } \quad  & \tr_A[V_k^\dagger V_k] \preceq \1_B \\
    & V_1 + V_1^\dagger \succeq 0 \\
    & \begin{pmatrix}
    \1 & V_i\\
    V_i^\dagger & \frac{(V_{i+1} + V_{i+1}^\dagger)}{2}
    \end{pmatrix} \succeq 0 \quad \forall i < k
    \end{split}
\end{equation}

In the context of QKD, the quantum state that we are interested in is the classical-quantum state $\rho_{\sA E} = \sum_{a \in \cA} \ketbra{a}{a}_{\sA} \otimes \tr_{A} [(M_{a|x^*}  \otimes \1_{E}) \ketbra{\psi}{\psi}_{AE}]$. By writing $V_i = \sum_{a,a' \in \cA} \ketbra{a}{a'}_{\sA} \otimes V_i^{(a,a')}$ and $V_{i,a} = V_i^{(a,a)}$, we can upper bound $Q^\uparrow_{(\alpha_k)}(\rho)$ using some straightforward algebra as
\begin{equation}
    \begin{split}
    \max_{\{V_{i,a}\}_{i,a}} \quad & \sum_{a \in \cA} \tr \left[ \left(M_{a|x^*} \otimes \left( \frac{V_{1,a} + V_{1,a}^{\dagger}}{2} \right)\right) \ketbra{\psi}{\psi}\right]  \\
    \text{s.t. } \quad  & \sum_{a \in \cA} V_{k,a}^\dagger V_{k,a} \preceq \1_E \\
    & V_{1,a} + V_{1,a}^\dagger \succeq 0, \quad \forall a\in \cA \\
    & V_{i+1,a} + V_{i+1,a}^\dagger \succeq 2 V_{i,a}^\dagger V_{i,a} \quad \forall i < k, a \in \cA
    \end{split}
\end{equation}

This allows us to bound $Q^\uparrow_{(\alpha_k)}(\rho)$ using the objective function of a non-commutative polynomial optimisation problem which can be relaxed using the SDP hierarchy. To implement the operator inequality constraints, we can use the localising matrix technique~\cite{pironio2010convergent}. Therefore, $Q^\uparrow_{(\alpha_k)}(\rho)$ (and hence the conditional entropy $H^\uparrow_{(\alpha_k)}(A_{x^*}|E)$) can be computed using the SDP hierarchy with the help of localising matrices. The SDP relaxations are then of the form
\begin{equation} \label{eq: BFF1 NPA}
    \begin{split}
    \max \quad & \sum_{a \in \cA} \bra{\psi} \left[M_{a|x^*} \left( \frac{V_{1,a} + V_{1,a}^{\dagger}}{2} \right) \right] \ket{\psi} \\
    \text{s.t. } \quad  & \1 - \sum_{a \in \cA} V_{k,a}^\dagger V_{k,a} \succeq 0 \\
    & V_{1,a} + V_{1,a}^\dagger \succeq 0, \quad \forall a\in \cA \\
    & V_{i+1,a} + V_{i+1,a}^\dagger - 2 V_{i,a}^\dagger V_{i,a} \succeq 0 \quad \forall i < k, a \in \cA
    \end{split}
\end{equation}
where additionally, we impose that $\{M_{a|x}\}_{a,x}$ and $\{V_{i,a}, V_{i,a}^\dagger\}_{i,a}$ commute. We also impose the constraints from the observed statistics. Finally, the conditional entropy associated to the iterated mean divergences are lower bounds on the conditional von Neumann entropy. \revision{As $k$ increases, the bound $H(\sA_{x^*}|E) \geq H^\uparrow_{(\alpha_k)}(\sA_{x^*}|E)$ (which was obtained by plugging in an upper bound on $Q_{(\alpha_k)}^{\uparrow}(\rho)$ from optimisation problem \eqref{eq: BFF1 NPA} to Eq.~\eqref{eq: iterated mean entropy}) becomes tighter.}

The main advantage of the iterated mean divergence method, as compared to the Gibbs-Golden-Thompson method, is that it applies the NPA hierarchy to low degree polynomials, at the cost of introducing more monomials and the localising matrix, which could reduce the overall computation time. Furthermore, it is proven that the lowest level of the iterated mean hierarchy (i.e., $k = 1$ or $\alpha_k = 2$) can be relaxed to the conditional min-entropy, which guarantees that the iterated mean divergence method would be at least as tight as the conditional min-entropy bound. The iterated mean method is also shown to be able to give better bounds than the Gibbs-Golden-Thompson method in the low noise regime, but the Gibbs-Golden-Thompson method gives better bounds in the high noise regime.

\subsubsection{Quasi-relative entropies}
Finally, the tightest bound that has been obtained so far in this family of methods was presented by Brown, Fawzi and Fawzi~\cite{brown2021device}. They showed that one can use the integral representation of the logarithm function and the \textit{Gauss-Radau quadrature}\footnote{The Gaussian quadrature is a family of numerical techniques of approximating definite integrals by taking a discrete weighted sum of its integrand evaluated at appropriately chosen nodes. The Gauss-Radau quadrature is a variant of the Gaussian quadrature with one of its nodes fixed to one of the endpoints of the integration. For more details, we refer the reader to~\cite[pg.103]{davis1984methods}} to obtain a rational lower bound for the logarithm function:
\begin{equation}
 \ln(x) \geq r_m(x) \coloneqq \sum_{i = 1}^m w_i f(t_i, x),
\end{equation}
with
\begin{equation}
 f(t, x) \coloneqq \frac{x - 1}{t(x-1) + 1}
\end{equation}
and
\begin{equation*}
 \lim_{m\to\infty} r_m(x) = \ln(x).
\end{equation*}
Here, $\{(t_i, w_i)\}_{i=1}^m$ are nodes and weights of the $m$-point Gauss-Radau quadrature in the interval (0,1] with a fixed node $t_m = 1$.

We next define
\begin{equation}
    F_t(x, y) \coloneqq yf(t, x/y) = \frac{y(x - y)}{t(x-y) + y},
\end{equation}
which gives the bound
\begin{equation}
    y \ln(y/x) \leq -\sum_{i = 1}^m \frac{w_i}{\ln 2} \underbrace{yf(t_i, x/y)}_{F_{t_i}(x,y)}.
\end{equation}
Now, observe that the quantum relative entropy
\begin{equation}
D(\rho||\sigma) \coloneqq \tr[\rho(\log \rho - \log \sigma)]
\end{equation}
can be expressed using $F(x, y) \coloneqq y\log(y/x)$ as
\begin{equation}\label{eq: BFF bound}
   D(\rho||\sigma) = \tr[F(\sigma, \rho)] \leq -\sum_{i = 1}^m \frac{w_i }{\ln 2} \underbrace{\tr[F_{t_i}(\sigma, \rho)]}_{\eqqcolon D_{F_{t_i}}(\rho||\sigma)},
\end{equation}
where $D_{F_{t}}(\rho||\sigma)$ is the \textit{quasi-relative entropy} defined using the function $F_t$. The key insight is that this quantity admits the variational expression
\begin{multline} \label{eq: quasi-relative entropy variational}
\frac{1}{t} \inf_{Z} \{ \tr\left[\rho (\1 + Z + Z^\dagger + (1-t)Z^\dagger Z) \right] \\
    + t \, \tr[\sigma Z^\dagger Z] \}
\end{multline}
where the infimum is taken over all bounded operators in the Hilbert space $\cH$ in which $\rho$ and $\sigma$ live. Since the objective function of this optimisation is a linear combination of moments, its optimal value can be approximated using the NPA hierarchy.

For protocols with a single key-generating basis, we can express the relevant conditional von Neumann entropy in terms of the quantum relative entropy as
\begin{equation}
    H(\sA_{x^*}|E) = - D(\rho_{\sA E}||\1_{\sA} \otimes \rho_E),
\end{equation}
where $\sA$ is Alice's classical register holding her raw bit, and the classical-quantum state $\rho_{\sA E}$ is given by
\begin{equation}
    \rho_{\sA E} = \sum_{a \in \cA} \ketbra{a}{a}_{\sA} \otimes  \tr_{AB}[(M_{a|x^*} \otimes \1_{BE}) \ketbra{\psi}{\psi}_{ABE}].
\end{equation}
This yields the lower bound
\begin{equation}
    H(\sA_{x^*}|E) \geq \sum_{i = 1}^m \frac{w_i}{\ln 2} D_{F_{t_i}}(\rho_{\sA E}||\1_\sA\otimes \rho_E)
\end{equation}
on the conditional von Neumann entropy.

Assuming that the alphabet for Alice's raw bit $\cA$ is finite, $Z$ can be written as $Z = \sum_{a,a' \in \cA} \ketbra{a}{a'} \otimes Z_{(a,a')}$ with $Z_{(a,a')}$ acting on the Hilbert space $\cH_E$. After some straightforward algebra, the variational problem in Eq.~\eqref{eq: quasi-relative entropy variational} for the quasi-relative entropy $D_{F_t}(\rho_{\sA E}||\1_\sA\otimes \rho_E)$ can then be solved using the SDP hierarchy. The objective function for each node $t$ is
\begin{multline} \label{eq: BFF2 NPA}
    \frac{1}{t} \inf_{\{Z_{a}\}} \sum_{a \in \cA} \bra{\psi} M_{a|x^*} (Z_{a} + Z_{a}^\dagger + (1-t) Z_{a}^\dagger Z_{a}) \\
    + t (Z_{a} Z_{a}^\dagger) \ket{\psi},
\end{multline}
where $Z_a = Z_{(a,a)}$.

In principle, one should perform the optimisation for all $t_i$ in a single SDP by including all $\{Z_{a,i}\}_{a,i}$ in the moment matrix and using the RHS of Eq.~\eqref{eq: BFF bound} as the objective function (instead of the quasi-relative entropy for a given $t_i$) to obtain the tightest bound. However, in practice, a faster solution can be obtained by solving the optimisation \eqref{eq: BFF2 NPA} for each Gauss-Radau node separately. This numerical trick provides a valid lower bound to the case in which we perform the optimisation for all $t_i$ in a single SDP. Furthermore, in protocols where tight analytical bounds are known, the bounds obtained by the faster algorithm are still very close to the analytical bounds for a sufficiently high number of nodes.

After solving \eqref{eq: BFF2 NPA} for each node $t_i$ in the Gauss-Radau quadrature, we can obtain a lower bound on the conditional von Neumann entropy $H(A_{x^*}|E)$. Moreover, as we increase the number of nodes $m$, the approximation converges to the conditional von Neumann entropy~\cite{brown2021device}.

Just like the Gibbs-Golden-Thompson~\cite{tan2021computing} and the iterated mean divergence methods~\cite{brown2021computing}, the quasi-relative entropy method can be easily adapted to other DI-QKD protocols that include noisy pre-processing, random post-selection, as well as random key bases. Importantly, for each optimisation, if we optimise the quasi-relative entropy for each node separately, the size of the moment matrix is smaller than the ones used in the Gibbs-Golden-Thompson and iterated mean divergence method, although one has to perform the optimisation~\eqref{eq: BFF2 NPA} $m$ times (with different values of $t$). As the required computation time scales linearly with $m$, in many cases, the quasi-relative entropy method allows us to obtain a tighter bound with less computation time.

\subsection{Upper bounds}
\begin{table*}[t]
    \centering
    \caption{Upper bounds on DI-QKD key rates, and the requirements on protocols for the bounds to apply}
    \begin{tabular}{>{\centering}p{0.35\linewidth}*5c}
    \toprule
      &          & distill  & fixed key  & both parties & specific\\
      & protocol & key from & generating & announce      & Bell\\
Bounds & dependent & outputs & inputs & inputs &  inequality\footnotemark[1]\\
    \midrule
    \hline
        Eq.~\eqref{eq: partial transpose upper bound}~\cite{christandl2021upper}; Eq.~\eqref{eq: general upper bound from REE}~\cite[Thm 4]{kaur2021dynamic} & \xmark & \xmark & \xmark & \xmark & \xmark\\
    \hline
        Eq.~\eqref{eq: intrinsic nonlocality}~\cite[Thm 22]{kaur_2020}; & \xmark & \cmark & \xmark & \cmark & \xmark\\
    \hline
        Eq.~\eqref{eq: cc attack bound}~\cite[Eq. 5]{farkas2021bell};
        Eq.~\eqref{eq: ccsq for broadcast}~\cite[Thm 11]{kaur2021dynamic}
                               & \xmark & \cmark & \xmark & \cmark & \xmark\\
    \hline
        Eq.~\eqref{eq: upper bound Bell diagonal}~\cite[Thm 9]{arnon2021upper}                               & \xmark & \cmark & \cmark & \cmark & \cmark\\
    \hline 
        Eq.~\eqref{eq: ccsq for fixed key inputs}~\cite[Eq. 4]{kaur2021dynamic}
                               & \xmark & \cmark & \cmark & \xmark\footnotemark[2] & \cmark\\
    \hline
        Eq.~\eqref{eq: tripartite extensions}~\cite{lukanowski2022upper}\footnotemark[3] & \cmark & \cmark & \cmark & \cmark & \cmark\\
    \bottomrule
    \end{tabular}
    \footnotetext[1]{This refers to whether the bound is applicable only to protocols that use a specific Bell inequality to test for nonlocality.}
    \footnotetext[2]{Requires whether a given round is used for testing or key generation to be announced at the classical post-processing stage.}
    \footnotetext[3]{While the other bounds apply independently of whether the protocol use one-way or two-way classical post-processing, \cite{lukanowski2022upper} also adapts the upper bound to the type of classical post-processing used in the protocol.}
    \label{tab: upper bound asumption}
\end{table*}


While security proofs require us to find lower bounds on the secret key length $\ell$ (and equivalently, the secret key rate $r$), upper bounds on the secret key rate of DI-QKD protocols indicate how much further they can be improved. For example, when the upper bound of the key rate for a specific protocol matches its lower bound, we can conclude that the security proof is tight and one needs to modify the protocol if one hopes to improve its key rate, instead of trying to improve the proof.

One can also derive upper bounds that hold for a \emph{family} of protocols, as was done in a number of recent works~\cite{kaur_2020,winczewski2019,arnon2021upper,christandl2021upper,kaur2021dynamic,lukanowski2022upper}, some of which we discuss further below. In the most general case, one could allow that family to include all possible DI-QKD protocols (based on some fixed honest behaviour), but the existing works consider slightly more restricted families, e.g.~only protocols where the secret key is generated from the device outputs (rather than inputs), or in some cases only protocols where the inputs are announced. In this section, we shall refer to such bounds as \textit{protocol-independent}, though with the implicit understanding that the bounds hold only for a corresponding family of protocols.

\revision{As with protocol-dependent bounds}, if such a bound were to be saturated by a protocol, one could conclusively rule out the possibility that any alternative protocols from the family can attain a higher key rate, and instead focus protocol design efforts on protocols outside that family. \revision{However, as such bounds cannot be optimised using features of one specific protocol, but must apply to all protocols from the family, they tend to be less tight than protocol-dependent bounds.} 


\subsubsection{Explicit attacks for specific protocols}\label{sub: explicit attacks specific protocols}
\revision{The simplest way to derive an upper bound for a specific protocol is to consider a particular \emph{attack}, consisting of the \emph{strategy} used to produce the behaviour, and possibly some post-processing that Eve does on her side information. In particular, we refer to pairs $(\rho, \cM)$ as \emph{quantum strategies}, where $\rho$ is the tripartite quantum state shared by Alice, Bob and Eve, and
    \begin{equation}
        \cM \coloneqq \{ \{M_{a|x}\}_{a,x}, \{N_{b|y}\}_{b,y} \}
    \end{equation}
    is the set of measurements applied to $\rho$ to produce the observed behaviour $P(a,b|x,y)$:
    \begin{equation}
        \forall x,y : \tr\left[ \left( M_{a|x} \otimes N_{b|y} \right) \rho_{AB} \right]  = P(a,b|x,y).
    \end{equation}}
We can then compute the achievable key rate of a given protocol under that attack using a tight formula, e.g.~the Devetak-Winter bound, which is optimal in a particular context~\cite{devetak2005distillation}.
This approach is inherently protocol-dependent, i.e.~the resulting upper bound is only applicable to the specific protocol under consideration. For example, the key rate originally computed in~\cite{pironio2009device} for the standard DI-QKD protocol \revision{is also an upper bound}, because the authors discovered an explicit attack on the protocol that saturated their lower bound.

One way to come up with explicit attacks is to use a heuristic numerical optimisation 
after assuming certain Hilbert space dimensions for the quantum systems. Since any valid quantum strategy would yield a valid upper bound on the secret key rate, certified global optimality is not needed, and hence a heuristic optimisation is valid here.

\revision{Recently, another method of deriving protocol-dependent upper bounds was considered in~\cite{lukanowski2022upper}. The authors developed an attack, called the \textit{convex combination attack}, which involves constructing a tripartite behaviour $P(a,b,e|x,y)$ that can be obtained from a quantum state and measurements, such that
\begin{equation}\label{eq: tripartite extensions}
    \forall x,y : \sum_{e} P(a,b,e|x,y) = P(a,b|x,y),
\end{equation}
where $P(a,b|x,y)$ is the behaviour observed by Alice and Bob. This was first introduced in \cite{farkas2021bell} in the context of deriving protocol-independent bounds. More details are provided in Section~\ref{sub: intrinsic info explicit attacks}, but in brief, this involves decomposing $P(a,b|x,y)$ as a \emph{convex combination} of local and nonlocal behaviours. 

Now, upper bounds on the key rate of any tripartite behaviour obeying Eq.~\eqref{eq: tripartite extensions} are also upper bounds on the DI-QKD key rate of $P(a,b|x,y)$. However, the key benefit of the convex combination attack is that tripartite behaviours constructed using it can be easily adapted to various protocols, and efficiently optimised to maximise the local weight in the decomposition of the behaviour using linear programming, thereby providing tighter upper bounds.

For example\footnote{While this is the class of protocols considered in~\cite{lukanowski2022upper}, it is only an example: the attack can be applied to any protocol~\cite[Sec. 2.1]{lukanowski2022upper}.}, consider DI-QKD protocols where Alice and Bob
\begin{enumerate}
    \item classically post-process the outputs in rounds where they have chosen the pre-agreed key generation inputs $x^*$ and $y^*$ to generate the key, and
    \item announce their inputs.
\end{enumerate}

We first define the joint distribution
\begin{equation}
    P^*(a,b,e) \coloneqq P(a,b,e|x^*,y^*),
\end{equation}
denote the random variables corresponding to $e$ by $\sE$, and recall that $\rawbit$ and $\rawbit'$ denote Alice and Bob's raw key bits, which are their pre-processed outputs $\sA$ and $\sB$ respectively. This pre-processing might involve, for example, binning of inconclusive outcomes or noisy pre-processing. Now, consider protocols of this type that use one-way classical post-processing, where Alice's message to Bob consists of information reconciliation, privacy amplification, and a message $\mathsf{M}$ stochastically generated from $\rawbit$ in each round. Then, the key rate is upper-bounded by the \textit{Csiszar-Korner bound}~\cite{csiszar1978broadcast,ahlswede1993common}
\begin{equation}
    r \leq H(\rawbit|\mathsf{M},\sE)_{P^*} - H(\rawbit|\rawbit',\mathsf{M})_{P^*},
\end{equation}
where the entropies are evaluated on the classical probability distributions $P^*$, after applying the pre-processing maps $\sA \to \rawbit$, $\sB \to \rawbit'$ and $\rawbit \to \mathsf{M}$ specified by the protocol.

On the other hand, for protocols of this type that use two-way classical post-processing, the key rate is upper-bounded by the \textit{intrinsic information}:
    \begin{equation}\label{eq: classical intrinsic info}
        r \leq I{(\sA;\sB \downarrow \sE)}_{P^*} \coloneqq \min_{\Lambda: \sE \rightarrow \sE'} I{(\sA;\sB|\sE')}_{P^*},
    \end{equation}
where $I{(\sA;\sB|\sE')}_{P^*}$ is the \emph{conditional mutual information} of $P^*$, and the minimisation is taken over all classical systems $\sE'$ and stochastic maps $\Lambda$ from $\sE$ to $\sE'$. The intrinsic information can, in turn, be upper-bounded by simple choices such as $\sE' = \sE$ and $\Lambda$ being the identity channel (i.e.\ the conditional mutual information itself), which is easy to compute given $P^*$.

One interesting application of this method was to show that the upper bounds thus found on the critical detection efficiency for various protocols with two-way post-processing are almost optimal. We consider the following protocols:
\begin{enumerate}
    \item a noisy-preprocessing protocol, evaluating security using \emph{generalised CHSH inequalities} which account for bias in the probability of the different outcomes~\cite{masini2021simple},
    \item a noisy-preprocessing protocol, evaluating security using 
 the NPA hierarchy, with constraints provided by the full behaviour~\cite{brown2021device}, and
    \item the random post-selection protocol~\cite{xu2021device}.
\end{enumerate}

In the table below, the best known upper bounds on the critical detection efficiency for the protocols are juxtaposed with the lower bounds on the critical efficiency obtained from the convex combination attack.

\begin{center}
\renewcommand*{\arraystretch}{1.25}
\begin{tabular}{>{\centering\arraybackslash}p{0.5\linewidth}*2c}
\toprule
Protocol & Upper & Lower \\ \midrule
Generalised CHSH inequalities & 80.26\% & 79.15\% \\ \hline
Full behaviour                & 80.00\% & 79.04\% \\ \hline
Random post-selection         & 68.5\%  & 66.66\% \\ \hline
\bottomrule
\end{tabular}
\end{center}

These results imply that there is limited room for improving the noise tolerance of these protocols through better security proofs. Hence, it may be more worthwhile to consider different protocols instead.

The authors also apply the techniques discussed here to develop upper bounds on a large variety of protocols, and on different classes of states. These useful results can be found in~\cite[App. C and D]{lukanowski2022upper}.}

\subsubsection{Intrinsic information and intrinsic nonlocality}\label{sub: intrinsic nonlocality}

\revision{In the context of deriving more general (protocol-independent) upper bounds on the achievable DI-QKD key rates, the \textit{quantum intrinsic information}, which generalises the classical intrinsic information~\eqref{eq: classical intrinsic info}, is frequently used as a starting point. The quantum intrinsic information of a state is an upper bound on its DD-QKD key rate, and is defined in terms of the \emph{quantum conditional mutual information.}} For a tripartite quantum state $\rho_{ABE}$, the quantum conditional mutual information $I{(A;B|E)}_\rho$ is defined as
\begin{multline}
    I(A;B|E)_\rho \coloneqq H(AE)_\rho + H(BE)_\rho \\
    - H(E)_\rho - H(ABE)_\rho,
\end{multline}
and the quantum intrinsic information $I(A;B \downarrow E)$ is then defined as~\cite{christandl2007unifying}
\begin{equation}
    I(A;B \downarrow E)_\rho \coloneqq \inf_{ \Lambda:\, E \rightarrow E'} I(A;B|E')_\sigma,
\end{equation}
where the infimum is taken over all states $\sigma_{ABE'} = (\1_{AB} \otimes \Lambda) [\rho_{ABE}]$, with $\Lambda$ being a quantum channel to be optimised.

\revision{A natural way to generalise the quantum intrinsic information to the device-independent setting is to optimise the conditional mutual information over all tripartite states giving a specified behaviour.\footnote{\revision{Since we are also optimising over Eve's subsystem, there is no need for a separate optimisation over quantum channels, as in the definition of the quantum intrinsic information.}} This was done in~\cite{kaur_2020}, which introduced the \textit{quantum intrinsic nonlocality}.}
For a given correlation $P(a,b|x,y) \in \mathcal{Q} $, this quantity can be understood as 
the amount of nonlocality present in $P(a,b|x,y)$ and is defined as:
\begin{equation}
	\label{eq: intrinsic nonlocality}
	N^{Q}(\sA;\sB)_P \coloneqq \sup_{P(x,y)} \inf_{\rho_{\sA \sB \sX \sY E}} I(\sA;\sB|\sX \sY E)_{\rho_{\sA \sB \sX \sY E}}
\end{equation}
where $P(x,y)$ is the joint probability distribution of the inputs $x$ and $y$, and the infimum is taken over all classical-quantum states $\rho_{\sA \sB \sX \sY E}$ (with $\sA \sB \sX \sY$ being classical registers corresponding to the outputs and inputs of Alice and Bob) that are ``compatible with'' the correlation $P(a,b|x,y)$. More precisely, this means that the optimisation is taken over states of the form
\begin{multline} \label{eq: feasible states}
    \rho_{\sA \sB \sX \sY E} = \sum_{a,b,x,y} P(x,y) \ketbra{a,b,x,y}{a,b,x,y}_{\sA \sB X Y} \\
    \otimes \tr_{AB}[(M_{a|x}\otimes N_{b|y} \otimes \1_E)\rho_{ABE}], 
\end{multline}
where $(\rho_{AB} = \tr_E[\rho_{ABE}], \cM)$ is some quantum strategy for $P(a,b|x,y)$.

\cite{kaur_2020} showed that, for an i.i.d.~device characterised by the correlation $P(a,b|x,y)$, the quantum intrinsic nonlocality $N^{Q}(\sA;\sB)_P$ gives a protocol-independent upper bound on the secret key rate that can be achieved using a large family of DI-QKD protocols (against a quantum adversary). The family of protocols to which this bound applies includes those that use pre-processing and advantage distillation, but is restricted to protocols where the measurement choices are announced and the secret key is extracted from the measurement outcomes. This covers the majority of existing DI-QKD protocols, but there are some exceptions, such as protocols where the secret key is extracted from the measurement settings (inputs) instead~\cite{DIQKDmeasureinput}.

Note that as we are concerned about upper bounds, one does not need to exactly solve for the infimum in the definition~\eqref{eq: intrinsic nonlocality} --- any feasible $\rho_{ABE}$ in Eq.~\eqref{eq: feasible states} will yield a valid upper bound (although $P(x,y)$ will still have to be optimised). This is helpful, as this optimisation problem may be difficult to solve.

\subsubsection{Intrinsic information under explicit attacks}\label{sub: intrinsic info explicit attacks}

\revision{Similarly, while the optimisation in the definition of the intrinsic information may be difficult to solve, in certain cases, there are upper bounds on it which can be explicitly computed, which would then in turn be upper bounds on the key rate in those cases. For instance, for the family of protocols where Alice and Bob
\begin{enumerate}
    \item classically post-process the outputs in rounds where they have chosen the pre-agreed key generation inputs $x^*$ and $y^*$ to generate the key,
    \item announce their inputs, and 
    \item use only the CHSH value $S$ and the QBER $Q$ to determine whether or not to abort the protocol (we refer to protocols fulfilling this condition as \emph{CHSH-based protocols}),
\end{enumerate}
\cite{arnon2021upper} found an explicit attack, for which the intrinsic information had an upper bound that was easy to compute.

The attack uses the following state}
\begin{multline}
    \ket{\psi}_{ABE} = \sqrt{\frac{1+C}{2}} \ket{\Phi^+}_{AB}\ket{0}_E \\
    + \sqrt{\frac{1-C}{2}} \ket{\Phi^-}_{AB}\ket{1}_E,
\end{multline}
and the following measurements
\begin{equation}
    \begin{split}
        A_0 &= \sigma_Z\\
        A_1 &= \sigma_X,\\
        B_0 &= \frac{1}{\sqrt{1+C^2}} \sigma_Z + \frac{C}{\sqrt{1+C^2}} \sigma_X,\\
        B_1 &= \frac{1}{\sqrt{1+C^2}} \sigma_Z - \frac{C}{\sqrt{1+C^2}} \sigma_X,
    \end{split}
\end{equation}
where $C = \sqrt{(S/2)^2 - 1}$. For Bob's key measurement $B_2$, he uses $\sigma_Z$ with probability $1-2Q$ and with probability $2Q$, he outputs a random bit. This strategy was an optimal attack on the DI-QKD protocol of~\cite{pironio2009device}; the observation made in~\cite{arnon2021upper} was essentially that it also yields upper bounds on a broader family of protocols by considering the intrinsic information. Explicitly, they obtained the upper bound
\begin{equation} \label{eq: upper bound Bell diagonal}
    1 + h_2(a_{S,Q}) - h_2(Q) - h_2\left( \frac{1+\sqrt{(S/2)^2-1}}{2} \right),
\end{equation}
where $a_{S,Q} = \frac{1}{2} (1 + \sqrt{1 + Q (1-Q) (S^2 - 8)})$~\cite[Thm 9]{arnon2021upper}.
However, as this bound comes from evaluating this attack on CHSH-based protocols, it does not directly apply to those using other Bell inequalities, such as~\cite{woodhead2021device, sekatski2021device}.

\revision{Like the quantum intrinsic nonlocality $N^Q$ from~\cite{kaur_2020}, this bound applies to key rates from protocols where the key is distilled from the outputs. However, it is further restricted to protocols using fixed key generation inputs for both parties ($x^*$ and $y^*$), while $N^Q$ also applies to protocols using multiple generation inputs. Therefore, this approach also does not apply, for example, to the random key basis protocol of~\cite{schwonnek2021device}.

However, as discussed in~\cite[App. B]{arnon2021upper}, this approach can be an upper bound for $N^Q$ in some settings. For a CHSH-based protocol, using the above quantum strategy, \eqref{eq: upper bound Bell diagonal} can be used to bound on $N^Q$ for the generated behaviour. Further, for any given $S\gtrsim 2.2$, this bound is tighter than the bound on $N^Q$ of a different quantum strategy (which can be parametrised by $S$) derived in~\cite[Sec. 7.1]{kaur_2020}. This approach therefore yields a tighter upper bound in this setting.}

Another significant subsequent development was the work~\cite{farkas2021bell}, which gave the first explicit example of a correlation that is nonlocal (in that it violates a Bell inequality) but cannot yield any secret key under 
a large family of DI-QKD protocols~\cite{farkas2021bell}; namely, those protocols where Alice and Bob
\begin{enumerate}
    \item classically post-process the outputs to generate the key, and
    \item announce their inputs.
\end{enumerate}
Note that most DI-QKD protocols --- in particular, all protocols covered in Section~\ref{sec: protocols} --- fall into this family.

The approach used in that work was to construct a so-called convex combination attack, which was first discussed in Section~\ref{sub: explicit attacks specific protocols}.
In such an attack, \revision{the DI-QKD system exhibits a local behaviour} $P^{\cL}(a,b|x,y)$, and a nonlocal behaviour $P^{\cN\cL}(a,b|x,y)$, with probabilities $q_{\cL}$ and $1-q_{\cL}$ respectively. When the local behaviour is used, it is assumed that Eve has full information about $a$ and $b$, but when the nonlocal behaviour is used, we assume that she is completely uncorrelated with Alice and Bob. Hence, the overall tripartite behaviour is given by
\begin{multline}\label{eq: cc attack behaviour}
    P(a,b,e|x,y) = q_{\cL} P^{\cL}(a,b|x,y) \delta_{(a,b),e} \\
    + (1-q_{\cL}) P^{\cN\cL}(a,b|x,y) \delta_{?, e},
\end{multline}
where $\delta$ is the Kronecker delta and $e =\mathrm{?}$ denotes the case in which Eve is not correlated to Alice and Bob. \revision{We assume that Eve has maximised the local probability $q_{\cL}$, given an appropriately chosen $P^{\cN\cL}(a,b|x,y)$, while constrained by Eq.~\eqref{eq: tripartite extensions}. Then, defining $P_{xy}(a,b,e) \coloneqq P(a,b,e|x,y)$ and $P(x,y)$ the joint probability of using inputs $x$ and $y$, we have the following upper bound on the key rate of such protocols on the behaviour $P$:
\begin{equation}\label{eq: cc attack bound}
    \sum_{x,y} P(x,y) I{(\sA;\sB\downarrow\sE)}_{P_{xy}},
\end{equation}
where we sum over all pairs $(x,y)$ where rounds which used $(x,y)$ are used to generate the key.}

In~\cite{farkas2021bell}, the above convex combination attack is applied to correlations arising from arbitrarily many projective measurements on two-qubit Werner state with visibility $v$~\cite{werner1989quantum}. More precisely, by choosing the nonlocal correlation $P^{\cN \cL}(a,b|x,y)$ to be the one attainable using the maximally entangled state (i.e., $v = 1$) and the local correlation to be the one obtained using a local deterministic strategy, one can derive a lower bound on the critical visibility, i.e.~the minimum visibility for positive intrinsic information. They thus obtained a lower bound on the critical visibility of $v^L_\text{crit} \approx 0.7263$. On the other hand, there exist projective measurements that give rise to nonlocal correlations whenever the visibility is higher than $v_{\cN \cL} \approx 0.6964$~\cite{divianszky2017qutrit}. Since $v^L_\text{crit} > v_\mathcal{NL}$, it can be concluded that nonlocality is not sufficient for this family of DI-QKD protocols: there exist nonlocal behaviours that do not allow such protocols to generate secret keys.

\subsubsection{Extensions via the cc-squashed entanglement}

\revision{\cite{kaur2021dynamic}, building on the fact that many of these results were effectively upper-bounding the quantum intrinsic information of the joint post-measurement state of Alice, Bob and Eve, sought to unify and compare these bounds. The authors defined} the \emph{classical-classical (cc) squashed entanglement}~\cite[Def. 10]{kaur2021dynamic}
\begin{multline}
    E^{cc}_{sq}(\rho,\cM,P(x,y)) \coloneqq \\
    \sum_{x,y} P(x,y) \inf_{\Lambda_E} I(\sA;\sB|E)_{\sigma(x,y)},
\end{multline}
where
\begin{equation}
    \sigma(x,y) \coloneqq (\Lambda_{A|x} \otimes \Lambda_{B|y} \otimes \Lambda_E)[\psi^{\rho}],
\end{equation}
with $\rho$ being the bipartite state shared by Alice and Bob, $\psi^{\rho}$ the tripartite purification\footnote{As explained in the proof of~\cite[Obsv. 4]{kaur2021dynamic}, any purification is sufficient for Eve's optimal attack.} thereof, $\Lambda_{A|x}$ the map $\rho_A \mapsto \sum_a \tr[M_{a|x} \rho_A] \ket{a}\bra{a}$, $\Lambda_{B|y}$ the analogous map for $\{N_{b|y}\}_b$, and $\Lambda_E$ an arbitrary CPTP map on Eve's subsystem. \revision{Note that the terms $\inf_{\Lambda_E} I(\sA;\sB|E)_{\sigma(x,y)}$ are equal to the intrinsic information of $\sigma(x,y)$. However, a key additional insight is that $E^{cc}_{sq}$ is convex in $\rho$~\cite[Lem. 6 and 9]{kaur2021dynamic}.

Using~\cite[Obsv. 4]{kaur2021dynamic}, which shows that any extension can be obtained from applying an appropriate map to a purification, the quantum intrinsic nonlocality $N_Q$ can then be expressed as
\begin{equation}
    N_Q{(\sA; \sB)}_P = \sup_{P(x,y)} \inf_{(\rho, \cM)} E^{cc}_{sq}(\rho, \cM, P(x,y)),
\end{equation}
with the infimum taken over all quantum strategies for $P(a,b|x,y)$. As discussed in Section~\ref{sub: intrinsic nonlocality}, $N^Q$ gives an upper bound on a large class of protocols, but optimising the cc-squashed entanglement over different feasible sets can give bounds for more restricted classes of protocols that are tighter than those known from other approaches.


In particular, for any protocol where Alice and Bob
\begin{enumerate}
    \item classically post-process the outputs in rounds where they have chosen the pre-agreed key generation inputs $x^*$ and $y^*$ to generate the key, and
    \item announce whether a given round is used for testing or key generation\footnote{
     \revision{
     Under these conditions, Eve could simply attack every round as if it used the inputs $x^*$ and $y^*$, and later identify the key-generating rounds based on the announcement. She would thus have the same amount of information on the key-generating rounds as if Alice and Bob had announced their inputs.
     }
    },
\end{enumerate}
\begin{equation}\label{eq: ccsq for fixed key inputs}
    \inf_{(\sigma, \cN)} E^{cc}_{sq} \left(\sigma, \cN,P(x,y) = \delta_{xx^{*}}\delta_{yy^{*}}\right)
\end{equation}
is an upper bound on the key rate achievable using a quantum strategy $(\rho, \cM)$, where the infimum is taken over all strategies $(\sigma, \cN)$ which are indistinguishable from $(\rho, \cM)$ in that protocol~\cite[Eq. 4]{kaur2021dynamic}. For example, we optimise over all quantum strategies with the same CHSH value and QBER for CHSH-based protocols, while we optimise over all quantum strategies with the same behaviour if the protocol uses the full behaviour.

As a concrete application, for any CHSH-based protocol where inputs are announced and the key is distilled from the outputs in rounds where a specific pair of inputs are used, consider the upper bound from the Werner state attack~\cite{farkas2021bell} and the upper bound from \eqref{eq: upper bound Bell diagonal}~\cite{arnon2021upper} for the corresponding CHSH and QBER values. These upper bounds are lower-bounded by~\eqref{eq: ccsq for fixed key inputs}~\cite[Cor. 5]{kaur2021dynamic}. This seems rather trivial: if we optimise over compatible states and use the intrinsic information instead of upper-bounding it with the conditional information, our bound will clearly be tighter. However, since $E^{cc}_{sq}$ is convex in $\rho$, \eqref{eq: ccsq for fixed key inputs} lies below the convex hull in $\rho$ of the above two bounds~\cite[Thm 9]{kaur2021dynamic}. Although \eqref{eq: ccsq for fixed key inputs} may not be easily computable, this means that the convex hull of the above two bounds, which is computable, is an upper bound for \eqref{eq: ccsq for fixed key inputs}. Symbolically, let $\conv{\cdot}$ denote the convex hull, and $r(\rho)$ the maximum key rate that can be obtained from $\rho$ using this class of device-independent protocols. Then,
\begin{equation}
    \conv{\text{\eqref{eq: upper bound Bell diagonal}, Werner state attack}} \geq \text{\eqref{eq: ccsq for fixed key inputs}} \geq r(\rho).
\end{equation}

Turning to a different application, the bound~\eqref{eq: cc attack bound}~\cite[Eq. 5]{farkas2021bell} applies to protocols using arbitrary Bell inequalities and multiple key-generating inputs, but still requires the key to be generated from the outputs, and additionally requires the inputs to be announced. However, it is superseded by~\cite[Thm 11]{kaur2021dynamic}, which shows that~\eqref{eq: cc attack bound} is lower-bounded by
\begin{equation}\label{eq: ccsq for broadcast}
    \inf_{(\sigma,\, \cN)} E^{cc}_{sq}(\sigma, \cN, P(x,y)),
\end{equation}
with the infimum taken over all strategies giving the marginal behaviour $P(a,b|x,y) = \sum_e P(a,b,e|x,y)$, and that this infimum is a convex upper bound on the key rate of this behaviour under such protocols. As above, this implies that, if we have a number of upper bounds on~\eqref{eq: cc attack bound}, their convex hull is in turn an upper bound on the key rate.
}

\subsubsection{General upper bounds on the DI-QKD key rate}

\revision{Another approach to finding upper bounds is to focus on the bipartite Alice-Bob quantum state $\rho$ which is measured to produce the behaviour in a DI-QKD protocol. We define $\rDD(\rho)$ as the maximum achievable DD-QKD secret key rate when Alice and Bob share $\rho$. Since DI-QKD protocols can be seen as a special case of DD-QKD protocols, $\rDD(\rho)$ is an upper bound for the DI-QKD rate of \emph{any} protocol which measures $\rho$ to obtain the behaviour.
However, general upper bounds on $\rDD$ (e.g.~those reviewed in~\cite{christandl2007unifying}) may be difficult to compute on states of interest. In the other direction, it may be difficult to find a state that can generate a behaviour of interest, but which has a low $\rDD$ with an easily computable and reasonably tight upper bound. Therefore, this approach may be difficult to use in practice.}

In~\cite{christandl2021upper}, an alternative quantum strategy for a given behaviour is constructed by considering the partial transpose of the state in an honest implementation. Consider a strategy $(\rho, \cM)$ giving behaviour $P(a,b|x,y)$. Let $\rho^\Gamma$ denote the state $\rho$ with the partial transpose applied to Bob's subsystem. Then, $P(a,b|x,y)$ can be obtained from $\rho^{\Gamma}$ using the measurements $\{M_{a|x}\}_{a,x}$ and $\{N_{b|y}^T\}_{b,y}$, i.e.
\begin{equation}
    \tr[(M_{a|x} \otimes N_{b|y}) \rho] = \tr[(M_{a|x} \otimes N_{b|y}^T) \rho^\Gamma].
\end{equation}

For the partial-transposed strategy to be valid, we need $\rho^\Gamma \succeq 0$, i.e.~the quantum state $\rho$ must have a positive partial transpose (PPT). Hence, for a bipartite PPT state $\rho$ shared between Alice and Bob, \revision{the key rate of any DI-QKD protocol which measures this state} is upper-bounded by
\begin{equation}\label{eq: partial transpose upper bound}
    \min \{ \rDD(\rho), \rDD(\rho^\Gamma) \}.
\end{equation}

An interesting application of these results was to show that there are PPT states $\rho$ where the \textit{lower bound} on $\rDD(\rho)$ is high but the \textit{upper bound} on $\rDD(\rho^\Gamma)$ is very low~\cite{christandl2021upper}\footnote{For this family of states, the lower bound on the key rate is given by the Devetak-Winter protocol~\cite{devetak2005distillation} while the upper bound was computed in~\cite{christandl2017private}.}. \revision{This implies a huge gap between the DD and DI-QKD rates of such PPT states.}

In~\cite[Thm 4]{kaur2021dynamic}, bounds on the DI-QKD key rate from a general bipartite state $\rho$ with measurements $\cM$ by decomposing $\rho$ as:
\begin{equation}
    \rho = p \rho^L + (1-p) \rho^{NL},
\end{equation}
such that $(\rho^L, \cM)$ gives a local behaviour and $(\rho^{NL}, \cM)$ gives a nonlocal behaviour.

Taking infima over strategies $(\sigma^L, \cN^L)$ and $(\sigma^{NL}, \cN^{NL})$ that reproduce the behaviour of $(\rho^L, \cM)$ and $(\rho^{NL}, \cM)$ respectively, we then have the following upper bound on the DI-QKD key rate of $(\rho, \cM)$:
\begin{equation}\label{eq: general upper bound from REE}
    p \inf_{ (\sigma^L,\, \cN^L) } E_R(\sigma^L) + (1-p) \inf_{ (\sigma^{NL},\, \cN^{NL}) } E_R(\sigma^{NL})
\end{equation}
where $E_R(\rho)$ is the \emph{relative entropy of entanglement}
\begin{equation}\label{eq: relent entanglement}
    E_R(\rho) \coloneqq \min_{\rho'\text{ separable}} D(\rho||\rho'),
\end{equation}
and $D(\cdot||\cdot)$ is the quantum relative entropy. This bound follows because $E_R(\rho) \geq \rDD(\rho)$~\cite{horodecki2009general}, and the strategies $(\sigma^{L},\, \cN^{L})$ and $(\sigma^{NL}, \,\cN^{NL})$ can be used to construct a strategy $(\sigma, \cN)$ for the behaviour of $(\rho, \cM)$, such that $E_R(\sigma)$ is equal to the optimal value of~\eqref{eq: general upper bound from REE}. Therefore, $E_R(\sigma)$ is an upper bound on the DI-QKD key rate from $(\rho, \cM)$.

\revision{While these optimisations are difficult to solve exactly, any local-nonlocal decomposition would give a valid upper bound.} In the simpler case of CHSH-based protocols, this technique gives tighter bounds than~\cite{kaur_2020} for all CHSH values, and~\cite{arnon2021upper} in the low CHSH value regime.

\section{Finite-key analyses}
\label{sec: finite key}

In the preceding sections, we have mainly discussed key rates in the asymptotic setting. However, in a practical implementation, the number of rounds will always be finite, and hence it is important to address the question of proving that a protocol is secure in such a setting. We now briefly discuss various techniques for achieving this. Readers who are interested in this topic can find a more detailed introduction in e.g.~\cite{tan_thesis}.

The main task is to ensure that the security definition~\eqref{eq: eps secure} holds, and as noted in the Introduction, it is convenient to do so by separately arguing that correctness (Eq.~\eqref{eq:correct}) and secrecy (Eq.~\eqref{eq:secret}) are satisfied. Let $\vb{S}$ denote Alice's raw key (possibly after sifting, noisy pre-processing or random post-selection\footnote{More precisely: for random post-selection in particular, it is possible in principle to perform finite-size analysis under an i.i.d.~assumption, but it is currently not known how to prove it is secure in the non-i.i.d.~case --- we return to this point after discussing the entropy accumulation theorem.}), and let $E$ denote all of Eve's quantum side-information over the course of the protocol (this is a change of notation from previous sections, where it only denoted her side-information in a single round). Let $\transcript$ be the transcript of the public communication, which in general would depend on the protocol description.
However, as in our discussion in Section~\ref{sec: techniques}, in many protocols this transcript can be viewed as consisting of two parts: a string $\vb{T}$ of announcements $T_j$ made in individual rounds (for instance, Alice and Bob's setting choices $X_j Y_j$ in each round), followed by some additional communication  $\transcript_{\text{EC}}$ at the end for error correction and verification. Furthermore, we shall again focus on the case where Bob directly produces a guess for Alice's string $\vb{S}$ (and they immediately apply privacy amplification on the resulting values), rather than having Alice and Bob use public communication to modify \emph{both} their raw strings to some common values. (The latter situation can be more complicated and we do not discuss it here.)

For such protocols, the length of secret key that can be extracted via privacy amplification can then be characterised using the conditional smooth min-entropy, as mentioned in the Introduction. More precisely, we can ensure the secrecy condition holds as long as the final key length is chosen to be slightly shorter than $H_\text{min}^{\epsilon_s}(\vb{S}|\transcript E)$ (where $\epsilon_s \in (0,1)$ is a smoothing parameter chosen depending on the desired value of the secrecy parameter $\varepsilon_\text{sec}$), so a key task in the security proof is to lower-bound this quantity. For transcripts  $\transcript$ with the structure mentioned above, we have $H_\text{min}^{\epsilon_s}(\vb{S}|\transcript E) = H_\text{min}^{\epsilon_s}(\vb{S}|\vb{T} \transcript_\text{EC} E)$.
To bound this, a convenient approach is to use a chain rule for smooth entropies, which states an intuitive lower bound
\begin{equation} \label{eq: chain rule}
    H_\text{min}^{\epsilon_s}(\vb{S}|\transcript E) \geq H_\text{min}^{\epsilon_s}(\vb{S}|\vb{T} E) 
    - \text{len}(\transcript_{\text{EC}}),
\end{equation}
where $\text{len}(\transcript_{\text{EC}})$ is the length of $\transcript_{\text{EC}}$ (in bits). In other words, publicly communicating the register $\transcript_{\text{EC}}$ decreases the smooth min-entropy by at most $\text{len}(\transcript_{\text{EC}})$ bits. We briefly remark that this chain rule is a fully general bound, but in the context of security proofs, it is most suited for application in protocols of the form we described above rather than more elaborate procedures. 

For convenience in this description, we shall focus on protocols where $\transcript_{\text{EC}}$ can be further broken down into two parts: first, a string of length at most some value $\text{leak}_\text{EC}$ to allow Bob to produce a guess for Alice's string $\vb{S}$, second, a hash of $\vb{S}$ with length $\ceil{\log_2 (1 / \varepsilon_\text{cor})}$ to allow error verification (by having Bob compare it to the hash of his guess). 
Importantly, if the latter is chosen from a two-universal hash family, then it is straightforward to show (from the properties of two-universal hashing) that this procedure immediately ensures the protocol is $\varepsilon_\text{cor}$-correct, without \emph{any} conditions on how Bob produced his guess or what the actual device behaviour was. Some earlier error-correction procedures, e.g.~in~\cite{renner_thesis}, used other approaches to ensure the correctness condition, but applying them requires some technical conditions that we do not discuss here.

Therefore, with such an error verification procedure, it is possible to simply choose the value $\text{leak}_\text{EC}$ based on the \emph{honest} behaviour of the devices --- essentially, this is only required in order to ensure {completeness} of the protocol (i.e.~that the abort probability is low in the honest case), and is not involved at all in proving correctness or secrecy.\footnote{It is worth stressing that since we are focusing on fixed-length protocols, and the length of the final key will depend on $\text{leak}_\text{EC}$, this value must be fixed before the protocol begins, \emph{not} chosen adaptively during the protocol. As noted above, the error verification procedure described here ensures the correctness condition holds even without choosing $\text{leak}_\text{EC}$ adaptively.}
For the same reason, performing information reconciliation in this fashion also implies that e.g.~for protocols where the generation rounds are characterised by a QBER value $Q$ in the honest case, Alice and Bob do not need to estimate the value of $Q$ in the actual implementation; they can simply use the honest value.
Since the honest behaviour is usually i.i.d., one can then apply well-known classical protocols for error correction in this form, which usually require 
\begin{equation}
\text{leak}_\text{EC} \approx n f_\text{EC}\, h_2(Q),
\end{equation}
for an ``efficiency'' prefactor $f_\text{EC}$ that lies between 1.1 and 1.2 for typical sample sizes in QKD implementations. A more detailed treatment of the finite-key case can be found in e.g.~\cite{tomamichel2017fundamental}, but in the asymptotic limit, we simply have $f_\text{EC} \to 1$.

With this form of information reconciliation procedure, we see that the length of $\transcript_{\text{EC}}$ is upper bounded by $\text{leak}_\text{EC} + \log_2 (2 / \varepsilon_\text{cor})$. 
By the chain rule~\eqref{eq: chain rule}, this yields the following bound:
\begin{multline}
    H_\text{min}^{\epsilon_s}(\vb{S}|\transcript E) \geq H_\text{min}^{\epsilon_s}(\vb{S}|\vb{T} E) \\
    - \text{leak}_\text{EC} - \log_2 \frac{2}{\varepsilon_\text{cor}}.
\end{multline}
Thus, to prove the secrecy of the protocol, it is roughly\footnote{Though as emphasised in~\cite{tomamichel2017largely}, a rigorous security proof would need to account for the fact that conditioning on different steps of the protocol accepting can change the entropy; see e.g.~Lemma~10 in that work for one approach to handle this.} sufficient to find a lower bound on the conditional smooth min-entropy $H_\text{min}^{\epsilon_s}(\vb{S}|\vb{T} E)$ --- with that, we would have a lower bound on $H_\text{min}^{\epsilon_s}(\vb{S}|\transcript E)$, which characterises the length of $\varepsilon_\text{sec}$-secret key that can be extracted in the privacy amplification process, e.g.~via the Quantum Leftover Hash Lemma~\cite{tomamichel2011leftover}.

However, finding a lower bound on $H_\text{min}^{\epsilon_s}(\vb{S}|\vb{T} E)$ is still very challenging 
when non-i.i.d.~behaviour\footnote{In the i.i.d.~case, one can use the quantum asymptotic equipartition property~\cite{tomamichel2009fully} to obtain a bound of the form~\eqref{eq: single round reduction}, after applying an appropriate analysis of the parameter-estimation aspect; see e.g.~\cite{renner_thesis,tan_thesis}.} has to be taken into account.
Fortunately, it turns out that for a large class of protocols, we can find bounds of the form
\begin{equation} \label{eq: single round reduction}
    H_\text{min}^{\epsilon_s}(\vb{S}|\vb{T} E) \geq nh - O(\sqrt{n}),
\end{equation}
where the constant $h$ can be constructed by analysing individual rounds of the protocol instead of analysing all the rounds at once. In the next subsection, we shall discuss a technique known as the entropy accumulation theorem (EAT) that allow us to establish bounds of the form \eqref{eq: single round reduction}, and after that we will briefly discuss other techniques that have been used to analyse the finite-key security of DI-QKD. 

\subsection{Entropy accumulation theorem}

The entropy accumulation theorem is designed to apply to ``sequential'' device behaviours, in the sense that in each round, the device measurements can act not only on the state received from Eve in that round, but also additional registers that store (quantum or classical) ``memory'' from previous rounds. In particular, this implies for instance that the outputs in each round can depend on the inputs in all earlier rounds; however, they cannot depend on inputs in future rounds, i.e.~there is a form of time-ordering condition. While some proof techniques have been constructed for correlations that do not even have such a time-ordering condition (as we shall discuss later), this sequential model should already be sufficiently general to cover all currently plausible realisations of DI-QKD, in which the measurements are performed round-by-round.

\newcommand{\fmin}{f_\mathrm{min}}
\newcommand{\regPE}{Z}
\newcommand{\alphPE}{\mathcal{Z}}

With this in mind, suppose that the parameter-estimation aspect of the protocol is described in the following fashion: in every round, some function of the input and output values is computed and recorded in a classical register $\regPE_j$ with alphabet $\alphPE$ (which is the same in each round), and the parameter-estimation step accepts if the frequency distribution computed from the string $\vb{\regPE}$ lies within some set of ``acceptable'' distributions on $\alphPE$. (As a simple example, for a protocol based on the CHSH game in the test rounds, $\regPE_j$ can simply be a register that records whether the round was a test round, and if so, whether the CHSH game was won or lost in that round. More generally, we could consider other functions of the input/output values, up to and including simply recording all the input and output values in full.)

The core component of the EAT is constructing a function defined on distributions on the alphabet $\alphPE$, called the \textit{min-tradeoff function}. Note that $\alphPE$ is the alphabet for a single round, i.e.~this construction is centred around analysing the rounds individually.
To give a qualitative description in the context of DI-QKD, a min-tradeoff function is roughly\footnote{There are other technical details in the full definition of a min-tradeoff function, such as requiring it to be convex or affine depending on the EAT formulation, and some issues regarding what values the $Z_j$ registers can depend on. However, we do not discuss them here.} a function $\fmin$ with the following main property: for every round $j$, if the state and measurements in that round were to produce some distribution $q$ on the register $\regPE_j$, then the von Neumann entropy $H(S_j| T_j R)$ against any purification $R$ (of the state before measurement) is at least $\fmin(q)$. In other words, for any round, $\fmin$ is a lower bound on the von Neumann entropy as a function of the distribution produced in that round --- this is basically the bound being formulated in Eq.~\eqref{eq:vNopt}, hence justifying our earlier claim that solving that optimization is also essentially sufficient to allow a finite-size analysis against general attacks. It is worth emphasising that in this context, $q$ is a distribution involving only an individual round. As such, it is an ``abstract'' quantity that cannot be directly observed, unlike the final frequency distribution on the full string $\vb{\regPE}$. 

This brings us to the main technical advance offered by the EAT, which is that it connects the ``abstract'' individual-round analysis to a bound on $H_\text{min}^{\epsilon_s}(\vb{S}|\vb{T} E)$ that is expressed in terms of the observed string $\vb{\regPE}$. Specifically, the EAT can be used to derive the following statement: for any event $\Omega$ on $\vb{\regPE}$, if there is a constant $h$ such that $\fmin(\mathrm{freq}_{\vb{\regPE}}) \geq h$ for all values of $\vb{\regPE}$ in $\Omega$, then the final state conditioned on $\Omega$ satisfies a bound of roughly the form~\eqref{eq: single round reduction}\footnote{In the full derivation, there is a small additional correction due to technicalities regarding Bob's registers, but this correction also vanishes in the asymptotic limit.}. (The implicit constants in the $O(\sqrt{n})$ term in this case are functions of $\Pr[\Omega]$, the smoothing parameter $\epsilon_s$, and some properties of $\fmin$ --- this final point is why we had to introduce an ``intermediate'' object $\fmin$ rather than directly introducing the lower bound $h$ on the von Neumann entropy.) Notice that $\fmin$ could be constructed by analysing only the individual rounds, whereas $\Omega$ is an event defined in terms of the actual observed value of $\vb{\regPE}$, and hence the EAT serves the crucial purpose of relating the two frameworks.

The EAT does come with a technical restriction, in the form of a particular Markov condition 
--- informally, this condition states that the publicly announced data in the $j^\text{th}$ round must not ``leak'' any information about the device outputs in preceding rounds. 
On the abstract level, some condition of roughly this form is necessary, because without it, the public announcement in the $j^\text{th}$ round might simply be a copy of the previous round's outputs, making the scenario completely insecure (even with a nontrivial min-tradeoff function).

Fortunately, this condition is trivially satisfied for most of the basic DI-QKD protocols, where the $j^\text{th}$-round public announcement consists only of the input values $X_j Y_j$, which are generated using trusted randomness and completely independent of previous rounds. However, some later protocols such as the random post-selection protocol~\cite{xu2021device, liu2021high} have public announcements that do not necessarily fulfill this condition, and it is currently unclear how to construct a security proof for such protocols against non-i.i.d.~attacks.

\revision{To understand why this is the case, let us recall that, \textit{a priori}, Eve could introduce correlations between the measurement outcomes of different rounds. 
Crucially, in the random post-selection protocol of~\cite{xu2021device, liu2021high}, the decision of whether each round is discarded/accepted is based on the measurement outcomes of that round (along with some additional randomness), and this decision is then publicly announced. 
This announcement could hence be correlated to the private data that was generated in the preceding rounds, which violates the Markov condition of the EAT. Therefore, it does not seem straightforwardly possible to use the EAT to prove the security of DI-QKD protocols with post-selection involving the measurement outcomes, whether it is random~\cite{xu2021device,liu2021high} or deterministic~\cite{thinh2016randomness}. In fact, the explicit non-i.i.d.~attack that was derived in \cite{thinh2016randomness} relies on multi-round correlation in the outcomes in the sense we have described here.

We also highlight that in contrast, this point is not a concern for the noisy pre-processing protocol~\cite{ho2020noisy,woodhead2021device}
(as described in Section~\ref{sub: noisy preprocessing}), 
or more broadly any protocols that involve applying stochastic maps on the measurement outputs to generate the raw key, as long as the protocol has the property that the announced data in each round is independent of the private data generated (in preceding rounds). In particular, for the noisy pre-processing protocol, the only information announced in each round is the same as in the standard DI-QKD protocol, namely the measurement settings of each party. These settings are taken from a trusted random number generator (and independent across different rounds), which in particular ensures that they are independent of the private data in preceding rounds. Therefore, the EAT can be applied to such protocols.}

\subsection{Other techniques}
Another technique that can be useful in deriving lower bounds on the conditional smooth-min entropy is called the \textit{quantum probability estimation} (QPE) technique~\cite{knill2018quantum,zhang2020efficient, zhang2020experimental}. Similar to the EAT, the technique relies on the fact that most protocols are implemented sequentially, 
and it reduces the analysis to
a single round of the protocol. However, when QPE is used, the mathematical object of interest is a quantity known as the \emph{sandwiched $\alpha$-R\'enyi power}, which is to be upper-bounded in the security proof. 
This is done by deriving the so-called \textit{quantum estimation factors} (QEF), which serve to yield a lower bound on the smooth min-entropy. Importantly, when the quantum Markov chain condition is satisfied, it is possible to analyse the QEF for all the rounds by analysing the QEF for a single round of the protocol~\cite{knill2018quantum, zhang2020efficient, zhang2020experimental}. However, so far, the quantum probability estimation technique has only been used in the context of randomness generation and not in QKD. Importantly, in the context of DI randomness generation, the QPE method offered some advantage in terms of the minimum number of rounds required as compared to the original EAT~\cite{arnon2018practical}, especially for systems with low CHSH violation. It would be interesting to see if the same improvement can be observed in the context of DI-QKD.

Yet another proof technique was recently developed using an argument based on complementary observables~\cite{zhang2021quantum}. Roughly, this technique focuses on protocols where a reduction to qubit systems can be achieved using Jordan's lemma. With such a reduction, there is a well-defined notion of a complementary observable (in the sense that a qubit $X$ measurement is complementary to a qubit $Z$ measurement) for Alice's key-generating measurement in each round. That work uses this notion to reduce the analysis to a situation similar to earlier device-dependent security proofs~\cite{shor2000simple, koashi2009simple} based on \textit{phase errors} (i.e.~the rounds in which Alice and Bob's outcomes would have disagreed \emph{if} they had both measured in the complementary basis to the key-generating measurement). A core technical step in their work was an argument to relate the observed CHSH score to the distribution of phase errors, which allowed them to apply those proof techniques.

We now also briefly highlight some earlier techniques used for finite-key proofs (against coherent attacks). All of these techniques had the limitation that they generally gave lower asymptotic key rates than those discussed above (which yield asymptotic key rates that match the i.i.d.~case).
For instance, an approach based on entropic uncertainty relations was put forward in~\cite{lim2013device} (for a somewhat modified form of DI-QKD protocol, although the technique could be generalised to the standard DI-QKD setting). Another approach was presented in~\cite{vazirani2014fully}, deriving a
lower bound on the conditional smooth min-entropy for devices with sequential behaviour (similar to the EAT, but with a lower asymptotic rate). 

There have also been security proofs~\cite{jain2020parallel, vidick2017parallel} in the more challenging scenario of parallel-input behaviour, where Alice and Bob supply all their inputs at once, and the outputs can depend on the input choices in all the rounds. This is a very general form of device behaviour, and it is in fact currently an open question whether the same asymptotic key rates as the i.i.d.~case can even be achieved at all in this setting. Nonetheless, the proof techniques in those works can, in principle, be generalised even further to a DI-QKD scenario where the devices can leak a small amount of information about the inputs~\cite{jain2021direct}. 
\section{Outlook}
In theory, DI-QKD offers an information-theoretically secure method to distribute secret keys across distant parties with minimal assumption. In practice however, implementations of DI-QKD are still restricted to the the confines of state-of-the-art laboratories with sophisticated experimental techniques and setups that are far more advanced than commonly available experimental devices, let alone practical devices. While there have been a couple of demonstrations~\cite{nadlinger2021device, zhang2021experimental, liu2021high} recently, the transmission distance and the key rate are severely limited. From here, there are multiple research directions to pursue.

\subsection{Other frameworks for finite-key analyses}
In this review article, we have focused on finite-key analyses in the framework of EAT. As we have mentioned earlier, other frameworks for security analyses in the finite-key regime exist. In particular, the QPE framework~\cite{knill2018quantum,zhang2020efficient,zhang2020experimental} has been shown to provide a significant improvement in terms of the minimum number of rounds as compared to the EAT when applied to device-independent quantum randomness generation or expansion. However, the main bottleneck in applying this framework is the fact that we have yet to find an efficient method to construct the optimal QEF, as noted in~\cite{zhang2020efficient}. In fact, at the moment of writing, there do not appear to be any known method for constructing QEFs suitable for DI-QKD. In contrast, we have many systematic methods (as listed in Section~\ref{sec: techniques}) to construct min-tradeoff functions, and hence the EAT can be more readily applied.

On another note, a generalised version of the EAT has been recently applied in the context of DD-QKD~\cite{metger2022security, metger2022generalised}. As compared to the versions of EAT discussed in this review~\cite{dupuis2019entropy, dupuis2020entropy}, the generalised EAT uses a somewhat weaker assumption than the original Markov condition, and hence could be applied to a broader class of scenarios. As this generalised version still has the property that it is independent of the dimension of the underlying Hilbert space, it can also be applied to DI-QKD.

\subsection{Finite-key analyses of some protocols}
The DI-QKD protocol with random post-selection~\cite{xu2021device} offers a promising direction in which a fully-photonic setup can distribute secret keys if one assume that the block size is asymptotically large and the devices behave in an i.i.d.~manner. Indeed, the required specifications are not beyond the experimental state-of-the-art, as shown in the recent demonstration of the protocol~\cite{liu2021high}. Importantly, as the protocol can be implemented using a fully-photonic setup, the setup can be made simpler as no heralding system is required\footnote{Although to achieve longer transmission distances, qubit amplifiers may be required. See Section~\ref{sec: qubit amplifiers} for more details.}. Additionally, large block sizes can be easily achieved due to the higher clockrate of the devices as compared to the heralded entanglement setups. However, to achieve fully device-independent security, one has to extend the security analysis to the finite-key regime and against general attacks (i.e., without the i.i.d.~assumption). At the moment of writing, it is unclear whether any of the available finite-key proof techniques can be used to analyse such protocols or whether the key rate under general attacks would asymptotically converge to the value under collective attacks. That being said, it has been shown that a non-i.i.d.~attack can be more powerful than any i.i.d.~attack (even in the asymptotic limit) when a \textit{deterministic} post-selection strategy is employed~\cite{thinh2016randomness}, hence indicating some difficulties in trying to prove such a result. Another protocol in which security against general attacks is still an open question is the advantage distillation protocol studied in~\cite{tan2020advantage}.

\subsection{Protocol design: more inputs/more outputs}
Another interesting direction is designing protocols beyond the two-input-two-output scenario~\cite{gonzales2021device}. Most of the well-studied protocols with tight security bounds are based on the two-input-two-output scenario, which allows reduction to a qubit analysis via Jordan's lemma. A couple of other techniques~\cite{tan2021computing, brown2021computing,brown2021device} that do not rely on Jordan's lemma were also mainly applied to protocols with two binary measurements on Alice and three binary measurements on Bob\footnote{With notable exception of \cite{brown2021computing} where a three-output protocol was studied.}.
Given that the technique in~\cite{brown2021device} is efficient and versatile, and the resulting bounds are reasonably tight, it would be interesting to analyse the performance of DI-QKD protocols with more inputs and/or outputs.

\subsection{Trade-off between security and practicality}
Last but not least, although device-independent security is an appealing feature, perhaps full device-independence is not necessary for practical QKD. If security can be achieved under justifiable assumptions that go beyond those strictly necessary for DI-QKD, that is often sufficient for all intents and purposes. Indeed, device-dependent QKD is still a significant and active area of research. As mentioned in the Introduction, the invention of DI-QKD has given rise to the new sub-field of semi-DI quantum information processing, in which the quantum devices are partially characterised.

In practice, a well-chosen semi-DI framework in which the assumed features can be reasonably characterised and enforced might be a good trade-off between security and practicality. For example, a power limiter~\cite{zhang2021securing} that can passively enforce the constraint required in the energy-based semi-DI framework~\cite{vanHimbeeck2017semidevice} has been proposed. On a similar note, one-sided device-independent (1SDI-) QKD~\cite{branciard2012onesided,tomamichel2012tight} removes the need for any characterisation on the detection device used by one of the honest parties. Numerical toolboxes enabling security analysis of generic 1SDI-QKD protocols have been developed~\cite{wang2019characterising}, paving the way for practical protocols~\cite{ioannou2021receiver, ioannou2021receiver_protocols}. Furthermore, MDI-QKD~\cite{lo2012measurement, braunstein2012side} closes all detector-side-channels, allowing users to employ untrusted measurement devices as long as the state preparation is well-characterised. MDI-QKD also has an advantage in long-distance QKD experiments, as it is able to overcome the fundamental limit on repeaterless quantum communications~\cite{pirandola2017fundamental}. Other than the detectors, one can also relax the assumptions made on the sources~\cite{navarrete2021practical, zhang2021securing} in QKD protocols. \revision{Ultimately, however, the trade-off depends on the specific use case for QKD, and different use cases may require different levels of characterisation in order to reach an acceptable balance between practicality and security.} 

Besides relaxing the requirements on the devices by \revision{characterising them more fully}, one could also introduce computational assumptions on the devices, as studied in~\cite{metger2021device}. Those computational assumptions allowed the protocol in that work to achieve device-independent security \revision{without enforcing the no-signalling condition}. 
This could hence serve as an alternative to closing the locality loophole (or otherwise ensuring that the device inputs do not leak), although it would still be important to impose the condition that the device outputs are not leaked.
\section*{Acknowledgement and disclaimer}
 We acknowledge funding support from the National Research Foundation of Singapore (NRF) Fellowship grant (NRFF11-2019-0001), NRF Quantum Engineering Programme 1.0 grant (QEP-P2), and National Quantum-Safe Network project (NRF2021-QEP2-04-P01).
E.~Y.-Z.~Tan was funded by the Natural Sciences and Engineering Research Council of Canada (NSERC) Alliance, and Huawei Technologies Canada Co., Ltd.
We thank V\'ictor Zapatero and Gl\'aucia Murta for their valuable feedback.\newline

\textit{Disclaimer:} Charles Lim, in his present role at JPMorgan Chase \& Co contributed to this work for information purposes only. This work is not a product of the Research Department of JPMorgan Chase \& Co, or its affiliates. Neither JPMorgan Chase \& Co nor any of its affiliates make any explicit or implied representation or warranty and none of them accept any liability in connection with this work, including, but limited to, the completeness, accuracy, reliability of information contained herein and the potential legal, compliance, tax or accounting effects thereof. This document is not intended as investment research or investment advice, or a recommendation, offer or solicitation for the purchase or sale of any security, financial instrument, financial product or service, or to be used in any way for evaluating the merits of participating in any transaction.

\bibliography{references}

\end{document}